\documentclass[12pt,letterpaper]{article}

\usepackage{amsfonts} 
\usepackage{amssymb}
\usepackage{graphicx}
\usepackage{latexsym}
\usepackage{amsmath}
\usepackage{enumerate}
\usepackage{color}
\newcommand\openone{\leavevmode\hbox{\small1\normalsize\kern-.33em1}}
\def\ee{\end{eqnarray}}

\def\nn{\nonumber}

\newcommand{\half}{{\textstyle{\frac12}}}

\newcommand{\<}{\langle}
\renewcommand{\>}{\rangle}

\newcommand{\be}{\begin{eqnarray}}
\newcommand{\en}{\end{eqnarray}}
\newcommand{\bea}[1]{\left(\begin{array}{#1}}
\newcommand{\ena}{\end{array}\right)}
\newcommand{\ba}{\begin{eqnarray}}
\newcommand{\ea}{\end{eqnarray}}
\newcommand{\CL}{\mathcal{L}}
\newcommand{\CM}{\mathcal{M}}
\newcommand{\CO}{\mathcal{O}}
\newcommand{\CH}{\mathcal{H}}
\newcommand{\qv}{\vec{q}}
\newcommand{\sv}{\vec{S}}

\newcommand{\vv}{\vec{v}}
\newcommand{\vperp}{\vec{v}^\bot}

\usepackage{hyperref}

\newcommand\lrnab{\raise .8ex\hbox{$^\leftrightarrow$} \hspace{-9.8pt}
\nabla}
\newcommand\lnab{\raise .8ex\hbox{$^\leftarrow$} \hspace{-9.8pt}
\nabla}
\newcommand\rnab{\raise .8ex\hbox{$^\rightarrow$} \hspace{-9.8pt}
\nabla}

\newcommand\lrpar{\raise .8ex\hbox{$^\leftrightarrow$} \hspace{-9pt}
\partial}
\newcommand\lpar{\raise .8ex\hbox{$^\leftarrow$} \hspace{-9pt}
\partial}
\newcommand\rpar{\raise .8ex\hbox{$^\rightarrow$} \hspace{-9pt}
\partial}

\begin{document} 

\vspace{5mm}
\vspace{0.5cm}
\begin{center}

\def\thefootnote{\fnsymbol{footnote}}

{\Large \bf The Effective Field Theory of Dark Matter Direct Detection}
\\[0.5cm]
{\large  A.~Liam Fitzpatrick$^1$,  Wick Haxton${^2}$, Emanuel Katz$^{1,3,4}$, Nicholas Lubbers$^3$, Yiming Xu$^3$}


{\it $^1$ Stanford Institute for Theoretical Physics, Stanford University, Stanford, CA 94305} \\
{\it $^2$ Dept. of Physics, University of California, Berkeley, 94720,
and Lawrence Berkeley National Laboratory} \\
{\it $^3$ Physics Department, Boston University, Boston, MA 02215, USA} \\
{\it $^4$ SLAC National Accelerator Laboratory, 2575 Sand Hill, Menlo Park, CA 94025, USA.}


\end{center}

\vspace{.8cm}

\hrule \vspace{0.3cm}
{\small  \noindent \textbf{Abstract} \\[0.3cm]
\noindent

We extend and explore the general non-relativistic effective theory of dark matter (DM) direct detection. 
We describe the basic non-relativistic building blocks of operators and discuss their symmetry properties,
writing down all Galilean-invariant operators up to quadratic order in momentum transfer
arising from exchange of particles of spin 1 or less.  Any DM particle theory
can be translated into the coefficients of an effective operator and any effective operator can be simply related to
 most general description of the nuclear response.  We find several operators which lead to novel nuclear
responses.  These responses differ significantly from the standard minimal WIMP cases in their relative coupling strengths 
to various elements, changing how the results from different experiments should be compared against each other.  
Response functions are evaluated for common DM targets - F, Na, Ge, I, and Xe -
using standard shell model techniques.   We point out that each of the nuclear responses is familiar 
from past studies of semi-leptonic electroweak interactions, and thus potentially testable in weak interaction
studies.  We provide tables of the full set of required matrix elements at finite momentum transfer for a range of common
elements, making a careful and fully model-independent analysis possible.  
Finally, we discuss embedding non-relativistic effective theory operators into UV models of dark matter.

\vspace{0.5cm}  \hrule
\def\thefootnote{\arabic{footnote}}}
\setcounter{footnote}{0}
 
\section{Introduction and Summary of the Effective Theory}
\label{sec:intro}

The nature of dark matter is a fascinating mystery that continues to be unsolved.  Direct detection
experiments offer the possibility of determining the precise interactions of a dark matter (DM) particle 
with nuclei.  The experiments use different targets, potentially testing various 
types of interactions between the dark matter particle and the nucleus.  Previously, there have been 
attempts to exploit these differences between targets, in order to reconcile potential signals seen at some experiments
\cite{DAMALIBRA,COGENT2010}
with the absence of a signal at others.  However, studies are typically performed in a model-driven way, 
with the goal of putting constraints on a specific particle model.  Recently, \cite{fanEFT} took a different
approach, by considering the leading non-relativistic operators coupling DM to nuclei, and placing
bounds on their coefficients.  This was done in the context of elastic scattering.  In this paper 
we would like to extend this approach by going beyond the leading operators to include the full set
of possible operators in the non-relativistic theory, including momentum- and velocity-dependent operators as well.  
The goal of this study is to identify all possible elastic DM-nucleus response functions that may
be exploited by experimentalists to characterize DM, and to relate these responses
to the underlying effective theory operators that mediate the DM-ordinary matter interaction.

Models with momentum-dependent interactions have received some attention lately, as they have
helped lessen the tension between apparently conflicting direct detection experiments \cite{feldsteinfitzpatrickkatz,mddm,Feldstein:2009np,An:2010kc,barger}.  
However, from a bottom-up point of view, their appeal is much more general.  At the present moment, almost nothing
is known about the non-gravitational interactions of dark matter with the Standard Model, and in general, assumptions
about couplings are driven almost totally by appeals to minimalism or specific models of the electroweak scale.
Such principles are not necessarily a good guide as to what we should expect for the nature of dark matter,
and if the dark matter is instead as complicated as the Standard Model itself, then we may expect much
richer possibilities for its structure and interactions.  In particular, if the dark matter is composite, like
most of the visible matter in the universe is, then one should expect dark matter form factors related to their compositeness
scale. From this perspective, momentum-dependent interactions are a 
compelling and well-motivated possibility, since they require only a small amount of structure in the dark matter
sector.  The usual argument against such dependence is that it will be suppressed, since any momentum-dependent
terms will necessarily vanish in the limit of zero momentum transfer at direct detection experiments.  This however
neglects the possibility that the leading, momentum-independent interactions can easily be suppressed or forced
to vanish, leaving the momentum-dependent interactions as the dominant ones.

As a simple example of this kind of model, imagine that we have a gauge boson $A'_\mu$ that mixes kinetically with the photon $\CL_{\rm kin} = \epsilon F_{\mu\nu} F'^{\mu\nu}$ \cite{holdom,nima,philnatalia,pospelovkinetic}.  
Now, take the dark matter $\chi$ to be Majorana, so that a charge interaction with $A'_\mu$ is forbidden, whereas the anapole operator
$\bar{\chi} \gamma^\mu \gamma^5 \chi i g' A'_\mu$ is not, and can be generated when the $A'_\mu$ gauge symmetry is broken
\cite{darkmoments}.  In such a case, the four-fermion operator that is generated by integrating out $A'_\mu$ is 
the interaction $e g' \epsilon \bar{\chi} \gamma^\mu \gamma^5 \chi \bar{N} \gamma_\mu N/m_A^2$.  In the non-relativistic limit
and at zero momentum transfer, this is equivalent to
\ba 
\frac{e g' \epsilon}{m_N m_A^2} ( \bar{\chi } \vec{S}_\chi  \chi) \cdot (\bar{N} \lrpar N ),
\ea
which manifestly is suppressed by powers of momenta.  Another,  perhaps even more mundane, example is for the dark matter
to be a Dirac fermion that is a neutral composite particle made up of constituents charged under the $A'_\mu$ gauge force.
The charged constituents will cause $\chi$ to have a magnetic dipole moment of order their charge $Q'$ times the radius $r$ of the
bound object, i.e. $\CL_{\rm DM} \sim g' Q' r  \bar{\chi} \sigma^{\mu\nu} \chi F'_{\mu\nu} $.  Integrating out
the $A'_\mu$ again generates a four-fermion operator with momentum-dependence, $e g' \epsilon Q' r \bar{\chi} \sigma^{\mu\nu} q_\nu \chi
\bar{N} \gamma_\mu N /m_A^2$.  This can be seen to vanish in the limit of zero momentum, which occurs physically because
at long wavelengths the interaction averages over the charge of the constituents and sees only a neutral object.  
Some of the earliest considerations of such scenarios are \cite{pospelov,dine}. 

Rather than inventing all such possible models one by one, it is more efficient to pass directly to an effective field theory description.
Generally, such a description is the most natural and efficient tool to perform bottom-up, model-independent analyses.  
In this case, the appropriate effective field theory for direct
detection experiments involves a set of four-fermion operators for the interactions of dark matter with a nucleon
 in the nucleus in the non-relativistic limit.  The full set of such operators, being higher-order
in the momentum, have also not been considered previously.  Interestingly, as we will see,
some of these operators lead to novel nuclear responses, and therefore new form-factors are needed to describe DM interactions
with the nucleus.    In particular, direct detection should include not just spin-independent (SI) and spin-dependent (SD) interactions, but also 
angular-momentum dependent (LD), as well as spin and angular-momentum dependent (LSD)
interactions.  Under this new framework, the various elements used for direct detection, couple
with different strengths, depending on their nuclear properties.
It therefore becomes important to check whether current direct detection experiments
have a ``blind spot'' when combined.  Namely, whether there are any operators (or combinations of operators),
which render dark matter less visible to the currently available targets \cite{usinthefuture}.

Before describing the possible nuclear responses, let us first provide a quick summary of the non-relativistic effective theory
of nucleon-DM interactions.
Since we are interested in elastic scattering direct detection, all effective operators will be four-field operators, of the form
\ba
\CL_{\rm int} &=& \chi \CO_\chi \chi N \CO_N N .
\ea
The generalization to ``inelastic dark matter'' would involve allowing $\chi_1 \CO_\chi \chi_2$, with $\chi_1$ and $\chi_2$ having different masses; such a generalization should be straightforward to include, but we will not consider it further here.  
Passing to momentum space, we will take the incoming (outgoing) momentum of $\chi$ to be $p$ ($p'$) and of $N$ to be $k$ ($k'$).  
The form of possible interactions is constrained by several symmetries. In particular Galilean invariance imposes that the only combinations
of momentum that may appear are those made from the two invariants momentum transfer   $\vec{q} = p'-p$ and relative incoming velocity
$\vec{v}= \vec{v}_{\chi, \rm in} - \vec{v}_{N, \rm in}$. Interactions can contain the nucleon spin $\vec{S}_N$ and, if the dark matter carries intrinsic spin (for instance, if it is a fermion), then $\vec{S}_\chi$ as well.  Because the interaction must be Hermitian, it is useful to work with Hermitian quantities,
the complete set of which is
\ba
 && i \vec{q}, \ \ \ \ \vec{v}^\perp \equiv \vec{v} + \frac{\vec{q}}{2\mu_N} , \ \ \ \ \vec{S}_\chi, \ \ \ \ \vec{S}_N, 
 \ea
where the notation $\vec{v}^\perp$ is introduced because, by energy conservation, $\vec{v}^\perp \cdot \vec{q} =0$.  

We will work up to second-order in the momentum exchanged between the dark matter particle and the nucleus.
Also, we will limit ourselves to operators which
arise due to exchange of particles of spin one or less (i.e. at most quadratic in either $\sv$ or $\vv$).
In any Lorentz-invariant local quantum field theory, CP-violation is equivalent to T-violation, so let
us first consider operators that respect time reversal symmetry.  These
operators are
\be
& \bold{1},~~\sv_\chi \cdot \sv_N,~~v^2 , ~ ~i (\sv_\chi \times \qv) \cdot \vv, ~ ~i \vv \cdot (\sv_N \times \qv),~~( \sv_\chi \cdot \qv) (\sv_N \cdot \qv) 
\label{eq:geneftops}\\ \nonumber
& \vperp \cdot \sv_\chi,~~\vperp \cdot \sv_N, ~~i \sv_\chi \cdot (\sv_N \times \qv).
\ee
The operators in the first line of eq. (\ref{eq:geneftops}) are parity conserving, while those of the second line are parity violating. 
In addition, there are T-violating operators:
\be
&i \sv_N \cdot \qv, ~i \sv_\chi \cdot \qv, 
\label{eq:geneftops2} \\ \nonumber
&~(i \sv_N \cdot \qv) (\vperp \cdot \sv_\chi),~(i \sv_\chi \cdot \qv) (\vperp \cdot \sv_N).
\ee

In order to determine the interaction of DM particles with the nucleus, the above operators need to be inserted between nuclear states.  
Experimentally, the relevant question is thus what sort of nuclear responses these operators illicit when DM couples to the nucleus.
We find that there are six basic responses corresponding to single-nucleon operators labeled $M_{J;p,n}, ~\Sigma^\prime_{J;p,n},~\Sigma^{\prime \prime}_{J;p,n},~\Delta_{J;p,n},~\tilde{\Phi}^\prime_{J;pn},~\Phi^{\prime \prime}_{J;p,n}$ in our discussion of section 3.  Five of these responses
($M_{J;p,n}, ~\Sigma^\prime_{J;p,n},~\Sigma^{\prime \prime}_{J;p,n},~\Delta_{J;p,n},~\Phi^{\prime \prime}_{J;p,n}$) arise in CP
conserving interactions (due to the exchange of spin one or less), and we therefore primarily focus on this smaller set.  
Although a certain CP-violating interaction can be viable (see section 6), finding a UV-model which will result in the response
$~\tilde{\Phi}^\prime_{J;pn}$ seems more challenging.
In this paper we provide form factors in detail for some commonly used elements, however,
it is useful to have a heuristic description for the responses.  $M$ is the standard spin-independent response.
$\Sigma',~\Sigma''$ are the transverse and longitudinal (with respect to the momentum transfer) components of the nucleon spin (either $p$ or $n$).  
They favor elements with unpaired nucleons.  A certain linear combination of them is the usual spin-dependent coupling.  
$\Delta$ at zero-momentum transfer measures the net angular-momentum of a nucleon (either $p$ or $n$).  This response can be an important 
contribution to the coupling of DM to elements with unpaired nucleons, occupying 
an orbital shell with non-zero angular momentum.   Finally, $\Phi''$, at zero-momentum transfer is related to $(\vec{L}\cdot\vec{S})_{n,p}$.  
It favors elements with large, not fully occupied, spin-partner angular-momentum orbitals (i.e. when orbitals $j=\ell\pm\frac{1}{2}$ are not fully occupied). 
 As all these responses view nuclei differently, a completely model independent treatment of the experiments requires data to be considered for each response separately (up to interference effects). 

Our paper is organized as follows.  In section 2, we describe in detail the effective field theory, emphasizing the non-relativistic
 building blocks of operators and their symmetry properties, and demonstrate that the operators
in (\ref{eq:geneftops},\ref{eq:geneftops2}) describe the most general low-energy theory given our assumptions. In section 3, we discuss the relevant nuclear physics, and in particular we thoroughly analyze the possible nuclear response function in a partial wave basis, which is the standard formalism for such physics.  In section 4, we  give an overview of the various new nuclear responses, with an emphasis on 
their relative strength at different elements.  In section 5, we summarize these results in a format that can be easily read off and used in analyses of constraints from direct detection experiments.  
In particular, section 5 and appendix \ref{sec:nuclFF} contain the approximate necessary nuclear matrix elements and
form factors at the most experimentally relevant elements.  In section 6, we discuss possible models leading
to operators which have not been considered previously.  We conclude in section 7.

\section{Non-relativistic Effective Theory}

\subsection{Preliminaries}

Consider the following non-standard dark matter example \cite{feldsteinfitzpatrickkatz}.  Let dark matter be a complex scalar particle $\phi$ that is a bound state of two fermions
$\bar{\psi}_1\psi_2$, and with compositeness scale $\Lambda$.  To couple this to the Standard Model, introduce a new $U(1)$ gauge boson $A'_\mu$
of mass $m_A$ that kinetically mixes with the photon, $\CL \supset \epsilon F'_{\mu\nu} F^{\mu\nu}$.  If the fermions have equal and opposite charge under the
$A'_\mu$ gauge field, then $\phi$ will be neutral.  However, it will interact with $A'_\mu$ through the lowest-dimensional interaction that is not forbidden, which in this case is a charge radius coupling:
$ g'\partial^\mu \phi \partial^\nu \phi^* F'_{\mu\nu}/\Lambda^2.$  Integrating out the $A'_\mu$ generates, at leading order in momenta, the following interaction:
\ba
\frac{g' e \epsilon }{m_A^2 \Lambda^2}  p_{[\mu} p'_{\nu]} q^\mu   \phi^* \phi \bar{N} \gamma^\nu N,
\ea
where in this case the nucleon $N$ is the proton, and the brackets ${}_{[\mu \nu]}$ indicate the anti-symmetric component.  In the limit of small momentum transfer $q$, the $\nu=0, \mu=1,2,3$ terms dominate, and one is left with the
interaction
\ba
&& C \left( \frac{q^2}{\Lambda^2}\right)  \left( \phi^+ \phi^- N^+ N^- \right) , \ \ \ \ \  C = \frac{g' e \epsilon m_\phi m_N}{m_A^2} .
\ea
Here, $N^\pm, \phi^\pm$ are non-relativistic fields involving only creation or annihilation fields, i.e.  
\ba
N^-(y)  \equiv \int \frac{ d^3 k}{(2 \pi)^3} \frac{1}{\sqrt{2 m_N}} e^{-i k \cdot y} a^\dagger_k, && N^+(y) \equiv (N^-(y))^\dagger,
\ea
This example illustrates a few points that will be useful to keep in mind when we turn to a systematic description of the full effective
field theory. 
First of all, the leading interaction in this case is momentum-dependent, the leading standard interactions having been eliminated by charge assignments.
Second, it is just the first term in an expansion in powers of 
$q$ over the compositeness scale $\Lambda$.  In order for this expansion to make sense, $q$ must be less than $\Lambda$ over
the range of momentum transfers relevant at direct detection experiments.

The minimum possible cut-off $\Lambda_{\rm UV}$ on the effective theory is dictated by the relevant experiments, and must be at least as large
as the experimentally probed region of momentum transfer $\vec{q}$ between dark matter and atomic nuclei.  Direct detection experiments
directly measure the recoil energy $E_R$ of atomic nuclei, and for a target nucleus of mass $m_T$ this corresponds to
a specific momentum transfer $q = \sqrt{2 m_T E_R}$.  Consequently, the momentum transfer of any event is known up to
any uncertainty in the recoil energy and possibly the identity of the atomic element.  
In general, then, every experiment has a maximum momentum 
transfer that it is sensitive to.  This follows from the fact that larger momentum transfers require the dark matter to be incident
with greater velocity, at least $v_{\rm min} = q/2 \mu_T$.  The velocity distribution of the dark matter halo is expected to
fall off exponentially at around $v\sim 10^{-3}$, and essentially shut off completely at the escape velocity $v_{\rm esc} \sim 2 \times 10^{-3}$.  
Since the reduced mass $\mu_T$ is always less than the target mass $m_T$, and typical target masses are $m_T \sim 100$ GeV,
momentum transfers will rarely if ever be larger than 
\ba
q_{\rm max} \sim 200 \textrm{MeV} . 
\ea
So, in order for the effective theory to be a reliable description of 
direct detection experiments, one should have $\Lambda_{\rm UV} \gtrsim \textrm{few} \cdot  q_{\rm max}$.    Larger cut-offs
are of course allowed, though they will have correspondingly smaller cross-sections for dark matter scattering with nuclei.

\subsection{The Effective Theory}

We will now explore the effective theory in more detail, describing the essential ingredients and the full set of possible non-relativistic interactions.   The kinetic action is just the usual, non-relativistic form:
\ba
\CL_{\rm kin} &=& 2 m_\phi \phi^+(y) \left( i \frac{ \partial}{\partial t} - \frac{\vec{\nabla}^2}{2 m_\phi} \right) \phi^-(y).
\ea
By momentum-conservation, the momentum transfer $q$ is both 
\ba
q=p'-p=k-k'.  
\ea
There are several important symmetries that restrict the possible form of interactions.  The first of these is Galilean invariance, which is just
a constant shift in all velocities.  Thus, all momenta must appear through Galilean invariant combinations.  Between $p,k$ and $q$, there are
therefore only two independent momenta that can arise in any interaction.  It is easy to see that the momentum transfer $\vec{q}$ is Galilean invariant,
as is the relative incoming velocity 
\ba
\vec{v} \equiv \vec{v}_{\chi, \rm in} - \vec{v}_{N, \rm in},
\ea
 which is just the velocity of the incoming dark matter particle in the nucleon rest frame.  The final kinematic constraint is energy conservation.  This is easiest to impose by passing to the center-of-mass system, where the total kinetic energy is $E=\frac{1}{2} \mu_N v_{\rm rel}^2$, with $\mu_N = \frac{m_N m_\chi}{m_N+m_\chi}$ the dark-matter-nucleon reduced mass and $v_{\rm rel} = v_\chi - v_N$.  For the initial state energy, this is just $E_{\rm in} = \frac{1}{2} \mu_N v^2$, whereas for the final state energy it is $E_{\rm out} = \frac{1}{2}\mu_N (\vec{v} + \frac{ \vec{q}}{\mu_N} )^2$.  Imposing $E_{\rm in} = E_{\rm out}$ is therefore equivalent to
\ba
\vec{v} \cdot \vec{q} &=& -\frac{q^2}{2\mu_N}.
\ea
The next major constraint is Hermiticity of the interaction. This is essentially equivalent to crossing symmetry, because Hermitian conjugation
exchanges incoming for outgoing particles, i.e. $(\phi^-)^\dagger = \phi^+$.  Consequently, the momentum transfer $\vec{q}$ is effectively
anti-Hermitian, and it will be more convenient to work with the Hermitian operator $i \vec{q}$.  Under exchange of incoming and outgoing particles,
$\vec{v}$ does not have definite parity: $\vec{v} \stackrel{\dagger}{\rightarrow} \vec{v}_{\chi, \rm out} - \vec{v}_{N, \rm out} = \vec{v} + \frac{\vec{q}}{\mu_N}$.
However, we can easily construct a similar quantity that {\em is} Hermitian:
\ba
\vec{v}^\perp \equiv \vec{v} + \frac{\vec{q}}{2 \mu_N} .
\ea
The reason for this notation is that, by the energy-conservation condition above, $\vec{v}^\perp \cdot \vec{q} =0$.  Finally, we must include
the particle spins.
In the relativistic limit, this is just the familiar fact that four-fermion operators can contain $\gamma$ matrices.  In the non-relativistic limit, 
we can write down the dark matter and nuclear spins $\vec{S}_\chi$ and $\vec{S}_N$ as operators directly. Different possibilities for the spin of the dark matter are thus treated in a unified way.
 If dark matter is a spin-1/2 particle, then these spins operators are simply $\frac{1}{2} \vec{\sigma}$, where $\sigma^i$ are Pauli sigma matrices, acting on the $\chi$ and $N$ spinors; for vector dark matter, they
 are spin-1 representations of the angular momentum generators $J^i$ acting on the $\chi$ vector; and for scalars, they simply do not appear.  These are invariant under Hermitian conjugation, so we have for our complete 
 set of Galilean, Hermitian invariants the following:
 \ba
 && i \vec{q}, \ \ \ \ \vec{v}^\perp, \ \ \ \ \vec{S}_\chi, \ \ \ \ \vec{S}_N. 
 \ea
In addition to the above symmetries, there are strong constraints on violations of CP symmetry.  Since ultimately our non-relativistic theory must be embedded in
a Lorentz invariant quantum field theory, this is equivalent to T symmetry.  Spins behave like angular momentum, and thus change sign under T.
Also, velocities all change direction under T, so $\vec{v}^\perp$ and $\vec{q}$ change sign as well.  Finally, although we will not impose P as a symmetry,
it will be helpful to classify all operators according to whether they are even or odd under P.  In this case, spins do not change sign, whereas $\vec{v}^\perp$ and $\vec{q}$ do.  Thus, the complete set of Galilean invariants has the following transformation table:
\begin{center}
\begin{tabular}{|c|ccc|}
\hline
& $\dagger$ & $T$ & $P$ \\
\hline
$\sv$ & +1 & $-1$ & +1 \\
$i \qv$ & +1 & +1 & $-1$ \\
$ \vv^\perp$ & +1 & $-1$& $-1$\\
\hline
\end{tabular}
\end{center}

Since we are interested in elastic scattering direct detection, all effective operators will be four-field operators, of the form
\ba
\CL_{\rm int} &=& \chi^+ \CO_\chi \chi^- N^+ \CO_N N^- \equiv \CO \  \chi^+ \chi^- N^+  N^-  .
\ea
Furthermore, the momentum-transfer-squared $q^2$ is a completely invariant scalar quantity that depends only on dark matter kinematic quantities, and thus if ${\cal O}$ is an operator allowed by all symmetries of the theory, then
$q^{2n} {\cal O}$ is as well.  It is therefore natural to classify all such operators as a single one with a $q^2$-dependent coefficient, or form factor:
\be
c_0 {\cal O} + c_2 q^2 {\cal O} + c_4 q^4 {\cal O} + \dots &\equiv & F_{\cal O}\left( \frac{q^2}{\Lambda^2} \right) {\cal O} .
\ee
Massless mediators can be incorporated by including a $F_{\cal O} \sim q^{-2}$ term, though strictly speaking this is not an effective operator.  A related point is that
at the upper range of momentum at experiments, the pion should be included in the effective theory and $\chi$-$\chi$-$\pi$ couplings allowed.  For instance, if the underlying DM model contains couplings such as $\bar{\chi} \gamma_\mu \gamma^5 \chi J_3^{\mu 5}$ of DM to the axial current $J_3^{\mu 5} = i \bar{q} \gamma^\mu \gamma^5 \tau_3 q$, then the effective theory will couple $\chi$'s to pions due to the overlap of $J^{\mu 5}$ with $\pi$.  
Such interactions would contribute to dark matter-nucleon scattering through $t$-channel pion exchange at tree-level, effectively producing $F_{\cal O} \propto \frac{1}{q^2 + m_\pi^2}$ form factors in $\chi$-$\chi$-$N$-$N$ interactions.

So far, we have mainly discussed momentum scales.  In addition, there is an energy scale associated with the scattering process, of size
$\omega_q \sim q^2/2 m_T \lesssim 200$ keV. This is usually negligible, as the binding energy $\omega$ of nucleons is about 10 MeV per nucleon
for most elements, and inelastic transitions are kinematically suppressed.  However, for nuclei with small splittings $\sim \omega_q$ between the ground state and an excited state, it could affect direct detection rates. 

We are now ready to present the possible non-relativistic interactions.  The general Lagrangian is
\ba
\CL_{\rm int} &=& \sum_{N =n,p} \sum_i c_i^{(N)}  \CO_i \chi^+ \chi^- N^+ N^-,
\ea
with the following set of operators. Of the T-even operators, we have

\begin{enumerate}

\item P-even, $S_\chi$-independent
\ba
\CO_1 = \mathbf{1}, \ \ \ \ \ \CO_2 = (v^{\perp})^2, \ \ \ \ \ \ \CO_3 = i \vec{S}_N \cdot (\vec{q} \times \vec{v}^\perp) ,
\ea
\item P-even, $S_\chi$-dependent
\ba
\CO_4 = \vec{S}_\chi \cdot \vec{S}_N, \ \ \ \ \ \CO_5 = i \vec{S}_\chi \cdot (\vec{q} \times \vec{v}^\perp) , \ \ \ \ \ \ \CO_6= (\vec{S}_\chi \cdot \vec{q}) (\vec{S}_N \cdot \vec{q}) ,
\ea
\item P-odd, $S_\chi$-independent
\ba
\CO_7 = \vec{S}_N \cdot \vec{v}^\perp,
\ea
\item P-odd, $S_\chi$-dependent
\ba
\CO_8 = \vec{S}_\chi \cdot \vec{v}^\perp, \ \ \ \ \ \ \CO_9 = i \vec{S}_\chi \cdot (\vec{S}_N \times \vec{q})
\ea
 \newcounter{enumi_saved}
 \setcounter{enumi_saved}{\value{enumi}}

\end{enumerate}

In addition, we also have T-violating operators:

\begin{enumerate}

\setcounter{enumi}{\value{enumi_saved}}

\item P-odd, $S_\chi$-independent:
\ba
\CO_{10} = i \vec{S}_N \cdot \vec{q} ,
\ea
\item P-odd, $S_\chi$-dependent
\ba
\CO_{11} = i \vec{S}_\chi \cdot \vec{q}. 
\ea
\end{enumerate}

It is convenient to separate the operators as we have done above because each of these six groups of operators will not interfere with each other.
In addition, there are four operators that are products of the ones above: 
\be
\CO_{10}\CO_5, \quad \CO_{10}\CO_8, \quad \CO_{11} \CO_3, \quad \textrm{  and  }  \quad \CO_{11}\CO_7.
\label{eq:productops}
\ee 
With these, the above operators provide the most general effective theory at the dark matter-nucleon level
that can arise from exchange of a spin-0 or spin-1. In the completely general effective theory for elastic scattering, one would relax this condition and include arbitrary powers of $\vec{v}$ and $\vec{S}_\chi$, which would allow products of the operators we have written here and one additional operator $ \CO_{12} = \vec{S}_\chi \cdot (\vec{S}_N \times \vec{v}^\perp)$.  For instance, $\CO_7 \CO_8$ is a local operator that we have not written down above. However, quadratic powers of $\vec{S}_N$ and beyond (and $\vec{S}_\chi$ as well, if $\chi$ is spin-1/2) can always be reduced to at most linear powers by using the multiplication table of sigma matrices. In appendix \ref{app:matching}, we present the non-relativistic reduction of all relativistic operators arising from  a spin-0 or spin-1 exchange (or more precisely, with at most a single-index field exchange at tree-level) in terms of the local interactions above.  The product operators in eq. (\ref{eq:productops}) are seen to arise from a spin-1 particle coupling to fermion bilinears of the form $\bar{N} \lrpar^\mu \gamma^5 N$, which, for model-building concerns to be discussed in section \ref{sec:models}, we will not focus on further.  Nevertheless, it should be kept in mind that the general effective theory possible without any such restrictions contains these operators.  

In order to obtain the size of scattering cross-sections relevant to experiments, we need to 
evaluate matrix elements of the nucleon-level operators from the effective theory inside of a target nucleus. 
From the point of view of the effective field theory we have constructed, an atomic nucleus is a heavy, many-body
bound state of nucleons.  For the purpose of computing nucleon matrix elements inside such a nucleus, 
it is important to separate
out $\vec{v}^\perp$ into a term $\vec{v}^\perp_T$ that acts on the coherent center-of-mass velocity of the atomic nucleus as a whole,  and a term $\vec{v}^\perp_N$
that acts only on the relative distances of the nucleons within the nucleus.  We can write
\ba
\vec{v}^\perp = \frac{1}{2} \left( \vec{v}_{\chi, \rm in} + \vec{v}_{\chi, \rm out} - \vec{v}_{N, \rm in} - \vec{v}_{N, \rm out} \right)
 =  \vec{v}^\perp_T + \vec{v}^\perp_N,
\ea
where 
\ba
\vec{v}_T^\perp = \frac{1}{2} ( \vec{v}_{\chi, \rm in} + \vec{v}_{\chi,  \rm out} - \vec{v}_{T, \rm in} - \vec{v}_{T, \rm out} ) = \vec{v}_T + \frac{\vec{q}}{2 \mu_T}
\ea
acts only on the center of mass motion of the nucleus (here, $\vec{v}_T = \vec{v}_{\chi, \rm in} - \vec{v}_{T, \rm in}$ is the incoming dark matter velocity
in the lab frame).  Also, $\vec{v}_N^\perp$ is just
\ba
\vec{v}_N^\perp &=& -\frac{1}{2} (\vec{v}_{N, \rm in} + \vec{v}_{N, \rm out}),
\ea
but where the $\vec{v}_N$'s act only on the separation distance between the nucleons.  The reason for this separation is that $\vec{v}_T^\perp$
and $\vec{v}_N^\perp$ behave qualitatively differently, and have parametrically different sizes.  The former is determined by the kinematics of the dark-matter-nucleus scattering process, and does not require any detailed knowledge of the internal structure of the nucleus.  Its approximate magnitude is given by
\ba
m_T v_T \sim q .
\ea
Indeed, for elastic scattering,  by kinematics, $v_T$ must be strictly greater than $\frac{q}{2 m_T}$, and the event rate tends to be suppressed by the halo distribution if $v_T$ is significantly greater.   
As stated above, the typical size for $v_T$ is $\sim 10^{-3}$.  

On the other hand, $v_N^\perp$ depends on the internal distribution of nucleons in the nucleons and thus is determined by 
\ba
m_N \vec{v}_N^\perp \sim q.
\ea
This will lead to a relative kinematic enhancement of $m_T / m_N = A$ for $\vec{v}_N^\perp$ compared to $\vec{v}_T^\perp$.  In many cases this is cancelled by the fact that $\vec{v}_T^\perp$ tends to sum coherently over nucleons whereas $\vec{v}_N^\perp$ often does not, making the two terms comparable, 
but we will see some important exceptions.

 At low momentum-transfer, the internal structure of atomic nuclei can be summarized in just a finite number of macroscopic
quantities. In the case of the standard spin-independent interaction $\CO_1$ or spin-dependent interaction $\CO_4$, these are the atomic number $A$ and charge $Z$ or nucleon spin expectation values $\< S_n\> , \< S_p\>$, respectively.  However, there are many more possible macroscopic quantities
that appear associated with our full table of interactions than just these usual ones.  Furthermore, at finite momentum-transfer, there are multiple possible form factors associated with the nuclear responses that are required for calculating event rates.  Still, there are fewer independent nuclear responses than the full 
set of operators in the effective theory, so that a small number of plots can roughly capture the full range of possible models.  

In order to obtain these nuclear responses, one needs detailed input from nuclear physics on the wavefunctions of nucleons inside the nucleus.  In the
next section, we will therefore turn to the standard framework for the results of these computations, and a thorough discussion of possible nuclear responses.  We will provide a mapping of the operators in the effective theory onto the nuclear response functions. 

In section \ref{sec:handles}, we will discuss how these nuclear responses favor different elements.   The reader whose immediate goal is to use the resulting form factors in order to compute specific experimental event rates will find the relevant results summarized in section \ref{sec:presentation} and appendix \ref{sec:tables}.  A series of nuclear physics calculations of moderate complexity have been carried out in order to illustrate the kinds
of variations among nuclear target responses one should expect, given the unknown nature of DM-nuclear
interactions and the range of effective theory possibilities.  While we would characterize our nuclear structure
calculations as reasonable -- based on the shell model, using realistic interactions that have been ``vetted" in
related electroweak studies, and employing bases of reasonable size (ranging up to $\sim$ 0.7M Slater determinants,
after applications of symmetries) -- we also hope our results will
motivate others to train
even more sophisticated nuclear structure technology on this problem.
For example, the relevant isotopes of Ge span a region in neutron number where a sharp spherical-to-deformed transition occurs,
accompanied by fascinating quantum-mechanical level-crossing phenomena and associated sharp changes 
in proton and neutron spectroscopic factors.   This paper will provide those specialists with tools necessary to tackle such problems an important additional motivation for undertaking new work, 
its relevance to ongoing experiments that address one of the most important open questions in particle astrophysics, 
the nature of DM.

\section{The Nuclear Responses}
\label{sec:nuclear}
The non-relativistic effective theory treatment of dark matter responses connects naturally to the standard
language of multipole expansions for nuclear electroweak responses that we summarize here. 
Such expansions allow one to exploit nuclear selection rules based
on rotational invariance, parity and time reversal.  We specialize here to the case of elastic dark
matter interactions, as the energy transfers in dark-matter scattering generally preclude inelastic excitations.
(However, the extension of the multipole formalism to inelastic dark matter interactions is straightforward and will
be presented elsewhere.)
The good approximate parity and CP of the nuclear ground state then impose important new selection
rules on the possible elastic operators,
restricting the multi-polarities that contribute as well as their interference.

\subsection{Nuclear Charges and Currents}
The usual construction of coordinate-space nuclear charge and current densities in electroweak
interactions begins with a covariant interaction that is reduced to produce the needed non-relativistic
operators.  The effective theory approach significantly simplifies this analysis, and also
provides important guidance to those who might want to follow a model-dependent analysis based
on some specified covariant interaction.  Our starting point is the interaction
\begin{eqnarray} 
\label{eq:A}
\CL_{EFT} =a_1 1 +a_2 \vec{v}^\perp \cdot \vec{v}^\perp+  a_3 \vec{S}_N \cdot (\vec{q}  \times \vec{v}^\perp) + a_4 \vec{S}_\chi \cdot \vec{S}_N +
i a_5 \vec{S}_\chi \cdot (\vec{q} \times \vec{v}^\perp) + a_6 \vec{S}_\chi \cdot \vec{q}   \vec{S}_N \cdot \vec{q} \nonumber \\
+ a_7 \vec{S}_N \cdot \vec{v}^\perp + a_8 \vec{S}_\chi \cdot \vec{v}^\perp +i a_9  \vec{S}_\chi \cdot (\vec{S}_N \times \vec{q})+ i a_{10} \vec{S}_N \cdot \vec{q}+ i a_{11} \vec{S}_\chi \cdot \vec{q}~~~~~~~~~~~~~~
\end{eqnarray}
 As we have discussed previously,  the Hermitian velocity $v^\perp$ can be divided into a target center-of-mass
piece $\vec{v}^\perp_T$ and components $\vec{v}^\perp_N$ associated with the relative velocities of target 
nucleons (and thus with the A-1 Jacobi momenta).  In combination with nuclear spins, these two velocities 
generate interactions that are separately invariant under Galilean transformations.   Before any models
of nuclear charges and currents are introduced, Eq. (\ref{eq:A}) shows that from the available nuclear degrees
of freedom - nuclear spins and relative momenta $\vec{v}^\perp_N$ -- one can construct the 
nuclear ``charges" 1, $\vec{v}^\perp_N
\cdot \vec{v}^\perp_N$, and $\vec{S}_N \cdot \vec{v}^\perp_N$ that transform under parity and time-reversal
as even-even, even-even, and odd-even, and nuclear ``currents" $\vec{v}^\perp_N$, $\vec{S}_N$, and 
$\vec{S}_N \times \vec{v}^\perp_N$ that transform as odd-odd, even-odd, and odd-even.  Given our 
assumption of a nuclear ground state with good parity and CP, this leads us to conclude that
there must be six independent nuclear response functions corresponding to the even multipoles of 
a vector-like charge operator $M_{JM}$, the odd multipoles of axial longitudinal $L^5_{JM}$, axial transverse
electric $T_{JM}^\mathrm{el5}$, and vector magnetic $T_{JM}^\mathrm{mag}$ operators, and the even
multipoles of vector-like longitudinal $L_{JM}$ and transverse electric $T_{JM}^\mathrm{el}$ operators.   
To go further -- to provide explicit forms for these multipole operators -- we must make nuclear model assumptions.
Our construction defines the spins and momenta 
 $S_N$ and $v^\perp_N$ as the local operators associated with nucleons.  This one-body
definition is the most common starting point for nuclear physics calculations.

We do an explicit example -- the axial-charge operator $\CO_7 = \vec{S}_N
\cdot \vec{v}^\perp$ --  to demonstrate the procedure for separating $v^\perp$ into its $v^\perp_T$ and
$v^\perp_N$ components and for constructing the associated nuclear operators.
First, in the elementary-particle point limit for a nucleus --
where the nucleus is characterized only by its macroscopic quantum numbers of charge, spin, and isospin --
we have
\begin{equation}
\label{eq:axialCM}
O_7 = \vec{v}^\perp \cdot \vec{S}_N  \xrightarrow[\mathrm{point}]{~} \vec{v}^\perp_T
\cdot \vec{S}_N = \vec{v}^\perp_T
\cdot {1 \over 2} \sum_{j=1}^A \vec{\sigma}(j) \mathrm{~where~} \vec{v}^\perp_T \equiv \frac{1}{2} \left(\vec{ v}_{\chi, \rm in} + \vec{v}_{\chi, \rm out} - \vec{v}_{T, \rm in} - \vec{v}_{T, \rm out} \right).
\end{equation}
(A factor of one-half is introduced in relating
the nuclear spin $\vec{S}_N$ to  nucleon Pauli spin operators.)
The center-of-mass nuclear velocities $\vec{v}_{T,\mathrm{in}}$ and $\vec{v}_{T,\mathrm{out}}$ can be
obtained by averaging over the velocities of the nucleons,
\begin{equation}
\vec{v}_{T,\mathrm{in}} = {1 \over A} \sum_{j=1}^A \vec{v}_{N,\mathrm{in}}(j) \mathrm{~~and~~}
\vec{v}_{T,\mathrm{out}} = {1 \over A} \sum_{j=1}^A \vec{v}_{N,\mathrm{out}}(j).
\end{equation}

The nuclear model-building assumption is that the underlying nuclear charge and current operators
are one-body and local, in this case the sum over the individual
(symmetrized and thus Hermitian) axial charge operators.  This operator can be explicitly separated into
its center-of-mass (which contributes to $\vec{v}^\perp_T$) and intrinsic components (the $\vec{v}_N^\perp$ contribution)
\begin{eqnarray}
\sum_{j=1}^A \vec{\sigma}(j) \cdot { \vec{p}(j)  \over 2 m_N} &=& {1 \over 2} \vec{v}_T \cdot \sum_{j=1}^A \vec{\sigma}(j)  + 
{1 \over 2 A m_N} \sum_{j>k=1}^A (\vec{\sigma}(j)-\vec{\sigma}(k)) \cdot ( \vec{p}(j)-\vec{p}(k) ) \nonumber \\
&\equiv& {1 \over 2} \vec{v}_T  \cdot \sum_{j=1}^A \vec{\sigma}(j)  +\left[  \sum_{j=1}^A \vec{\sigma}(j) \cdot { \vec{p}(j)  \over 2 m_N} \right]_{intrinsic}
\end{eqnarray}
The first (target recoil) term is already identified in Eq. (\ref{eq:axialCM}) while the second provides
an explicit definition for the
contribution associated with the A-1 relative Jacobi three-momenta and thus with $\vec{v}^\perp_N$.
 While in principle the first, explicitly Galilean invariant form for the intrinsic axial-charge operator -- a
two-body operator in relative coordinates -- could be used in calculations, in many cases the 
simpler one-body form can be employed provided the center-of-mass motion of the nucleus is properly treated.
Thus the interpretation of the $intrinsic$ subscript
on the axial-charge operator above is an instruction that such steps should be taken, if this form of the
operator is employed. This is typically done by working  in a 
translationally-separable (e.g., full shell) harmonic oscillator Slater determinant basis, then numerically
forcing the center-of-mass to reside in the $1s$ state: the limitations of this approach are discussed in the appendix.  

An advantage of the effective theory treatment is that it immediately identifies the translationally invariant 
recoil axial charge contribution to dark matter scattering proportional to $\vec{v}^\perp_T$.  For elastic scattering, this is the only contribution of the
axial-charge: matrix elements of the intrinsic operator vanish for even multipoles by parity and for
odd multipoles by time reversal.  This result would be more difficult to obtain in conventional treatments
that begin with a covariant interaction.   Such calculations would need to extract the recoil term
from the axial-charge operator (in contrast to having it manifestly in the point-nucleus limit of the
effective operator), a task often requiring the combining of charge and current contributions.  For example, consider 
the example of  a V-A four-fermion contact operator between dark matter and a nucleus,
$ \bar{\psi}_\chi \gamma_\mu \psi_\chi~\bar{\psi}_N \gamma^\mu \gamma^5 \psi_N$.
Defining  $l_\mu^V = \bar{\psi}_\chi\gamma_\mu  \psi_\chi$, the charge and three-current contributions to scattering are
\begin{eqnarray}
&&l_0^V  \cdot  \sum_{j=1}^A \sigma(j) \cdot { \vec{p}_f(j)  + \vec{p}_i(j)\over 2 m_N} \rightarrow \frac{\vec{v}_{T,\rm in} + \vec{v}_{T, \rm out}}{2} \cdot \sum_{i=1}^A \vec{\sigma}(i)+ \left[  \sum_{j=1}^A \sigma(j) \cdot { \vec{p}_f(j)  + \vec{p}_i(j)\over 2 m_N} \right]_{intrinsic} \nonumber \\
&&-\vec{l}^V \cdot   \bar{\psi}_N \vec{\gamma} \gamma_5 \psi_N \rightarrow   - \frac{\vec{v}_{\chi, \rm in} + \vec{v}_{\chi, \rm out}}{2} \cdot \sum_{i=1}^A \vec{\sigma}(i),
\end{eqnarray}
where we have inserted the spin operator as the non-relativistic limit of the axial three-current operator.  Indeed
we get the right answer: summing the two terms yields a
contribution proportional to $\vec{v}^\perp_T$ as well as the intrinsic operator.  But in
contrast to the effective theory treatment --where the
target contribution is immediate from the point-nucleus limit and appears as one term -- a certain degree of care is
needed to locate and regroup terms into Galilean invariants.

We can now handle the general case of Eq. (\ref{eq:A}), first arranging the various terms as follows
\begin{eqnarray}
&&\CL_{ET} = l_0~ 1 + l_0^A~ [-2 \vec{v}^\perp_N \cdot \vec{S}_N] + \vec{l}_5 \cdot [2\vec{S}_N ]+\vec{l}_M \cdot [-\vec{v}_N^\perp]+\vec{l}_E \cdot [2i ~\vec{v}_N^\perp \times \vec{S}_N] \nonumber \\
&&=   l_0  1 + l_0^A \left( {\vec{p}_i + \vec{p}_f \over 2 m_N}\right) \cdot \vec{\sigma}  + \vec{l}_5 \cdot \vec{\sigma} 
+ \vec{l}_M \cdot \left( {\vec{p}_i  + \vec{p}_f  \over 2m_N} \right)+ \vec{l}_E \cdot \left(-i~ {\vec{p}_i + \vec{p}_f \over 2m_N} \times \vec{\sigma} \right)
\end{eqnarray}
where the coefficients of the charge ($l_0$), axial charge ($l_0^A$), axial vector ($\vec{l}_5$),  
vector magnetic ($\vec{l}_M$), and vector electric ($\vec{l}_E$) densities, determined from Eq. (\ref{eq:A}), will be
given below.  This expression follows Eq. (\ref{eq:A}) exactly apart from one simplification, the elimination of the term
proportional to $\vec{v}_N^\perp \cdot \vec{v}_N^\perp$ within  $\CO_2$.  
This interaction transforms
as a parity- and time-reversal-even charge of $o(v/c)^2 \sim$ 1\%, and thus will be overwhelmed by the 
$o(v/c)^0$ coherent spin-independent response, if the latter is present. Furthermore, from a model-building point of
view, it is difficult to see how one could generate the former while avoiding the latter, without significant
fine-tuning.

The charge and current operators can be transformed to coordinate space via the substitution
\begin{eqnarray}
{\vec{p}_i + \vec{p}_f \over 2m_N} \rightarrow {1 \over 2m_N}\left( -{1 \over i} \overleftarrow{\nabla} \delta(\vec{x}-\vec{x}_i)+\delta(\vec{x}-\vec{x}_i) {1 \over i} \overrightarrow{\nabla} \right)
\end{eqnarray}
Thus we determine the Hamiltonian density
\allowdisplaybreaks
\begin{eqnarray}
\label{eq:currents}
\CH_{ET}(\vec{x}) &=&
\sum_{i=1}^A  l_0(i)~ \delta(\vec{x}-\vec{x}_i) 
+ \sum_{i=1}^A l_0^A(i)~ {1 \over 2M} \left[-{1 \over i} \overleftarrow{\nabla}_i \cdot  \vec{\sigma}(i)\delta(\vec{x}-\vec{x}_i) +\delta(\vec{x}-\vec{x}_i) \vec{\sigma}(i)  \cdot  {1 \over i} \overrightarrow{\nabla}_i \right]  \nonumber \\
 &+& \sum_{i=1}^A  \vec{l}_5(i) \cdot \vec{\sigma}(i) \delta(\vec{x}-\vec{x}_i) +   \sum_{i=1}^A \vec{l}_M(i) \cdot {1 \over 2M} \left[-{1 \over i} \overleftarrow{\nabla}_i\delta(\vec{x}-\vec{x}_i) +\delta(\vec{x}-\vec{x}_i){1 \over i} \overrightarrow{\nabla}_i \right]  \nonumber \\
&+& \sum_{i=1}^A \vec{l}_E (i) \cdot {1 \over 2M} \left[ \overleftarrow{\nabla}_i \times \vec{\sigma}(i) \delta(\vec{x}-\vec{x}_i) +\delta(\vec{x}-\vec{x}_i)  \vec{\sigma}(i) \times \overrightarrow{\nabla}_i \right] 
\end{eqnarray}
where the dark-matter amplitudes $l_0(i)$ and $\vec{l}(i)$ appear within the sum over nucleons because we will
allow the various couplings in Eq. (\ref{eq:A}) to have a nontrivial isospin dependence, e.g., $a_1 \rightarrow 
( a_1^0 + a_1^1\tau_3(i))$  (so that $a_1^0=a_1^1=a_1/2$ will correspond to a coupling only to protons of
strength $a_1$ while $a_1^0=-a_1^1=a_1/2$ will correspond a similar coupling only to neutrons).

The Hamiltonian for Eq. (\ref{eq:currents}) has the familiar form
\begin{equation}
\label{eq:planewave}
\int d \vec{x} ~e^{-i \vec{q} \cdot \vec{x}}~ \left[ l_0  \langle J_i M_i | \hat{\rho}(\vec{x}) | J_i M_i \rangle - \vec{l} \cdot
\langle J_i M_i | \hat{\vec{j}}(\vec{x}) | J_i M_i \rangle \right]
\end{equation}
where $\vec{q} $ 
is the three-momentum transferred from the nucleus
to the scattered DM particle.
One substitutes Eq. (\ref{eq:currents}) into
Eq. (\ref{eq:planewave}) and uses the spherical harmonic and vector spherical harmonic identities
\allowdisplaybreaks
\begin{eqnarray}
\label{eq:spherical}
e^{i \vec{q} \cdot \vec{x}_i} &=& \sum_{J=0}^\infty \sqrt{4 \pi} ~[J]~ i^J j_J(q x_i) Y_{J0}(\Omega_{x_i}) \nonumber \\
\hat{e}_\lambda e^{i \vec{q} \cdot \vec{x_i}} &=&\left\{  \begin{array}{lr} \displaystyle\sum_{J=0}^\infty \sqrt{4 \pi}~ [J]~ i^{J-1} {\vec{\nabla}_i \over q} j_J(qx_i) Y_{J0}(\Omega_{x_i}), &\lambda = 0 \\ ~ \\ \displaystyle\sum_{J \ge 1}^\infty \sqrt{2 \pi} ~[J]~ i^{J-2} \left[
\lambda j_J(qx_i)  \vec{Y}_{JJ1}^\lambda(\Omega_{x_i}) +{ \vec{\nabla}_i \over q} \times j_J(qx_i) \vec{Y}_{JJ1}^\lambda
(\Omega_{x_i}) \right], &\lambda=\pm 1 \end{array} \right.
\label{eq:vecsphharmId}
\end{eqnarray}
to project out charge multipoles and longitudinal, transverse magnetic, and transverse electric current multipoles,
respectively.  This defines the operators that generate the nuclear form factors 
describing the scattering of dark matter from nuclei.  Here $[J] \equiv \sqrt{2J+1}$ and $\hat{e}_\lambda, ~\lambda=1,0,-1$, are spherical unit vectors defined with respect to a
quantization z-axis along $\hat{q} \equiv \vec{q}/q$.  As detailed in the appendix, the multipole 
operators transform simply under parity and time reversal, allowing one to exploit 
selection rules to simplify the diagonal nuclear matrix elements of interest, assuming CP- and parity-violating
components in  nuclear ground-state wave functions are negligible.   Consequently, averaging over 
initial nuclear spins and summing over
final, one finds the general form of the dark-matter elastic scattering probability:
\allowdisplaybreaks
\begin{eqnarray}
\label{eq:Ham}
&&{1 \over 2J_i+1} \sum_{M_i,M_f} | \langle J_i M_f | ~H~ | J_i M_i\rangle |^2 =   {4 \pi \over 2J_i + 1} \left[   \sum_{J=1,3,,...}^\infty| \langle J_i||~\vec{l}_{5} \cdot \hat{q}~ \Sigma^{\prime \prime}_J(q)~ || J_i \rangle|^2 \right. \nonumber \\
&+& \left.\sum_{J=0,2,...}^\infty  ~\left\{ ~ | \langle J_i||~l_0~ M_{J}(q)~ || J_i \rangle|^2 +  |\langle J_i ||~ \vec{l}_E \cdot \hat{q}~ {q \over m_N}\Phi^{\prime \prime}(q)~ ||J_i \rangle |^2 \right. \right. \nonumber \\
&+&\left. \left.  2 \mathrm{Re} \left[  \langle J_i ||~ \vec{l}_E \cdot \hat{q}~{q \over m_N}\Phi^{\prime \prime}(q)~ ||J_i \rangle \langle J_i ||~ l_0~ M_J (q) ~|| J_i \rangle^* \right]  \right \} \right.~~ \nonumber \\
&+& \left. {q^2 \over 2m_N^2 }   \sum_{J=2,4,...}^\infty \left(  \langle J_i || ~ \vec{l}_E~\tilde{\Phi}_J^\prime(q)~ || J_i \rangle \cdot  \langle J_i ||  ~\vec{l}_E~ \tilde{\Phi}^\prime_J(q)~ || J_i \rangle^*-| \langle J_i || ~ \vec{l}_E \cdot \hat{q}~\tilde{\Phi}^\prime_J(q)~ || J_i \rangle |^2 \right) \right. \nonumber \\
&+&\left. \sum_{J=1,3,...}^\infty \left. \Big\{ {q^2 \over 2m_N^2}  \left(\langle J_i ||~ \vec{l}_M ~ \Delta_J(q)~ || J_i \rangle \cdot \langle J_i ||~ \vec{l}_M ~ \Delta_J(q)~ || J_i \rangle^*- | \langle J_i ||~ \vec{l}_M \cdot \hat{q}~  \Delta_J(q)~ || J_i \rangle |^2 \right) \right. \right. ~\nonumber \\
&+&\left. \left. {1 \over 2}  \left( \langle J_i ||~ \vec{l}_{5}~ \Sigma_J^\prime(q)~ || J_i \rangle \cdot  \langle J_i || ~\vec{l}_{5}~ \Sigma_J^\prime(q)~ || J_i \rangle^*- | \langle J_i ||~ \vec{l}_{5} \cdot \hat{q}~ \Sigma^\prime_J(q)~ || J_i \rangle |^2 \right) \right. \right.\nonumber \\ 
&+&\left. \left. 2 \mathrm{Re} \left[ i \hat{q} \cdot \langle J_i || ~\vec{l}_M~{q \over m_N} \Delta_J(q) ~||J_i \rangle \times  \langle J_i || ~\vec{l}_{5}~\Sigma^\prime_J(q)~ || J_i  \rangle^* \right]  \right. \Big\} \right.  \Bigg]
\end{eqnarray}
All nuclear matrix elements are intrinsic: contributions proportional to $v_T^\perp$ reside entirely
in the dark-matter amplitudes $l_0$, $\vec{l}_5$, $\vec{l}_E$, and $\vec{l}_M$, by virtue of the Galilean invariant
effective theory.
In Eq. (\ref{eq:Ham}) $||$ denotes a nuclear matrix element reduced in angular momentum. The expression is 
somewhat schematic in that
\begin{equation}
\langle J_i||~l~ O_{J}(q)~ || J_i \rangle \equiv \langle J_i||~\sum_{i=1}^Al(i)~ O_{J}(q\vec{x}_i)~ || J_i \rangle,
\end{equation}
The notation is a reminder that the dark matter amplitude in general cannot be moved outside the nuclear
matrix element because that amplitude may contain several effective theory couplings with different
isospin dependences.  (There are many cases where this expression does factor, however, and we give the
simpler form appropriate for those cases below.)

Equation (\ref{eq:Ham}) shows that there are six distinct nuclear response functions governing dark-matter
responses, corresponding to six single-particle operators.  Each of these operators
is familiar from standard treatments of weak interactions \cite{Walecka75,Donnelly79} or,
in the case of $\tilde{\Phi}^\prime$ and $\Phi^{\prime \prime}$, from extensions \cite{Serot78} that have been made to account for
currents of order $1/m_N^2$.  They are constructed from the Bessel spherical harmonics
and vector spherical harmonics,
$M_{JM}(q \vec{x}) \equiv j_J(q x) Y_{JM}(\Omega_x)$ and $\vec{M}_{JL}^M \equiv j_L(q x) \vec{Y}_{JLM}$,
\allowdisplaybreaks
\begin{eqnarray}
\label{eq:operators}
M_{JM}(q \vec{x}) &&\nonumber \\
\Delta_{JM}(q \vec{x}) &\equiv& \vec{M}_{JJ}^M(q \vec{x}) \cdot {1 \over q} \vec{\nabla} 
 \nonumber \\
\Sigma^\prime_{JM}(q \vec{x}) &\equiv& -i \left\{ {1 \over q} \vec{\nabla} \times \vec{M}_{JJ}^M (q \vec{x}) \right\} \cdot \vec{\sigma} =  [J]^{-1} \left\{ -\sqrt{J}~\vec{M}_{JJ+1}^M(q \vec{x}) + \sqrt{J+1}~\vec{M}_{JJ-1}^M(q \vec{x}) \right\} \cdot \vec{\sigma} \nonumber \\
 \Sigma^{\prime \prime}_{JM}(q \vec{x}) &\equiv& \left\{ {1 \over q} \vec{\nabla}  M_{JM} (q \vec{x}) \right\} \cdot \vec{\sigma} =  [J]^{-1} \left\{ \sqrt{J+1}~\vec{M}_{JJ+1}^M(q \vec{x}) + \sqrt{J}~\vec{M}_{JJ-1}^M(q \vec{x}) \right\} \cdot \vec{\sigma}\nonumber  \\
\tilde{\Phi}^{\prime}_{JM}(q \vec{x}) &\equiv& \left( {1 \over q} \vec{\nabla} \times \vec{M}_{JJ}^M(q \vec{x}) \right) \cdot \left(\vec{\sigma} \times {1 \over q} \vec{\nabla} \right) + {1 \over 2} \vec{M}_{JJ}^M(q \vec{x}) \cdot \vec{\sigma} \nonumber \\
\Phi^{\prime \prime}_{JM}(q \vec{x}) &\equiv& i  \left( {1 \over q} \vec{\nabla}  M_{JM}(q \vec{x}) \right) \cdot \left(\vec{\sigma} \times {1 \over q} \vec{\nabla} \right)
\end{eqnarray}
The multipole operators have been defined to have a simple behavior under time reversal, transforming with
a $\pm$1, as discussed in the appendix.  Time reversal and parity impose important constraints on allowed
responses:  for reasons noted previously, there is no elastic dark matter coupling to the intrinsic axial charge density
(though the axial charge contribution due to $v_T^\perp$ remains and contributes through the spin density).
Those responses that do appear involve sums over either even or odd multipoles, again because of the
parity/time reversal constraints.
The long-wavelength limits of these operators, showing explicitly the character of the nuclear response
(charge or current;  transverse magnetic, transverse electric, or
longitudinal; vector-like or axial-vector like) are given in Table \ref{table:LWL}.  

\begin{table}
\begin{tabular}{|c|c|c|c|}
\hline
\hline
& && \\
Response $\times \left[ {4 \pi \over 2J_i+1} \right]^{-1}$ & Leading & Long-wavelength & Response   \\
 & Multipole & Limit & Type \\
\hline
$\displaystyle \sum_{J=0,2,...}^\infty |\langle J_i|| M_{JM}|| J_i \rangle|^2$ & $M_{00}(q \vec{x}_i)$ &$ {1 \over \sqrt{4 \pi}} 1(i) $ & $ M_{JM}:  \mathrm{Charge}$ \\
$\displaystyle \sum_{J=1,3,...}^\infty |\langle J_i|| \Sigma^{\prime \prime}_{JM}|| J_i \rangle|^2$ & $\Sigma^{\prime \prime}_{1M}(q \vec{x}_i)$ &$ {1 \over 2\sqrt{3 \pi}} \sigma_{1M}(i) $ &$\begin{array}{c}  L_{JM}^5: \mathrm{Axial} \\ \mathrm{Longitudinal} \end{array}$\\
$\displaystyle \sum_{J=1,3,...}^\infty |\langle J_i|| \Sigma^{ \prime}_{JM}|| J_i \rangle|^2$ & $\Sigma^{\prime }_{1M}(q \vec{x}_i)$ &$ {1 \over \sqrt{6 \pi}} \sigma_{1M}(i) $ &$\begin{array}{c} T_{JM}^{\mathrm{el5}}: \mathrm{Axial} \\\mathrm{Transverse~Electric} \end{array}$ \\
$\displaystyle \sum_{J=1,3,...}^\infty |\langle J_i||{q \over m_N} \Delta_{JM}|| J_i \rangle|^2$ & ${q \over m_N}\Delta_{1M}(q \vec{x}_i)$ &$-{q \over 2m_N\sqrt{6 \pi}} \ell_{1M}(i) $ &$\begin{array}{c} T_{JM}^\mathrm{mag}: \\ \mathrm{Transverse~Magnetic} \end{array}$ \\
$\displaystyle \sum_{J=0,2,...}^\infty |\langle J_i||{q \over m_N} \Phi^{\prime \prime}_{JM}|| J_i \rangle|^2$ & ${q \over m_N}\Phi^{\prime \prime}_{00}(q \vec{x}_i)$ &$-{q \over 3m_N\sqrt{4 \pi}} \vec{\sigma}(i) \cdot \vec{\ell}(i) $ &$\begin{array}{c} L_{JM}:\\ \mathrm{ Longitudinal} \end{array}$ \\
 &${q \over m_N}\Phi^{\prime \prime}_{2M}(q \vec{x}_i)$  & $-{q \over m_N\sqrt{30 \pi}} [x_i \otimes (\vec{\sigma}(i) \times {1 \over i} \vec{\nabla})_1]_{2M}  $ & \\
 $\displaystyle \sum_{J=2,4,...}^\infty |\langle J_i||{q \over m_N} \tilde{\Phi}^{ \prime}_{JM}|| J_i \rangle|^2$ & ${q \over m_N} \tilde{\Phi}^{ \prime}_{2M}(q \vec{x}_i)$ &$-{q \over m_N\sqrt{20 \pi}} [x_i \otimes (\vec{\sigma}(i) \times {1 \over i} \vec{\nabla})_1]_{2M} $ &$\begin{array}{c} T^\mathrm{el}_{JM}:\\ \mathrm{Transverse~ Electric} \end{array}$ \\
 \hline
\end{tabular}
\caption{The response dark-matter nuclear response functions, their leading order behavior, and the response type.
The notation $\otimes$ denotes a spherical tensor product, while $\times$ is the conventional cross product.}
\label{table:LWL}
\end{table}

Table \ref{table:LWL} includes the standard spin-independent response governed by even multipoles of the
(generalized in isospin) charge operator $M_{JM}$, and two spin-dependent responses that,
though proportional in the long-wavelength limit, are characterized by different nuclear form factors and 
couple to dark matter in distinct ways, with $\Sigma^{\prime \prime}_{JM}$ being longitudinal and
$\Sigma^\prime_{JM}$ transverse.   The associated form factors involve sums over all allowed odd
multipoles.  There is a third $J=1$ response, a transverse magnetic 
response governed by odd multipoles of $\Delta_{JM}$.  This and two other new responses are explicitly associated with nuclear
substructure.  $\Delta_{JM}$ is generated by the convection current (the nucleon velocity term), which in the
long-wavelength limit
produces a coupling to the nuclear orbital angular momentum operator $\vec{\ell}(i)$.
This provides a third  interaction -- an interaction like the two spin-dependent ones -- that will
transform under rotations as $\langle J_i | \vec{J}_M | J_i \rangle$.   The two other responses arising
from the constituent nature of the nucleus transform as longitudinal and electric projections of the 
density $\sim \delta(\vec{x}-\vec{x}_i) \vec{\sigma}(i) \times \vec{\nabla}$.  The first of these is quite
interesting as its long-wavelength limit produces a scalar proportional to the spin-orbit interaction $\vec{\sigma}(i) \cdot \vec{\ell}(i)$  as well as a tensor contribution.  The full form factor involves a sum over all even multipoles of $\Phi_J^{\prime \prime}$.
Because of the leading $\vec{\sigma}(i) \cdot \vec{\ell}(i)$ contribution, this response is
present for all nuclei (that is, regardless of ground-state spin, like the usual spin-independent
charge coupling), but because of its form factor (leading-order behavior is proportional to $q/m_N$) 
and spin-orbit nature, its properties are quite different from those of the usual spin-independent
scalar operator $M_{00}$.  $\Phi_{00}^{\prime \prime}$ can be important in heavy nuclei because 
$\vec{\sigma}(i) \cdot \vec{\ell}(i)$ produces a coherent isoscalar contribution over closed spin-orbit partner 
shells, e.g., the closed $1f_{7/2}$
shell for Ge isotopes and the closed $1g_{9/2}$ shell for Xe or I.   The operator
$ \Phi_J^{\prime \prime}$ transforms as the longitudinal projection of a vector current (denoted
$L_{JM}$ in Table \ref{table:LWL}).
Just as in the case of the nuclear spin density, where 
two distinct spin-dependent form factors are generated,  corresponding to the axial-like longitudinal
and transverse electric nuclear responses
(denoted $L^5_J$ and $T^\mathrm{el5}_J$ in Table \ref{table:LWL}), the density responsible for
$\Phi_J^{\prime \prime}$ also generates a transverse-electric response (denoted $T^\mathrm{el}_{JM}$)
that consequently transform as a $J=2$ operator in the long-wavelength limit.   From the specific form,
$ [ x_i \otimes (\sigma(i) \times {1 \over i} \nabla)_1]_{2M}$, one can see that this operator is
closely related to $\Phi_{00}^{\prime \prime} \rightarrow \sigma(i) \cdot \vec{\ell}(i)$ and is in fact proportional in the long-wavelength limit
to its tensor partner $\Phi^{\prime \prime}_{2M}$.  Thus the relationship of $\Phi_J^{\prime \prime}$
to $\tilde{\Phi}_J^\prime$ -- $L_J$ and $T^\mathrm{el}_J$ operators -- is analogous to that of $\Sigma_J^{\prime \prime}$
to $\Sigma_J^\prime$ -- $L^5_J$ and $T^\mathrm{el5}_J$ operators --
except that the transverse nature of
 $\tilde{\Phi}^{\prime}_{JM}$  excludes the possibility of a J=0 multipole.
We will see below that $\tilde{\Phi}^{\prime}_{JM}$ is an exotic response, arising only for dark matter
with unusual couplings.

The three composite operators arise from the fact that there are interesting current densities in the
nucleus that can mediate dark matter interactions, but fail to have the proper parity and time-reversal
properties to contribute in the point-nucleus limit.  As momentum transfers in dark
matter interactions are not small compared to the inverse nuclear size, these new responses can be
numerically quite important.  More important, we will see below that in many cases, these new responses
can provide the dominant coupling of dark matter to nuclei, depending of the effective theory operator.
 The long-wavelength limits of these three new responses are determined by operators that transform properly
under parity and time reversal because a factor of $\vec{q} \cdot \vec{x}_i$ has been convolved with
the underlying bare nuclear densities.  Consequently these operators have a leading-order form-factor behavior
proportional to $q/m_N$ and an explicit dependence on $\vec{x}_i$, and thus on the nuclear size.

\subsection{The Effective Theory Content}
Thus we can proceed to the dark-matter physics, which is encoded in the amplitudes $l_0$, $\vec{l}_5$,
$\vec{l}_M$ and $\vec{l}_E$ that are determined by our effective theory through Eq. (\ref{eq:A}).  We find
\begin{eqnarray}
\label{eq:ls}
l_0 &=& (a_1^0 + a_1^1 \tau_3(i)) -i (\vec{q} \times \vec{S}_\chi) \cdot \vec{v}_T^\perp ~(a_5^0 + a_5^1 \tau_3(i))
+ \vec{S}_\chi \cdot \vec{v}_T^\perp ~(a_8^0 + a_8^1 \tau_3(i))\nonumber \\
 &+& i \vec{q} \cdot \vec{S}_\chi ~(a_{11}^0 + a_{11}^1 \tau_3(i)) \nonumber \\
\vec{l}_5 &=& {1 \over 2} \left[ i \vec{q} \times \vec{v}_T^\perp~ (a_3^0 + a_3^1 \tau_3(i)) + \vec{S}_\chi ~(a_4^0 + a_4^1 \tau_3(i))
+ \vec{S}_\chi \cdot \vec{q} ~ \vec{q}~(a_6^0 + a_6^1 \tau_3(i)) + \right. \nonumber \\
&+& \left.  \vec{v}_T^\perp ~(a_7^0+a_7^1\tau_3(i))+ i \vec{q} \times \vec{S}_\chi ~(a_9^0 + a_9^1 \tau_3(i)) + i \vec{q}~(a_{10}^0 +a_{10}^1 \tau_3(i)) \right] \nonumber \\
\vec{l}_M &=& i \vec{q} \times \vec{S}_\chi ~(a_5^0+a_5 \tau_3(i)) - \vec{S}_\chi ~(a_8^0+a_8^1) \tau_3(i) \nonumber \\
\vec{l}_E &=& {1 \over 2} \vec{q}~ (a_3^0+a_3^1 \tau_3(i))
\end{eqnarray}

We observe that there is no coupling to the $T_J^\mathrm{el}$ nuclear density associated with the tensor operator $\tilde{\Phi}^\prime_{J=2}$.  
$\CO_3$ generates a nonzero $\vec{l}_E$, but it is longitudinal.  None of our eleven effective theory operators generates
a transverse component to $\vec{l}_E$.
This point relates to our decision to limit the effective theory to interactions that could arise from spin-0 or spin-1
exchanges.  Indeed, we pointed out the existence of an additional operator linear in dark-matter and nuclear spins
\begin{equation}
\CO_{12} = \vec{S}\chi \cdot (\vec{S}_N \times \vec{v}^\perp)
\end{equation}
that would need to be included if we were to relax conditions on the nature of the exchange.  With the inclusion of this
term,
\begin{equation}
\vec{l}_E \rightarrow {1 \over 2} \left[ \vec{q}~ (a_3^0+a_3^1 \tau_3(i))+i \vec{S}_\chi (a_{12}^0 + a_{12}^1 \tau_3(i))\right]
\end{equation}
leading to a contribution to the sixth potential nuclear response, the tensor one, governed by $\tilde{\Phi}^\prime$.  
$\CO_{12}$ would also produce a recoil contribution to the spin-dependent response functions.   Another example
of a term that produces a contribution to $\tilde{\Phi}^{\prime}$ is the
 four-fermion tensor/axial tensor interaction
\begin{eqnarray}
\label{eq:A1}
 \CL_\mathrm{tensor} = \bar{\psi}_\chi\sigma_{\mu \nu} (a_T^\chi -i a_{AT}^\chi \gamma_5) \psi_\chi~ \bar{\psi}_N \sigma^{\mu \nu} ( a_T^N -i a_{AT}^N \gamma_5) \psi_N
\end{eqnarray}

Inspecting Eq. (\ref{eq:ls}) one sees that there are several common situations in which our general result for the scattering
probability, Eq. (\ref{eq:Ham}), can be simplified by factoring the dark-matter amplitudes from the
nuclear matrix elements.   If we are interested in any one interaction $\CO_i$, then clearly its
associated isospin dependence can be written $a_i (1 + \alpha_i \tau_3)$.  The overall strength  $a_i$ could be associated 
with the dark-matter amplitudes in Eq. (\ref{eq:ls}), while the isospin factor could be included in the definition of the single-particle
operators of Eq. (\ref{eq:operators}).   Alternatively, several couplings might be nonzero, but all might share a
common behavior in isospin, e.g., all interactions coupling to protons.  The overall couplings could again be
incorporated into Eq. (\ref{eq:ls}), with the isospin dependence $(1+\tau_3(i))/2$ absorbed into the single particle
operators.  In such cases the scattering probability simplifies, taking the form

\allowdisplaybreaks
\begin{eqnarray}
\label{eq:Ham2}
&\longrightarrow&  ~{4 \pi \over 2J_i + 1} \left[   \sum_{J=1,3,,...}^\infty  \vec{l}_{5} \cdot \hat{q}~ \vec{l}_{5}^* \cdot \hat{q}~ | \langle J_i||~ \Sigma^{\prime \prime}_J(q)~ || J_i \rangle|^2 \right. \nonumber \\
 &+&\left. \sum_{J=0,2,...}^\infty  ~\left\{ ~l_0~ l_0{}^*~ | \langle J_i||~ M_{J}(q)~ || J_i \rangle|^2 + \vec{l}_E \cdot \hat{q} ~\vec{l}_E^*\cdot \hat{q}~ |\langle J_i || ~ {q \over m_N} \Phi_J^{\prime \prime}(q)~ ||J_i \rangle |^2 \right. \right. \nonumber \\
&+&\left. \left.  2 \mathrm{Re} \left[ \vec{l}_E \cdot \hat{q}~l_0^* ~\langle J_i || ~{q \over m_N} \Phi_J^{\prime \prime}(q)~ ||J_i \rangle \langle J_i || ~ M_J (q)~ || J_i \rangle^* \right]  \right \} \right. \nonumber \\
&+& \left. {q^2 \over 2m_N^2 } \left(\vec{l}_E \cdot \vec{l}_E^*-\vec{l}_E \cdot \hat{q}~ \vec{l}_E^* \cdot \hat{q} \right)  \sum_{J=2,4,...}^\infty  | \langle J_i ||~\tilde{\Phi}_J^\prime(q)~ || J_i \rangle |^2  \right.\nonumber \\
 &+& \left. \sum_{J=1,3,...}^\infty \left. \Big\{ {q^2 \over 2m_N^2}  \left( \vec{l}_M \cdot \vec{l}_M^*-\vec{l}_M\cdot \hat{q}~\vec{l}_M^* \cdot \hat{q} \right) |\langle J_i || ~\Delta_J(q)~ || J_i \rangle|^2  \right. \right.~~~~~~~~~~~~~~~~~~~~~~~~~~ ~\nonumber \\
&+& \left. \left. {1 \over 2}  \left( \vec{l}_{5} \cdot \vec{l}_{5}^* - \vec{l}_{5} \cdot \hat{q}~\vec{l}_{5}^* \cdot \hat{q} \right) |\langle J_i || ~ \Sigma_J^\prime(q)~ || J_i \rangle|^2 \ \right. \right. \nonumber \\ 
 &+&  \left. \left.2 \mathrm{Re} \left[ i \hat{q} \cdot \left( \vec{l}_M \times \vec{l}_{5}^* \right) \langle J_i || ~{q \over m_N} \Delta_J~ ||J_i \rangle  \langle J_i || ~\Sigma_J^\prime(q)~ || J_i  \rangle^* \right]  \right. \Big\} \right.  \Bigg]~~~~~~~~~~~~
\end{eqnarray}

As the multipole expansion is conventionally done in a coordinate system aligned along $\vec{q}$, another 
useful result is the expression for the Hamiltonian in a rotationally invariant form, in the long wavelength 
limit.  One finds
\begin{eqnarray}
\label{eq:interaction}
&\hat{H} = \displaystyle\sum_{i=1}^A \left\{  l_0 ~1(i) -{q \over 3M} ~\vec{l}_E \cdot \hat{q} ~ \vec{\sigma}(i)
\cdot \vec{\ell}(i)   -   \vec{l}_5  \cdot  \vec{\sigma}(i) \right. \nonumber \\
&\left.  -i {q \over 2M} ( \vec{l}_M \times \hat{q}) \cdot \vec{\ell}(i)
 +   {q \over M} (\vec{l}_E \otimes \hat{q} )_2  \cdot \left[ \vec{x}_i \otimes [ \vec{\sigma}(i) \times {1 \over i} \vec{\nabla}(i)]_1 \right]_2  \right\}
\end{eqnarray}
The third and fifth terms, proportional to $\vec{l}_5$ and $(\vec{l}_E \otimes \hat{q})_2$, can each
be divided into longitudinal and transverse electric pieces, which would be associated with
distinct nuclear form factors once one goes beyond the long wavelength limit.

\subsection{Response Function Evaluation}
The various response functions described above were evaluated in the shell model for
several of the key isotopes now used in dark-matter detectors.  Calculations
were performed for $^{19}$F, $^{23}$Na,  $^{70,72,73,74,76}$Ge, $^{127}$I, and
$^{128,129,130,131,132,134,136}$Xe.  Response functions were evaluated by
summing over the contributing isotopes, weighted according to their natural abundances.
(For F, Na, and I, there is a single stable isotope.)
Consequently, while all isotopes take part in scalar or spin-independent responses,
only those with ground-state spins $\ge$1/2 ($^{19}$F(1/2$^+$), $^{23}$Na(3/2$^+$), 
 $^{73}$Ge(9/2$^+$), $^{127}$I (5/2$^+$), $^{129}$Xe(1/2$^+$), and
$^{131}$Xe(3/2$^+$)) contribute to $J=1$ (or spin-dependent) responses, and 
only those with spins $\ge$ 1 ($^{23}$Na, $^{73}$Ge, $^{127}$I, and $^{131}$Xe)
contribute to the $J=2$ tensor response.  By defining the
nuclear responses per target atom for a detector made up of unenriched isotopes, 
we take into account the reduced efficiency of detectors for $J=1$ and $J=2$ responses 
due to noncontributing isotopes. The Ge calculations include the five stable isotopes, while
the Xe calculations summed over the seven principal isotopes, ignoring the 
trace ($\lesssim 0.1\%$) contributions from  $^{124,126}$Xe.

As our focus is a broad survey -- to understand the degree to which targets can vary
in the relative sensitivity to dark matter, given the broad range of response functions
that may govern that sensitivity -- the structure calculations we undertook were limited
to relatively small bases, and thus should be considered exploratory.  They
were performed in m-scheme bases on which we placed a limit of no more than 0.65 million
Slater determinants (after application of symmetries like time reversal to reduce basis 
dimensions).  The sd-shell calculations for $^{19}$F and $^{23}$Na are then unrestricted.
The interaction used was that of Brown and Wildenthal \cite{Brown88}.  The Ge isotopes
were treated in the standard $1f_{5/2} 2p_{1/2} 2p_{3/2} 1g_{9/2}$ model space above a $^{56}$Ni
core.  The basis truncation was based on limiting occupation of the $1g_{9/2}$ shell
to no more than two nucleons above the minimum occupation for all isotopes.  An
interaction developed by the Madrid/Strasbourg group was used \cite{Poves08,Poves09}.  These Ge
model spaces take into account some of the polarization effects that accompany 
the rather sharp spherical-to-deformed transition that occurs near neutron numbers 40-42
($^{72}$Ge, $^{74}$Ge), though we recognize the need for further expansion of the
basis in follow-up calculations.  Some of the motivating physics for more ambitious 
calculations -- particularly the rather complex interactions among spherical and 
deformed $0^+$ bands in the even isotopes -- is discussed in \cite{Haxton92}.

The $^{127}$I and Xe isotopes were treated in the $3s_{1/2} 2d_{3/2} 2d_{5/2} 1g_{7/2} 1h_{11/2}$
model space above a $^{100}$Sn core.  The interaction used was one developed some time ago by
Baldridge and Vary \cite{BV78} and employed in double beta decay studies in this mass region.  
The interaction is based on a G-matrix from the Reid soft-core potential augmented by phenomenological
pairing and multipole forces.
While the $^{134}$Xe and $^{136}$Xe calculations
were unrestricted, significant truncations became necessary for lighter Xe isotopes
where the neutron occupation of the  $1h_{11/2}$ shell drops.  The bases 
for $^{128,130,132}$Xe  and  $^{127}$I
were limited by fixing the $1h_{11/2}$ occupation to the minimum allowed
nucleon number.  Basis for the odd-neutron isotopes $^{131}$Xe and $^{129}$Xe were further 
restricted by limiting valence protons to the energetically favored $2d_{5/2}$ and
$1g_{7/2}$ shells, and by requiring neutrons to fully occupying these same shells (a choice that
preserves good isospin).  Less restrictive calculations can and should be done, but are beyond
the scope of the current survey.

From these wave functions the  ground-state to ground-state one-body density matrices can be generated
\begin{equation}
  \Psi^{J;T}_{i;i}(|\alpha|,|\beta|) = {1 \over [J] [T]} \langle J_i; T_f~\vdots \vdots \left[ a^\dagger_{|\alpha|} \otimes \tilde{a}_{|\beta|} \right]_{J;T} \vdots \vdots~J_i;T_i \rangle
  \end{equation}
  where $\tilde{a}_{|\beta|, m_j ,m_t }\equiv (-1)^{j_\beta -m_j + 1/2-m_t} a_{|\beta|,-m_j,-m_t}$,
  $|\alpha|$ denotes are nonmagnetic quantum numbers, $\otimes$ denotes a spherical tensor 
  product, and
  $\vdots \vdots$ indicates reduction in both angular momentum and isospin.  As we include
  all contributing multipoles, density matrices are needed for $0 \le J \le 2J_i$ and for $T=0,1$.  They provide
  the single-particle amplitudes needed for evaluating many-body matrix elements of any
  one-body operator,
 \begin{eqnarray}
 \label{eq:densitymatrix}
 \langle J_i; T_i ~\vdots \vdots~ \sum_{i=1}^A  \hat{O}_J(i) ~ \vdots \vdots ~J_i; T_i  \rangle &=& \sum_{|\alpha|,|\beta|} \Psi^{J;T=0}_{i;i} (|\alpha| ,|\beta|)~ \langle |\alpha| ~\vdots \vdots \hat{O}_J \vdots \vdots~ |\beta| \rangle \nonumber \\
 &=& \sqrt{2} \sum_{|\alpha|,|\beta|} \Psi^{J;T=0}_{i;i} (|\alpha|,|\beta|)~\langle |\alpha|~ || \hat{O}_J ||~ |\beta| \rangle \nonumber \\
  \langle J_i; T_i ~\vdots \vdots~ \sum_{i=1}^A  \hat{O}_J(i) \tau(i) ~ \vdots \vdots ~J_i; T_i  \rangle &=& \sum_{|\alpha| ,|\beta|} \Psi^{J;T=1}_{i;i}(|\alpha| ,|\beta|)~ \langle |\alpha| ~\vdots \vdots \hat{O}_J \tau \vdots \vdots~ |\beta| \rangle \nonumber \\
   &=& \sqrt{6} \sum_{|\alpha|,|\beta|} \Psi^{J;T=1}_{i,i}(|\alpha|,|\beta|)~ \langle |\alpha| ~|| \hat{O}_J ||~ |\beta| \rangle
  \nonumber \\
  ~~
  \end{eqnarray}
  where $\langle |\alpha| ~|| \hat{O}_J ||~ |\beta| \rangle$  is a single-particle space/spin matrix
  elements reduced in angular momentum.  By adopting a harmonic oscillator single-particle
  basis, these matrix elements can be evaluated analytically, yielding forms $\sim e^{-y} p(y)$
  where $p(y)$ is a polynomial in $y=(q b/2)^2$,
  $b$ is the oscillator parameter and $q$ the magnitude of the three-momentum transfer.  
  Consequently analytic expressions for nuclear form factors 
  can be provided, so that response functions can be easily evaluated 
  for changing experimental conditions, such as different WIMP masses, or for other choices of
  $b$.  Our numerical results were generated with the choices $b$=1.833, 1.835, 2.108, 2.282, and 2.292 fm
  for $^{19}$F, $^{23}$Na, the Ge isotopes, $^{127}$I, and the Xe isotopes, respectively.
  Harmonic oscillator matrix elements for four of the operators arising in dark-matter elastic scattering, 
  $M_{JM}(q \vec{x})$,  $\Delta_{JM}(q \vec{x})$,  $\Sigma^\prime_{JM}(q \vec{x})$, and  
  $\Sigma^{\prime \prime}_{JM}(q \vec{x})$ can be evaluated from a publicly available Mathematica
  script \cite{Cecilia08} (as these operators also arise in standard treatments of weak interactions).
  A generalization of this script that includes the two additional operators  $\tilde{\Phi}^\prime_{JM}(q \vec{x})$
  and $\Phi^{\prime \prime}_{JM}(q \vec{x})$ is available from the authors.

\section{Comparing the novel responses for different elements of interest }
\label{sec:handles}

The most important lesson that the general EFT of dark matter-nucleon elastic scattering has to teach us is
exactly what are all the phenomenological properties that distinguish nuclear recoil rates at different experiments.
Specific models or effective operators for dark matter interactions will be proportional to some particular combination
of these properties, which can enhance or diminish the relative sensitivity of different experiments.  Coupling through
atomic number $A$ or charge $Z$ in the standard spin-dependent case or through 
the proton or neutron spin in the standard spin-dependent case are by far the best-known examples. 
However, as we have seen in the previous section there are other possible nuclear responses.  In this section
we would like to explore the less familiar responses, $\Delta$ and $\Phi''$.

\subsection{$\Delta_p$ and $\Delta_n$}

\begin{figure}
\begin{center}
\includegraphics[width=0.9\textwidth]{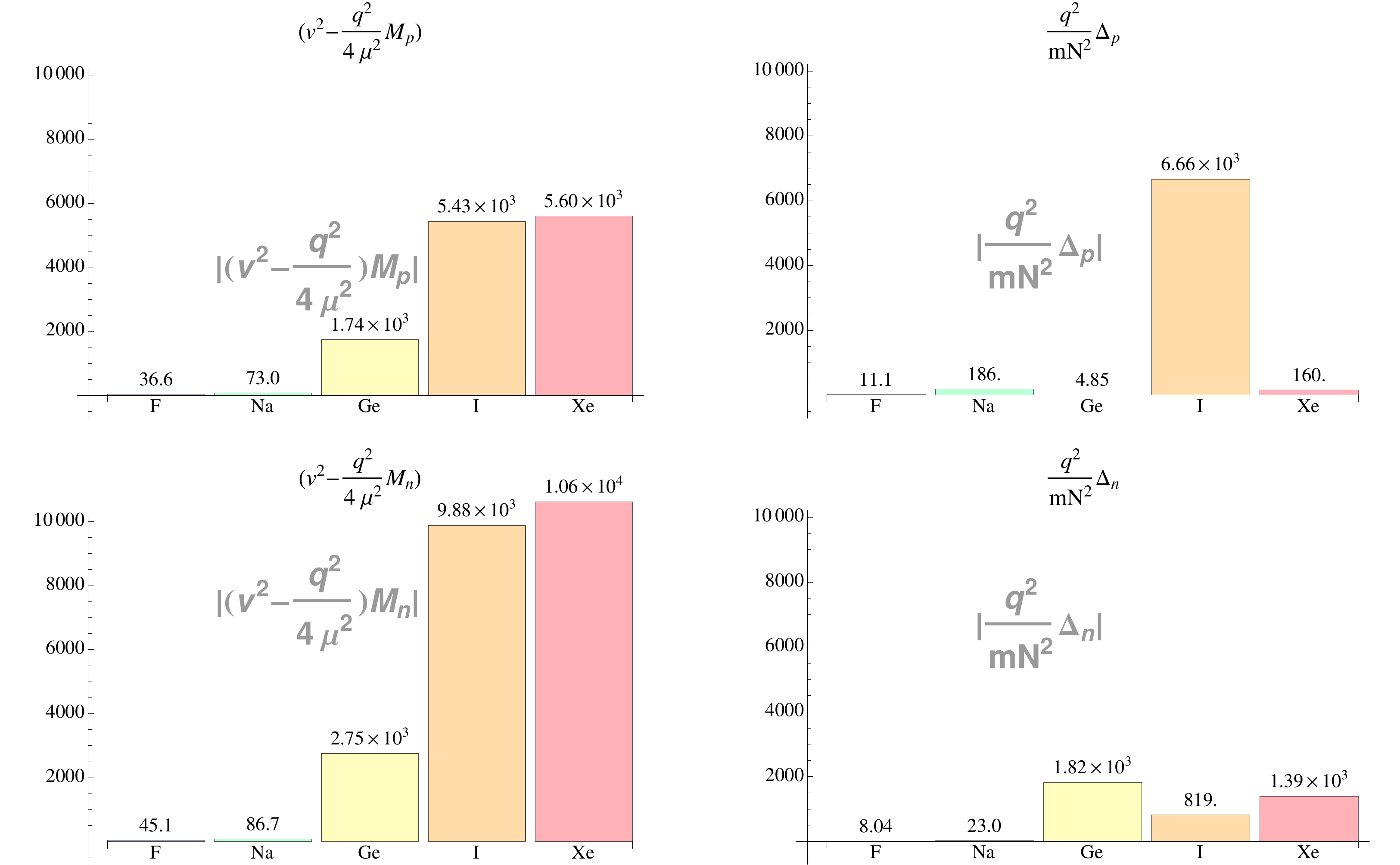}
\end{center}
\caption{Size of integrated responses $\int_{v_{\rm min}=\frac{q}{2\mu_T}}  d^3 v \frac{f(v)}{v}  \int_0^{\infty} q dq F(q^2)$  for the coherent vs. angular-momentum-dependent pieces indicated, with $m_\chi = 100$ GeV. 
\label{fig:MvsDelta}}

\end{figure}

\begin{figure}
\begin{center}
\includegraphics[width=0.9\textwidth]{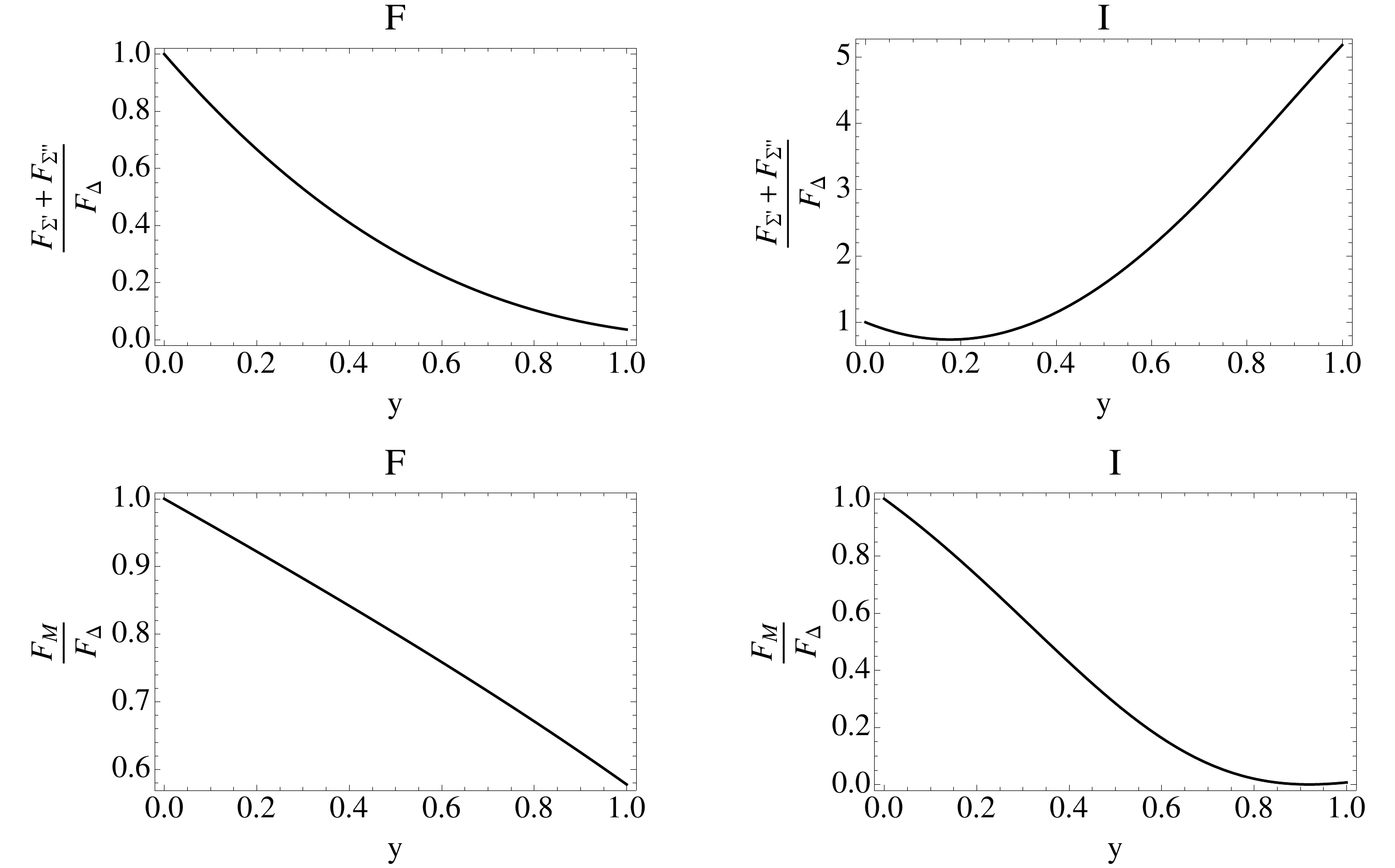}
\end{center}
\caption{Comparison of the $\Delta$ form factor to the standard spin-dependent ($F_{\Sigma'} + F_{\Sigma''}$), and to
the standard spin-independent ($F_M$) ones, for protons in iodine and fluorine.  Ratios have been normalized to 1 at $y=0$. 
\label{fig:comparison-of-shapes}}

\end{figure}

The $\Delta$ responses in the zero momentum transfer limit simply measure the nucleon angular momentum content 
of the nucleus.  Therefore, elements which have an unpaired nucleon (either $n$ or $p$), in a non s-shell orbital 
are favored.  For the proton response, this includes $^{23}$Na and $^{127}$I.  $^{19}$F, whose 
proton is approximately in the $2s_{1/2}$ orbital, is disfavored, however.  
Among odd-neutron nuclei of interest, $^{73}$Ge and $^{131}$Xe exhibit the strongest response, with other isotopes less favored.
In Fig. \ref{fig:spinsizes} at finite momentum transfer, $q$, we provide a comparison of the strength of the response for various elements (integrated over
a representative range of $q$, and weighted by the natural abundances of isotopes).  As explained in section 2, the $\Delta$ response receives a kinematic enhancement of $A$ that
is competitive with the coherent enhancement factor associated with the center of mass of motion of the nucleus.  Thus,
the $\Delta$ response contribution to operators such $\CO_5$ and $\CO_8$ can be important, and can become dominant
for elements with unpaired nucleons in large angular-momentum orbitals.  To illustrate this point we have included in Fig. \ref{fig:MvsDelta}
a comparison of the $\Delta$ response to the standard SI response, $M$, as they occur for operator $\CO_8$.  Finally, we would also like to point out, in case
spectral data becomes available in future experiments, that as a function of recoil energy (or momentum transfer), the 
$\Delta$ responses has different behavior from either the more standard SI ($M$) or the SD ($\Sigma'$, $\Sigma''$) responses.   
This is shown in Fig. \ref{fig:comparison-of-shapes}.  

\subsection{$\Phi''_p$ and $\Phi''_n$}

\begin{figure}
\begin{center}
\includegraphics[width=0.825\textwidth]{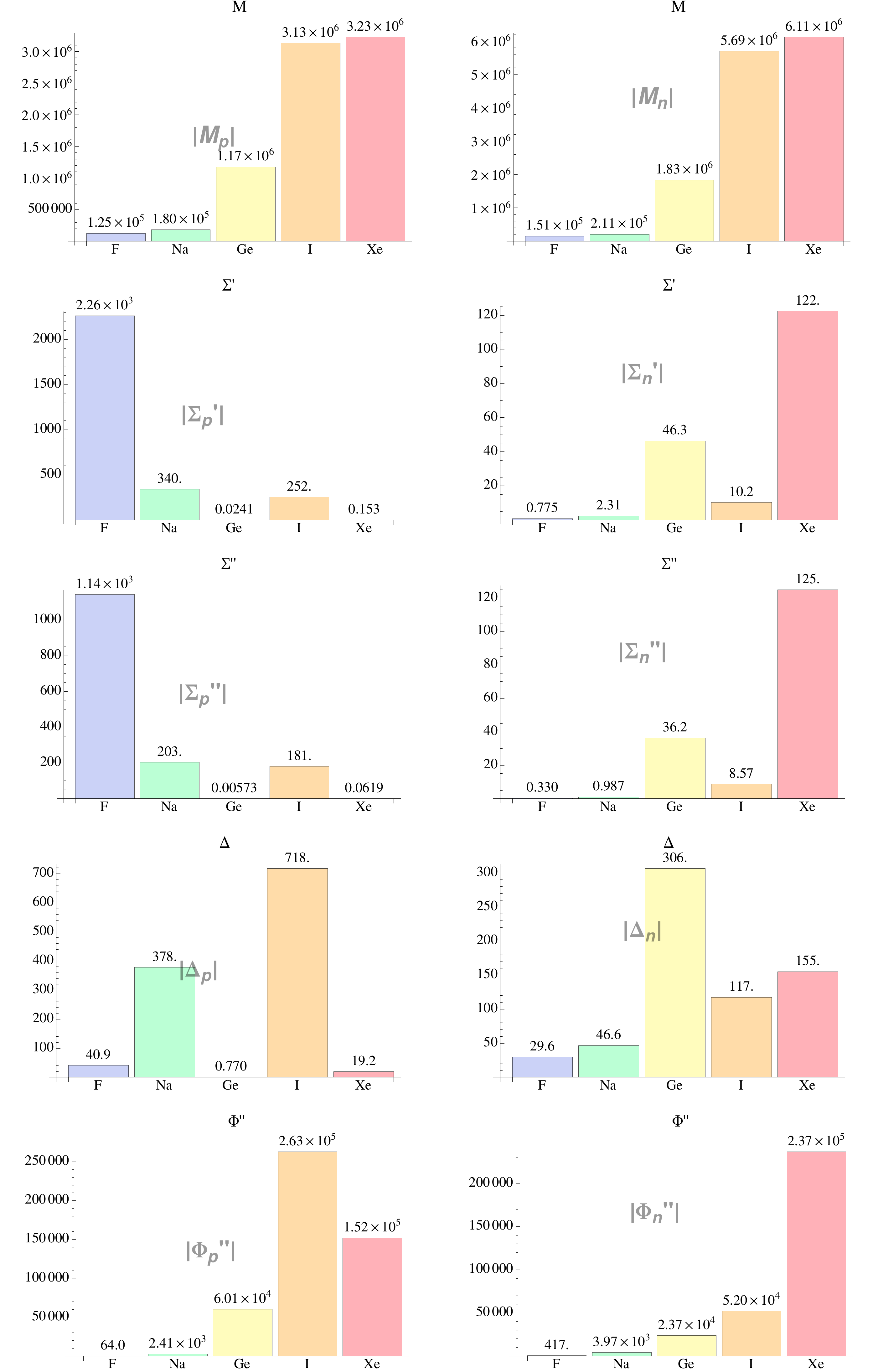}
\end{center}
\caption{Size of integrated form factors $\int_0^{100 \textrm{MeV}} \frac{1}{2}q dq F(q^2)$  for different nuclear responses,  weighted by the natural abundances of isotopes.
\label{fig:spinsizes}}

\end{figure}

\begin{figure}
\begin{center}
\includegraphics[width=0.825\textwidth]{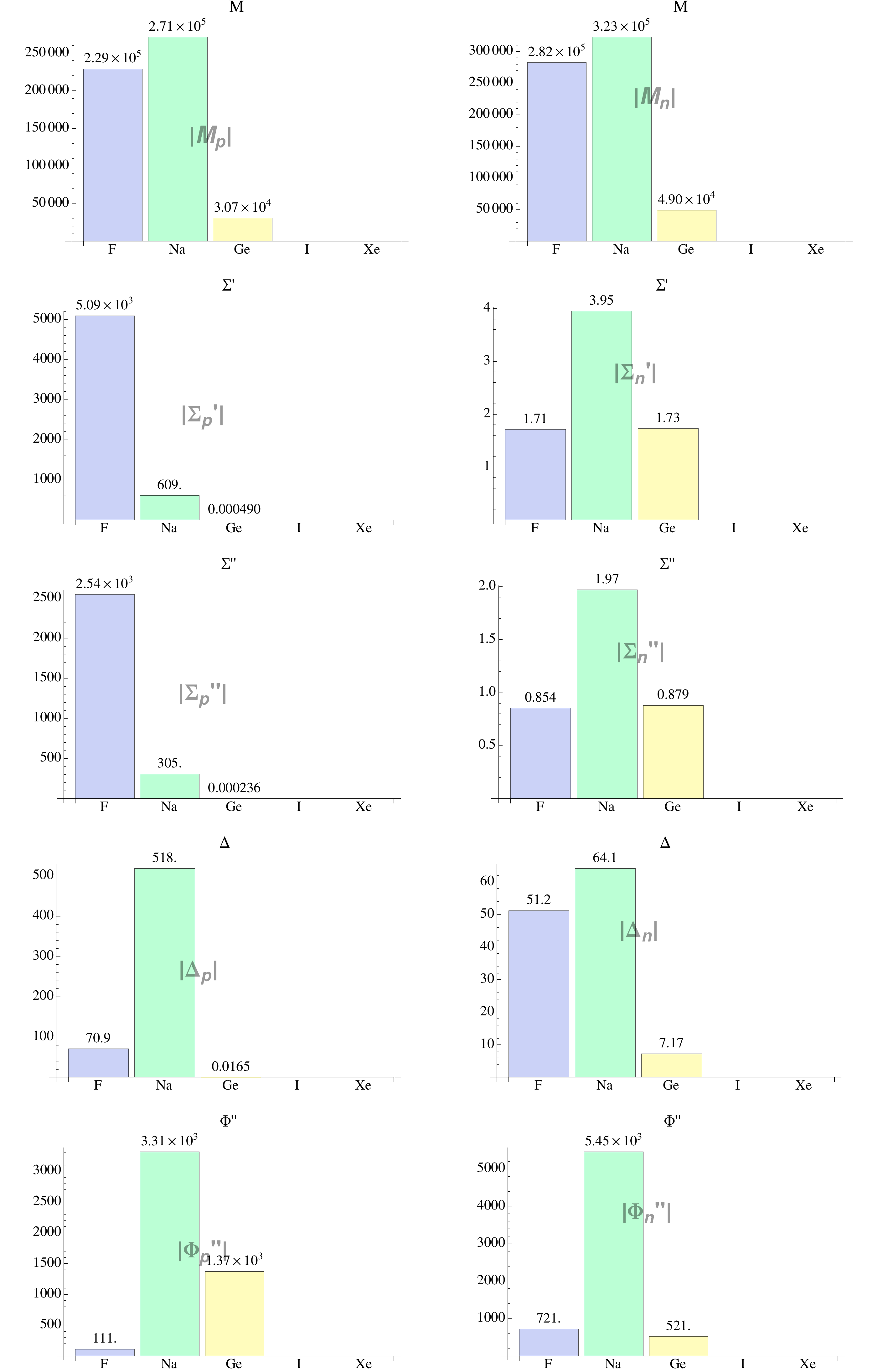}
\end{center}
\caption{Size of integrated form factors $\int_{v_{\rm min}=\frac{q}{2\mu_T}}  d^3 v \frac{f(v)}{v}  \int_0^{\infty} q dq F(q^2)$ for a DM of 3 GeV (integrated against the velocity distribution) for different nuclear responses.
\label{fig:lightDM}}

\end{figure}

The $\Phi''$ responses at zero momentum transfer are sensitive to a product of the nucleon spin and its angular momentum.
The dominantly coherent part of this product, is the scalar inner product $(\vec{L} \cdot \vec{S})_N$.  For completely filled
angular momentum orbitals, this dot product vanishes.  Namely, when all $2(\ell+1)$ states of the spin-aligned ($j=\ell+\frac{1}{2}$) subshell
and all $2\ell$ states of the spin-anti-aligned ($j=\ell-\frac{1}{2}$) subshell are occupied, this dot product vanishes.
%
%
In general, however, as the $\ell \pm \frac{1}{2}$ orbitals have 
different energies and so the highest occupied orbital for a given element will not be filled.  Let $n_\pm(\ell)$ be the approximate occupation 
numbers of the $\ell \pm \frac{1}{2}$ orbitals.   In terms of these, the dot product is proportional to $ (\ell+1) ~n_+(\ell)- \ell ~n_-(\ell) $.
Usually, the least energetic orbital will be filled first, and so one expects a mismatch
between $n_\pm(\ell)$ of order $\ell$ (the strong spin-orbit nuclear force in nuclei moves the spin-aligned orbit lower in energy).  Consequently, $\langle(\vec{L} \cdot \vec{S})\rangle \sim \ell_{highest}^2$ for most 
elements.  The $\Phi''$ responses tend to favor heavier elements, as these have larger $\ell$ orbitals not fully occupied.
Much as in the case of the delta responses, $\Phi''$ receives a kinematic enhancement of $A$, and can be important.
It is the dominant response for the operator $\CO_3$.  As we will see in section ~\ref{sec:models}, models which contain
$\CO_3$ can also typically contain $\frac{q^2}{m_N} \CO_1$.  Due to the $\ell_{highest}^2$ enhancement, for heavier elements,
the $\Phi''$ response can easily be of order the $M$ response in a large portion of the parameter space of such models.  In Fig. \ref{fig:spinsizes} we
show the $\Phi''$ responses for the various elements.  This response is particularly interesting in the context of 
light DM, as the sodium coupling strength can be more than ten times bigger than that of fluorine (see Fig.\ref{fig:lightDM}).

\section{Presentation of Results}
\label{sec:presentation}

Ultimately, we are interested in the prediction for the differential scattering rate $\frac{dR}{d E_R}$ (per unit time per unit recoil energy) with respect to nuclear recoil
energy.  This is related to the differential cross-section through
\ba
\frac{d R}{d E_R} &=& \left\< \frac{\rho_\chi m_T}{\mu_T^2 m_\chi v} \frac{d\sigma}{d \cos \theta} \right\> ,
\ea
where $\rho_\chi$ is the dark matter density, and $\< \dots \>$ indicates average over the halo velocity distribution.\footnote{
This formula follows straightforwardly from the recoil energy in terms of the velocity and scattering angle in the center-of-mass frame, $E_R = \frac{\mu_T^2}{m_T} v^2 (1-\cos \theta)$, and the fact that the rate per unit time is $R= \< n_\chi \sigma v\>$.  A canonical review of dark matter direct detection is \cite{lewinsmith}.}  The 
differential cross-section depends on the matrix-elements-squared in the usual way,
\ba
\frac{d \sigma}{d \cos \theta} &=&  \frac{1}{2j_\chi +1} \frac{1}{2j+1} \sum_{\rm spins} \frac{1}{32 \pi} \frac{|{\cal M} |^2}{(m_\chi + m_T)^2} ,
\ea
where we have averaged over $2j_\chi+1$ and $2j+1$ initial dark matter and nuclear spins, and summed over the final spins. 
The matrix-elements-squared in general contain interference terms between the different operators, and this leads to a large
number of possible different form factors. A general Lagrangian of the form
\ba
\CL &=& \sum_{i=1}^{12} c_i^{(n)} \CO_i^{(n)} + c_i^{(p)} \CO_i^{(p)} ,
\label{eq:genLag}
\ea
will therefore lead to a matrix-elements-squared that can be written
\ba
\frac{1}{2j_\chi+1} \frac{1}{2j+1} \sum_{\rm spins} |{ \cal M}|^2 &\equiv &  \frac{m_T^2}{m_N^2} \sum_{i,j=1}^{12} \sum_{N,N'=p,n} c^{(N)}_i c^{(N')}_j F^{(N,N')}_{ij}(v^2, q^2),
\label{eq:FFdefn}
\ea
where the form factors $F^{(N,N')}_{ij}(q^2)$ are {\it defined} as the coefficients of the $c_i$'s in this relation, and are defined to be symmetric in $(i,N)\leftrightarrow (j,N')$.
We give approximations for them at the most relevant nuclei in appendix \ref{sec:nuclFF}.  We have factored out the generic kinematic term
$\frac{m_T^2}{m_N^2}$ which arises due to the conventional relativistic normalization of states.   Because the operators fall into sectors that
do not interfere with each other due to symmetry and WIMP spin, only a few of the off-diagonal ($i \ne j$) form factors $F_{ij}$'s are non-zero.
In summary, the master formula for the detector event rate $\frac{d R_D}{d E_R}$ (per unit time per unit detector mass per unit recoil energy) in terms of the form factors $F_{ij}^{(N,N')}$ and operators coefficients $c_i^{(N)}$ in the effective theory is
\ba
\frac{dR_D}{dE_R} &=& N_T \frac{\rho_\chi m_T}{32 \pi m_\chi^3 m_N^2} \left\< \frac{1}{v} \sum_{ij} \sum_{N,N'=p,n} c_i^{(N)} c_j^{(N')} F_{ij}^{(N,N')}(v^2, q^2) \right\>,
\label{eq:masterformula}
\ea
where $N_T$ is the number of target nuclei per detector mass.  

Finally, let us give the explicit connection between these general form factors and the convention for the form factors in the standard spin-dependent case, which is our $\CO_4$.  In the standard spin-dependent interaction, the usual convention is to write the coefficients of the operators $\CO_4^{(p)}$ and $\CO_4^{(n)}$ in terms of isospin-respecting and isospin-violating parameters $a_0=a_n+a_p, a_1=a_p-a_n$ respectively, which are related to the coefficients $c_4^{(p)}, c_4^{(n)}$ in
eq. (\ref{eq:genLag}) by
\ba
c_4^{(N)} &=& (32\sqrt{2}) m_N m_\chi G_F a_{N}, \ \ \ \ \ (N=n,p)
\ea
where $G_F$ is Fermi's constant and DM spin $j_\chi = \frac{1}{2}$ is assumed.  It is also conventional to define form factors $S_{00}, S_{11}, S_{01}$:
\ba
S_{00} &=& \frac{1}{4\pi} \sum_{\rm spins} | \< \vec{S}_n + \vec{S}_p \> |^2,  \ \ \ \ \ \ S_{11} = \frac{1}{4\pi} \sum_{\rm spins} | \< \vec{S}_p - \vec{S}_n \> |^2, \ \ \ \ \ \ S_{01} = \frac{1}{2\pi} \sum_{\rm spins} | \< \vec{S}_p\>|^2 - |\< \vec{S}_n \> |^2, \nn\\
\ea
Consequently, they are related to our $F_{44}^{(N,N')}$'s with DM spin $j_\chi = \frac{1}{2}$ according to
\ba
F_{44}^{(p,p)} &=&  \frac{\pi}{4(2j+1)} (S_{00} + S_{11} + S_{01}), \ \  F_{44}^{(n,n)} =  \frac{\pi}{4(2j+1)}(S_{00} + S_{11} -S_{01}) , \\
F_{44}^{(n,p)} &=& F_{44}^{(p,n)} =  \frac{\pi}{4(2j+1)} (S_{00} - S_{11}) .
\ea

\section{Models}
\label{sec:models}
One of the more interesting operators we have found is $\CO_3$, as it leads to non-trivial dependence on the nucleon
angular momentum.  Since this operator is somewhat unusual, let us provide a sketch of a model where such an operator
might arise.  $\CO_3$ can be obtained from the non-relativistic limit of (see appendix)
\ba
\bar{\chi} \gamma^\mu\chi \bar{N} i \sigma_{\mu \nu} q^\nu N &\rightarrow& (2 m_\chi) q^2 + 8 m_N m_\chi i v \cdot (q \times S_N)  = 2 m_\chi q^2 \CO_1 - 8 m_N m_\chi \CO_3.
\ea

Our goal will then be to provide a model which generates this type of operator and describe its parameter space.
An example of a model of this type is one which contains a new Dirac pair of colored fermions, $U, \tilde{U}$ and
$D,\tilde{D}$, which also carry charges $\pm Q_u$ and $\pm Q_d$ under a new gauge boson, $A'_\mu$.  We can
imagine that there is some UV sector which couples the left and right handed quarks of the SM ($q$ and $\tilde{u}$,$\tilde{d}$) 
to our new fermions.  If this UV sector respects a new-particle parity symmetry, then upon integrating the UV sector out we get
a Lagrangian containing the terms:

\ba
\CL &=& \frac{1}{\Lambda_1^2} [ y_u (U\tilde{u})(U\tilde{u})^\dagger + y_d (D\tilde{d})(D\tilde{d})^\dagger ] \\ \nonumber
&+& \frac{1}{\Lambda_2^3} [ y'_u (q_\alpha H \tilde{u}_\beta \tilde{U}^\alpha U^\beta + c.c) + y'_d (q_\alpha H^\dagger \tilde{d}_\beta \tilde{D}^\alpha D^\beta + c.c) ] +\cdots
\ea
Here $\alpha,\beta$ are two-component, left-handed, spin indices, and we assumed some structure in the UV theory
which couples $U$'s to the up sector and $D$'s to the down sector.
\footnote{For example, the four-fermion terms above can arise from the exchange of scalar fields charged under hyper-charge and $A'$.}

\begin{figure}
\begin{center}
\includegraphics[width=0.9\textwidth]{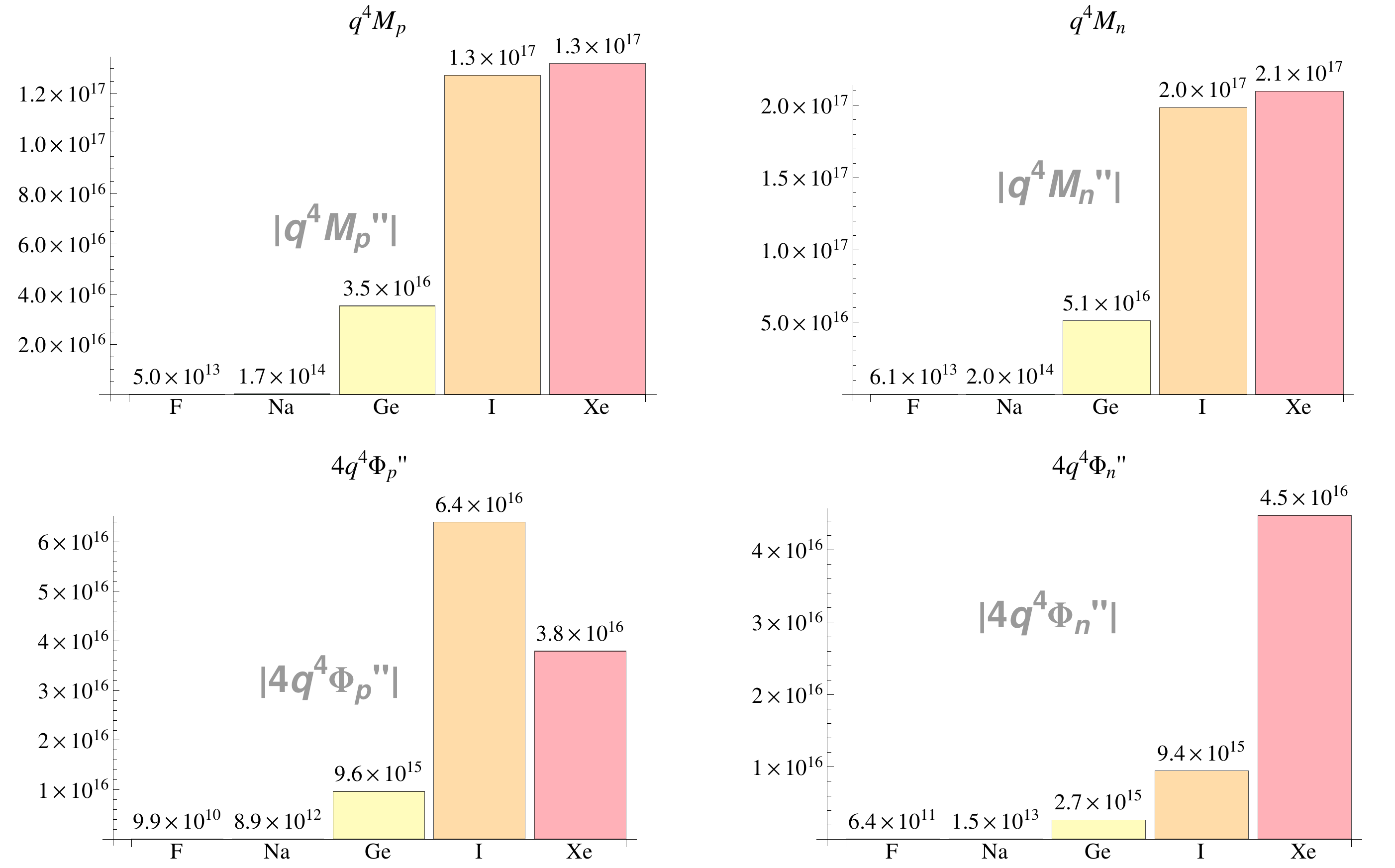}
\end{center}
\caption{ Comparison of the $\Phi''$ form factor to the standard spin-independent ($F_M$) form factor, with relative coefficients as given by the moment operator of the model in section \ref{sec:models}.
\label{fig:Phi-vs-M}}

\end{figure}

Upon integrating out the heavy fermions, and including the Higgs vev, $v$, the induced coupling of quarks to the 
new gauge boson take the form
\ba
\CL &=&  \kappa_1 \frac{g_{A'}}{ (4\pi)^2\Lambda_1^2} F'_{\mu\nu} \partial^\nu[ y_uQ_u \bar{u}_R\gamma^\mu  u_R + y_dQ_d \bar{d}_R \gamma^\mu d_R] \\ \nonumber
&+&  \kappa_2 \frac{g_{A'}  v}{ (4\pi)^2\Lambda_2^3} F'_{\mu\nu} [ M_u Q_u \bar{u}\sigma^{\mu\nu}u + M_d Q_d \bar{d} \sigma^{\mu\nu} d],
\ea
where $\kappa_i $ are numbers of order one, and the $M_i$ are of order the heavy fermion masses.  
We assume that the dark matter is also charged under the new gauge boson, and so after integrating the gauge
boson out, there is an induced local interaction between dark matter and nucleons of the form

\ba
\CL &=&  \frac{g_{A'}^2 Q_\chi}{ (4\pi)^2\Lambda_1^2 M_{A'}^2} \bar{\chi} \gamma_\mu \chi~ \left(c_{1p} ~\partial^2 (\bar{p}\gamma^{\mu}p) + c_{1n}~ \partial^2 (\bar{n} \gamma^{\mu}  n) + \cdots \right) \\ \nonumber
&+& \frac{g_{A'}^2 Q_\chi M_u v}{ (4\pi)^2\Lambda_2^3 M_{A'}^2} \bar{\chi} \gamma_\mu \chi~ \partial_\nu \left( c_{2p} ~\bar{p}\sigma^{\mu\nu}p + c_{2n} ~\bar{n} \sigma^{\mu\nu}  n\right) + \cdots,
\ea
where the various $c$'s depend on the parameters of the model and can be adjusted separately, for example by
changing the fermion masses and charges.  Adjusting the masses and charges (both of which are radiatively stable) one
can reduce the part of the interaction which couples to $q^2 \CO_1$, and vary the coupling to protons and neutrons
through $\CO_3$ independently.  Note, that for the heavier elements the $\Phi''$ response can be of the same order
as the $M$ response (see Fig. \ref{fig:Phi-vs-M} ).  Therefore, in a significant portion of parameter space of this model, the new
LS-response can dominate the more standard SI response for some elements.

Another interesting operator is the CP violating spin operator  $\CO_{10}=iq\cdot S_N$ as it couples to only a portion of the nucleon spin (resulting in the response $\Sigma''$).  A model generating such an operator is as follows.  Consider a scalar $\phi$ whose couplings violate CP, but preserve CP in any given sector.  After EW breaking, its couplings take the form  (schematically):
\ba
\CL = y_\chi \phi \bar{\chi}\chi + y_q \phi i\bar{q}\gamma_5q.
\ea
Upon integrating $\phi$ out, we get the following Lagrangian
\ba
\CL =  \frac{y_\chi y_q}{m_\phi^2} \bar{\chi}\chi i\bar{q}\gamma_5q + c \frac{y_\chi^2 y_q^2}{(4\pi)^2m_\phi^2}\bar{\chi}\chi \bar{q}q 
+ \kappa \frac{y_\chi^3 y_q^3}{(4\pi)^4m_\phi^2}\bar{q}q ~i\bar{q}\gamma_5q + \cdots 
\ea
The first term will lead in the non-relativistic limit to $\CO_{10}$ and can dominate the interaction between $\chi$ and quarks for small Yukawa couplings, while the last term produces CP violating effects in the SM.     In particular, the last term can mediate direct CP violating decays in the kaon system.   Experimental constraints place a bound on the mass and couplings of $\frac{m_\phi^2}{y_\chi^3 y_q^3} > (300~ \textrm{GeV})^2$.  This still allows for significant rates at direct detection experiments.  

Finally, let us discuss some of the restrictions we made on the effective theory.  Specifically, we have neglected operators that require contracting two indices from the dark matter part of the operator with the nucleon part.  In order for such interactions to be generated by tree-level exchange of a mediator, the exchanged field would have to be spin-2 or higher (or, at least a non-standard anti-symmetric $B^{\mu\nu}$ spin-1 field).  Higher-spin fields are subject to strong theoretical constraints and typically arise as composites fields with size comparable to their mass; this is the case for instance with spin-2 resonances of QCD.  There is nothing wrong with such particles, however they will usually be accompanied by spin-0 or spin-1 resonances as well, and it is unlikely that the higher-spin exchanges will dominate the interactions.  We have made one additional truncation which is to neglect the product operators in eq. (\ref{eq:productops}). This is motivated by the fact that any spin-1 exchange leading to such operators would necessarily couple to the dark matter sector as a CP-even field and to the nucleons as a CP-odd field, or vice versa.  In a theory where CP is broken, this is not necessarily forbidden.  However, it requires more work to see if such a framework can be UV-completed, in particular in a manner where the spin-1 mediator couples dominantly to a current in the UV in CP-violating way.

\section{Discussion and Future Directions}

The exact nature of dark matter remains as yet unknown, and little can be said for certain about its interactions
with the Standard Model.  Especially when comparing the results from different experiments, it is therefore
prudent to keep an open mind about what form such interactions can take, and to avoid prejudices about underlying
models when possible.  The most efficient tool for separating out assumptions about UV physics and parametrizing
the relevant low-energy possibilities is effective field theory, which we have applied here to the direct detection of
dark matter.  Effective field theory has been considered in the context, the most thorough analysis probably being
\cite{fanEFT}; however, the full set of possible interactions, including all derivative couplings and momentum-suppressed
interactions, as well as the possibility of  interference between different operators, has not previously been explored.
Since direct detection experiments are sensitive to interactions with finite momentum transfer, it is entirely possible and
well-motivated for momentum-independent operators to be absent, and for momentum-dependent interactions to be
responsible for the leading source of direct detection scattering.  It turns out that the full set
of possibilities is much richer than the standard cases, and can favor atomic nuclei in a qualitatively different way.

We have systematically constructed the low-energy, non-relativistic effective field theory describing direct detection scattering.
We have paid special attention to the basic non-relativistic building blocks, that connect directly to relevant experimental observables,
and to their symmetry properties.  This simplifies the identification of all possible non-relativistic operators and makes their interpretation
more physical, as well as explaining patterns in the form of non-relativistic operators that arise when taking the non-relativistic limit of
standard four-fermion operators.  It furthermore eliminates the need for embedding the field operators in full representations of the Lorentz group, 
and thus gives a unified description of all possible spins for the dark matter particle.  

This effective field theory is necessarily for interactions between dark matter and nucleons.  To make contact with experiment, 
one requires the matrix elements of these operators between atomic nuclei.  This marriage of an effective theory treatment of 
DM interactions with a treatment of the nuclear
response shows that there are six independent nuclear response functions characterizing DM elastic
interactions with nuclei, and that these response functions are associated with six single-particle operators
having the requisite transformation properties under parity and time reversal.  The new response
functions are associated with the nuclear convection current and related spin-velocity currents
that depend explicitly on nuclear compositeness.  In addition, two interference terms arise, adding 
additional complexity.  Such complexity is helpful,
providing more diagnostic handles for experimentalist to exploit, as they seek to determine the nature of
DM.  Our effective theory of DM interactions that include CP preserving exchanges of spin-1 or less utilizes five of the six
possible responses.

We have implemented this formalism by completing shell-model calculations of moderate complexity for
several of the critical targets - ${}^{19}$F, ${}^{23}$Na, ${}^{70,72,73,74,76}$Ge, ${}^{127}$I, 
${}^{128,129,130,131,132,134,136}$Xe - using realistic effective interactions.  Operator matrix elements
were evaluated in a harmonic oscillator basis, a choice that allows one to express the needed 
form factors as polynomials in the square of the three-momentum transfer.  The results show a wide
range of sensitivities to underlying effective theory interactions, highly dependent on the choice of nuclear target.
In a follow-up paper, we will analyze experimental constraints on the full effective theory using these form factors, though
as more accurate form factors for heavier elements become available in the literature, such analyses should become increasingly reliable.  

Aside from the improvement of nuclear form factors, the results here can be continued in several directions.  Models with inelastically-scattering dark matter can be well-motivated and lead to qualitatively interesting predictions, and it would be useful to extend the effective theory analysis to include such operators and any additional possible nuclear responses.  Also, while some combinations of the operators here arise easily from UV models, for others it is less obvious whether or not tuning is required.  It would be interesting to understand better if natural models for the full effective theory can be constructed or not.

\section*{Acknowledgments}
We would like to acknowledge useful conversations with Andrew Cohen, Jared Kaplan, and Jay Wacker.  We also thank Eugenio Del Nobile and Jussi Virkaj\"arvi for a careful reading of a previous version of the paper that led to several corrections and improvements. ALF was partially supported by ERC grant BSMOXFORD no. 228169. WH is supported by the US Department of Energy under contract DE-SC00046548.  EK is supported by DOE grant DE-FG02-01ER-40676, NSF CAREER grant PHY-0645456, and also by an Alfred P. Sloan Fellowship.  NL and YX are supported by DOE grant DE-FG02-01ER-40676.

\appendix

\section*{Appendix}

\section{Nuclear Matrix Elements and Form Factors}
\label{sec:nuclFF}

\subsection{Partial Wave Decomposition of Operators}

To make connection with results in nuclear physics, in section \ref{sec:nuclear} we decomposed effective theory interactions into partial waves.  Such a decomposition allows one to package the matrix elements-squared into reduced matrix elements through the Wigner-Eckart theorem.  In this appendix, we will review some of the details of these computations.  We will then in section \ref{sec:reduction} give the form factors for the general effective theory in terms of a smaller set of independent form factors.  Finally, in section \ref{sec:tables}, we will give numerical approximations for those independent form factors.

We will begin with a few examples, starting with $\CO_1=1$. Expanding $e^{iq \cdot x}$ in partial waves, the corresponding DM-nucleus scattering matrix element takes the form
\be
{\cal M}_{r r', ss'} =  \frac{m_T}{m_N} \delta_{s s'}  \sum_{J=0}^\infty    \sqrt{4 \pi} (-i)^J [J] \< r | \int d^3 x j_J(q x) Y_{J0}(\hat{q} \cdot \hat{x})  \hat{\rho}_p(x) | r' \>,
\ee
where $[J] \equiv \sqrt{2J+1}$.  
The expansion of $e^{i q \cdot x}$ in this way is what leads to the introduction of the operator $M_{J M}(q \vec{X}_i) \equiv j_J(q X_i) Y_{J M} (\hat{q} \cdot \hat{X}_i)$:
\be
\< r |  M_{JM}(q \vec{X}_i) | r' \> = \int d^3 x_i \psi^*_i(x_i) \psi_i(x_i) 
j_J(q x_i) Y_{J M} (\hat{q} \cdot \hat{x}_i),
\ee
where $\psi_i(x_i)$ is just the harmonic oscillator wavefunction for the nucleon corresponding to $i$.  Rotational invariance has been
used in order to pick a fiducial direction for $\vec{q}$, so that $M_{JM}$ depends only on $q \vec{x}$.

Consider for the next simple example the T-violating operator $\CO_{10} = i \vec{S}_N \cdot \vec{q}$.  The decomposition into partial waves takes the form
\ba
e^{i q \cdot x} \CO_{10} \cong -\frac{1}{2} \vec{\sigma} \cdot \vec{\nabla}_x e^{i q \cdot x}  \cong -\frac{1}{2}  \sum_{J=0}^\infty   \sqrt{4 \pi}(i)^J [J] \vec{\sigma} \cdot \vec{\nabla}_x M_{JM}(q \vec{x}) ,
\ea
where we have integrated by parts. 
More generally, in the partial wave analysis, one treats vectors like $\vec{S}_N$ by decomposing $\vec{A} = \sum_{\lambda=0, \pm 1} A_\lambda \vec{e}_\lambda{}^\dagger$, and using the identities in eq. (\ref{eq:vecsphharmId}).
Then, an arbitrary operator of the form $\vec{\ell} \cdot \vec{S}_N$ with $\vec{\ell}$ a constant vector can be decomposed as 
\be
\vec{\ell} \cdot \vec{S}_N e^{i q \cdot x} \cong \frac{1}{2} \sum_J \left( \sqrt{4 \pi} (i)^J [J] \right) \left( \ell_0 i \Sigma''_{J0} (q \vec{x}) - \sum_{\lambda = \pm 1} 
\frac{\ell_\lambda}{\sqrt{2}} \left( \lambda \Sigma_{J \lambda}(q \vec{x}) + i \Sigma'_{J \lambda} (q \vec{x})\right) \right) , 
\ee
where $\Sigma_{JM}( q \vec{x}) \equiv \vec{M}_{JJM}(q \vec{x}) \cdot \vec{\sigma}$. $\Sigma_{JM}$ has the wrong parity to contribute to elastic scattering, so we may discard it in the following. Consequently, $\CO_4$ decomposes as 
\ba
e^{i q \cdot x} \CO_4 &\cong& \frac{1}{2} \vec{S}_\chi \cdot \sum_J \left( \sqrt{4 \pi} (i)^J[J] \right)  \left(  \vec{e}_0 i \Sigma''_{J0}(q \vec{x}) - \sum_{\lambda=\pm 1} \frac{\vec{e}_\lambda^{\phantom{l}*}}{\sqrt{2}} (\lambda \Sigma_{J\lambda} + i\Sigma'_{J\lambda}) \right) .
\ea

Such manipulations can be performed for all the operators in the effective theory, for which one obtains the following matching:

\begin{center}
\begin{tabular}{l|c}
\hline
\hline
$  \CO_1= \mathbf{1} $ & $ \sum_J  c_J M_{J0}$\\
$\CO_3= i \vec{S}_N \cdot (\vec{q} \times \vec{v})$ & $ \sum_J c_J \left( \frac{q^2}{2m_N}\Phi''_{J0}- (i \vec{q} \times \vec{v}_T^\perp) \cdot \left(\sum_{\lambda=\pm 1} 2^{-\frac{1}{2}} \vec{e}_\lambda^{\phantom{l}*} ( i\Sigma'_{J\lambda}) \right) \right)$  \\
$\CO_4= \vec{S}_\chi \cdot \vec{S}_N$ & $ \frac{1}{2} \vec{S}_\chi \cdot \sum_J c_J\left(  \vec{e}_0 i \Sigma''_{J0} - \sum_{\lambda=\pm 1} 2^{-\frac{1}{2}} \vec{e}_\lambda^{\phantom{l}*} ( i\Sigma'_{J\lambda}) \right)$ \\
$\CO_5=i \vec{S}_\chi \cdot (\vec{q} \times \vec{v})$ & $ (\vec{S}_\chi \times i \vec{q}) \cdot \sum_J c_J \left( \vec{v}_T^\perp M_{J0}(q \vec{x})
  - \frac{i q}{ m_N } \sum_{\lambda = \pm 1} 2^{-\frac{1}{2}} \vec{e}_\lambda^{\phantom{l}*} (\lambda\Delta_{J \lambda} ) \right) 
  $ \\
  $ \CO_6= (\vec{S}_\chi \cdot \vec{q}) (\vec{S}_N \cdot \vec{q}) $ & $(\vec{q} \cdot \vec{S}_\chi) \frac{\vec{q}}{2} \cdot \sum_J c_J \left( \vec{e}_0 i \Sigma''_{J0} - \sum_{\lambda = \pm 1} 2^{-\frac{1}{2}}\vec{e}_\lambda^{\phantom{l}*} (i \Sigma'_{J\lambda} ) \right) $\\
  $ \CO_7 = \vec{S}_N \cdot \vec{v}^\perp$ & $  \sum_J c_J \left( -\frac{1}{2} \vec{v}_T^\perp \cdot \sum_{\lambda=\pm 1} 2^{-\frac{1}{2}} \vec{e}_\lambda^{\phantom{l}*} ( i \Sigma'_{J\lambda}) \right) $ \\
  $\CO_8= \vec{S}_\chi \cdot \vec{v}^\perp$ & $ \vec{S}_\chi \cdot \sum_J c_J \left( \vec{v}_T^\perp M_{J0} - \frac{i q }{m_N} \sum_{\lambda=\pm 1} 2^{-\frac{1}{2}} \vec{e}_\lambda^{\phantom{l}*} (\lambda \Delta_{J\lambda}  \right) $ \\
  $ \CO_9 = i \vec{S}_\chi \cdot (\vec{S}_N \times \vec{q})$ & $-\frac{1}{2} (i \vec{q} \times \vec{S}_\chi )\cdot \sum_J c_J \left(  \sum_{\lambda=\pm 1} 2^{-\frac{1}{2}} \vec{e}_\lambda^{\phantom{l}*} (i \Sigma'_{J \lambda}) \right) $ \\
  $ \CO_{10}= i \vec{S}_N \cdot \vec{q}$ & $ -\frac{1}{2}\sum_J c_J  q \Sigma''_{J0} $ \\ 
  $ \CO_{11}= i \vec{S}_\chi \cdot \vec{q} $ & $ (i \vec{S}_\chi \cdot \vec{q})  \sum_J c_J M_{J0} $ \\
\hline
\end{tabular}
\end{center}
Here, $c_J \equiv \sqrt{4 \pi} (i)^J [J] $, and in all cases, the scattering amplitude $\CM_{rr',ss'}$ is given by $\frac{m_T}{m_N}$ times the matrix element $\< s,r| \dots | s',r'\>$ of the 
nuclear response function in the table above.  Our convention for the reduced matrix elements $\< j || T_J || j'\>$ is
\ba
\<j'm' | T_{J M} | j m \> &=& (-)^{j'-m'} \left( \begin{array}{ccc} j' & J & j \\ -m' & M & m \end{array} \right) \< j' || T_J || j \>.
\ea
For elastic scattering, $\Sigma_{J\lambda}, \tilde{\Omega}_{J0}$ and $\tilde{\Delta}'_{J\lambda}$ do not contribute due to their parity, and all
form factors can be written in terms of a small set of independent form factors that depend only on the nuclear responses:
\ba
F^{(N,N')}_{X}(q^2) &\equiv&  \frac{4\pi}{2j+1} \sum_{J=0}^{2j+1} \< j || X_J^{(N)} || j \> \< j || X_J^{(N')} || j \>, 
\label{eq:FFfirst} 
\ea
for $X=M,\Sigma', \Sigma'', \Delta, \Phi''$, are required for the diagonal matrix elements.  Additionally, 
\ba
F^{(N,N')}_{X,Y}(q^2) &\equiv& \frac{4\pi}{2j+1} \sum_{J=0}^{2j+1} \< j || X_J^{(N)} || j \> \< j || Y_J^{(N')} || j \>,
\label{eq:FFsecond}
\ea
for $(X,Y) = (M, \Phi'')$ and  $(\Sigma', \Delta)$ appear when there is interference between different responses.   We emphasize that these independent nuclear form factors depend only on the nuclear
physics and are {\it not} all special cases of the form factors $F_{ij}$ for the effective theory operator coefficients $c_i$.  In particular, the $F_{ij}$'s can depend on the dark matter spin.  

If one prefers a basis of isoscalar $c^{(0)}=c^{(n)}+c^{(p)}$ and isovector $c^{(1)}=c^{(p)}-c^{(n)}$ couplings, rather than the basis of neutron
($N=n$) and proton ($N=p$) couplings we have chosen here, then one can use an isoscalar-isovector form of the general event rate formula eq. (\ref{eq:masterformula}),
\ba
\frac{dR_D}{dE_R} &=& N_T \frac{\rho_\chi m_T}{32 \pi m_\chi^3 m_N^2} \left\< \frac{1}{v} \sum_{ij} \sum_{a,b=0,1} c_i^{(a)} c_j^{(b)} F_{ij}^{ab}(v^2, q^2) \right\>,
\ea
with 
\ba
F^{00}_{i,j} &=& \frac{1}{4} \left( F^{(n,n)}_{i,j} + F^{(p,p)}_{i,j} + F^{(p,n)}_{i,j} + F^{(n,p)}_{i,j} \right), \nn\\
F^{11}_{i,j} &=& \frac{1}{4} \left( F^{(n,n)}_{i,j} + F^{(p,p)}_{i,j} - F^{(p,n)}_{i,j} - F^{(n,p)}_{i,j} \right), \nn\\
F^{01}_{i,j} &=& \frac{1}{4} \left( -F^{(n,n)}_{i,j} + F^{(p,p)}_{i,j} - F^{(p,n)}_{i,j} + F^{(n,p)}_{i,j} \right), \nn\\
F^{10}_{i,j} &=& \frac{1}{4} \left(-F^{(n,n)}_{i,j} + F^{(p,p)}_{i,j} + F^{(p,n)}_{i,j} - F^{(n,p)}_{i,j} \right).
\ea

\subsection{Form Factors for the General Effective Theory}
\label{sec:reduction}
 The full set of form factors $F_{i,j}^{(N,N')}$ (defined by eq. (\ref{eq:FFdefn})) necessary
for a general model can be written in terms of the basic independent nuclear form factors (defined by eq. (\ref{eq:FFfirst}) and (\ref{eq:FFsecond})) as follows:

\begin{subequations}
\ba
F_{1,1}^{(N,N')} &=& F_{M}^{(N,N')} , \\
F_{3,3}^{(N,N')} &=&  \left( \frac{q^4}{4 m_N^2} F_{\Phi''}^{(N,N')} + q^2 \left( v^2 - \frac{q^2}{4 \mu_T^2} \right) F_{\Sigma'}^{(N,N')} \right), \\
F_{4,4}^{(N,N')} &=& C(j_\chi) \frac{1}{16}  \left( F_{\Sigma''}^{(N,N')} +  F_{\Sigma'}^{(N,N')} \right), \\
F_{5,5}^{(N,N')} &=& C(j_\chi) \frac{1}{4}  \left( q^2 \left(v^2 - \frac{q^2}{4 \mu_T^2} \right) F_{M}^{(N,N')} +  \frac{q^4}{m_N^2} F_{\Delta}^{(N,N')} \right), \\
F_{6,6}^{(N,N')} &=& C(j_\chi) \frac{q^4}{16} F_{\Sigma''}^{(N,N')}, \\
F_{7,7}^{(N,N')} &=& \frac{1}{8} \left( v^2 - \frac{q^2}{4 \mu_T^2} \right)  F_{\Sigma'}^{(N,N')}, \\
F_{8,8}^{(N,N')} &=& C(j_\chi) \frac{1}{4} \left(  \left(v^2 - \frac{q^2}{4 \mu_T^2} \right) F_{M}^{(N,N')} +  \frac{q^2}{m_N^2} F_{\Delta}^{(N,N')} \right) ,\\
F_{9,9}^{(N,N')} &=& C(j_\chi) \frac{q^2}{16} F_{\Sigma'}^{(N,N')} ,\\
F_{10,10}^{(N,N')} &=& \frac{q^2}{4}  F_{\Sigma''}^{(N,N')} ,\\
F_{11,11}^{(N,N')} &=& C(j_\chi) \frac{q^2}{4}  F_{M}^{(N,N')}, \\
F_{1,3}^{(N,N')} &=& \frac{q^2}{2m_N} F_{M, \Phi''}^{(N,N')}, \\
F_{4,5}^{(N,N')} &=& -C(j_\chi) \frac{q^2}{8m_N}  F_{\Sigma', \Delta}^{(N,N')} ,\\
F_{4,6}^{(N,N')} &=&  C(j_\chi) \frac{q^2}{16} F_{\Sigma''}^{(N,N')} ,\\
F_{8,9}^{(N,N')} &=& C(j_\chi) \frac{q^2}{8m_N} F_{\Sigma', \Delta}^{(N,N')} ,
\label{eq:FFlast}
\ea
\end{subequations}
where $C(j_\chi) = ( 4 j_\chi(j_\chi+1)/3)$ is a prefactor that depends on the DM spin $j_\chi$ and has been normalized to $C(\frac{1}{2}) =1$.  
All interference terms not given explicitly above can be seen to vanish.  We have not included $\CO_2$ here since it does not appear at leading order from any relativistic interaction without cancellations.

\subsection{Approximate Form Factors}
\label{sec:tables}

Here, we provide form factors for the basic responses defined in eq. (\ref{eq:FFfirst}) and (\ref{eq:FFsecond}).  They depend on the momentum transfer through the
dimensionless variable $y=(qb/2)^2$, where $b$ is the harmonic oscillator parameter, $b\approx \sqrt{41.467/(45A^{-1/3}-25A^{-2/3})}$ fm.  For the interference-type
responses, $F_{X,Y}^{(p,n)} \ne F_{X,Y}^{(n,p)}$, so these are given separately, whereas for the non-interference-type ones, we have $F_X^{(p,n)} = F_X^{(n,p)}$.

${}^{19}$F:
\ba
F_M{}^{(p,p)}&=&e^{-2 y} \left(81.-96. y+36. y^2-4.7 y^3+0.19 y^4\right)\nn \\
F_M{}^{(p,n)}&=&e^{-2 y} \left(90.-110. y+48. y^2-7.5 y^3+0.37 y^4\right)\nn \\
F_M{}^{(n,n)}&=&e^{-2 y} \left(100.-130. y+61. y^2-11. y^3+0.73 y^4\right)\nn \\
F_{\Sigma '}{}^{(p,p)}&=&e^{-2 y} \left(1.81-4.85 y+4.88 y^2-2.18 y^3+0.364 y^4\right)\nn\\
F_{\Sigma '}{}^{(p,n)}&=&e^{-2 y} \left(-0.0331+0.0815 y-0.0511 y^2-0.00142 y^3+0.00602 y^4\right)\nn \\
F_{\Sigma '}{}^{(n,n)}&=&e^{-2 y} \left(0.000607-0.00136 y+0.000266 y^2+0.000550 y^3+0.0000997 y^4\right)\nn \\
F_{\Sigma ''}{}^{(p,p)}&=&e^{-2 y} \left(0.903-2.37 y+2.35 y^2-1.05 y^3+0.175 y^4\right)\nn \\
F_{\Sigma ''}{}^{(p,n)}&=&e^{-2 y} \left(-0.0166+0.0509 y-0.0510 y^2+0.0199 y^3-0.00237 y^4\right)\nn \\
F_{\Sigma ''}{}^{(n,n)}&=&e^{-2 y} \left(0.000303-0.00107 y+0.00114 y^2-0.000348 y^3+0.0000320 y^4\right)\nn \\
F_{\Delta}{}^{(p,p)}&=&e^{-2 y} \left(0.0251-0.0201 y+0.00401 y^2\right)\nn \\
F_{\Delta}{}^{(p,n)}&=&e^{-2 y} \left(-0.0213+0.0170 y-0.00341 y^2\right)\nn \\
F_{\Delta}{}^{(n,n)}&=&e^{-2 y} \left(0.0181-0.0145 y+0.00290 y^2\right)\nn \\
F_{\Phi''}{}^{(p,p)}&=&e^{-2 y} \left(0.0392-0.0314 y+0.00627 y^2\right)\nn \\
F_{\Phi''}{}^{(p,n)}&=&e^{-2 y} \left(0.100-0.0800 y+0.0160 y^2\right)\nn \\
F_{\Phi''}{}^{(n,n)}&=&e^{-2 y} \left(0.255-0.204 y+0.0408 y^2\right)\nn \\
F_{M,\Phi''}{}^{(p,p)}&=&e^{-2 y} \left(-1.78+1.77 y-0.509 y^2+0.0347 y^3\right)\nn \\
F_{M,\Phi''}{}^{(p,n)}&=&e^{-2 y} \left(-4.55+4.51 y-1.30 y^2+0.0884 y^3\right)\nn \\
F_{M,\Phi''}{}^{(n,p)}&=&e^{-2 y} \left(-1.98+2.11 y-0.697 y^2+0.0675 y^3\right)\nn \\
F_{M,\Phi''}{}^{(n,n)}&=&e^{-2 y} \left(-5.05+5.39 y-1.78 y^2+0.172 y^3\right)\nn \\
F_{\Sigma ',\Delta }{}^{(p,p)}&=&e^{-2 y} \left(-0.213+0.371 y-0.210 y^2+0.0382 y^3\right)\nn \\
F_{\Sigma ',\Delta }{}^{(p,n)}&=&e^{-2 y} \left(0.181-0.315 y+0.178 y^2-0.0325 y^3\right)\nn \\
F_{\Sigma ',\Delta }{}^{(n,p)}&=&e^{-2 y} \left(0.00390-0.00592 y+0.000163 y^2+0.000632 y^3\right)\nn \\
F_{\Sigma ',\Delta }{}^{(n,n)}&=&e^{-2 y} \left(-0.00331+0.00503 y-0.000138 y^2-0.000537 y^3\right)\nn
\ea
\newpage
${}^{23}$Na:
\ba
F_M{}^{(p,p)}&=&e^{-2 y} \left(120.-180. y+87. y^2-17. y^3+1.2 y^4\right)\nn \\
F_M{}^{(p,n)}&=&e^{-2 y} \left(130.-200. y+100. y^2-20. y^3+1.5 y^4\right)\nn \\
F_M{}^{(n,n)}&=&e^{-2 y} \left(140.-220. y+120. y^2-25. y^3+1.8 y^4\right)\nn \\
F_{\Sigma '}{}^{(p,p)}&=&e^{-2 y} \left(0.273-0.824 y+1.19 y^2-0.477 y^3+0.0593 y^4\right)\nn\\
F_{\Sigma '}{}^{(p,n)}&=&e^{-2 y} \left(0.0219-0.0578 y+0.0360 y^2-0.00300 y^3-0.000363 y^4\right)\nn \\
F_{\Sigma '}{}^{(n,n)}&=&e^{-2 y} \left(0.00176-0.00396 y+0.00228 y^2+0.0000195 y^3\right)\nn \\
F_{\Sigma ''}{}^{(p,p)}&=&e^{-2 y} \left(0.136-0.267 y+0.458 y^2-0.112 y^3+0.00828 y^4\right)\nn \\
F_{\Sigma ''}{}^{(p,n)}&=&e^{-2 y} \left(0.0110-0.0300 y+0.0217 y^2-0.00897 y^3+0.000592 y^4\right)\nn \\
F_{\Sigma ''}{}^{(n,n)}&=&e^{-2 y} \left(0.000882-0.00310 y+0.00399 y^2-0.00203 y^3+0.000409 y^4\right)\nn \\
F_{\Delta}{}^{(p,p)}&=&e^{-2 y} \left(0.231-0.185 y+0.0502 y^2\right)\nn \\
F_{\Delta}{}^{(p,n)}&=&e^{-2 y} \left(0.0812-0.0650 y+0.0138 y^2\right)\nn \\
F_{\Delta}{}^{(n,n)}&=&e^{-2 y} \left(0.0286-0.0228 y+0.00462 y^2\right)\nn \\
F_{\Phi''}{}^{(p,p)}&=&e^{-2 y} \left(1.48-1.19 y+0.275 y^2\right)\nn \\
F_{\Phi''}{}^{(p,n)}&=&e^{-2 y} \left(1.89-1.53 y+0.334 y^2\right)\nn \\
F_{\Phi''}{}^{(n,n)}&=&e^{-2 y} \left(2.43-1.95 y+0.413 y^2\right)\nn \\
F_{M,\Phi''}{}^{(p,p)}&=&e^{-2 y} \left(-13.+15. y-5.3 y^2+0.58 y^3\right)\nn \\
F_{M,\Phi''}{}^{(p,n)}&=&e^{-2 y} \left(-17.+19. y-6.7 y^2+0.70 y^3\right)\nn \\
F_{M,\Phi''}{}^{(n,p)}&=&e^{-2 y} \left(-15.+17. y-6.3 y^2+0.71 y^3\right)\nn \\
F_{M,\Phi''}{}^{(n,n)}&=&e^{-2 y} \left(-19.+22. y-8.0 y^2+0.86 y^3\right)\nn \\
F_{\Sigma ',\Delta }{}^{(p,p)}&=&e^{-2 y} \left(-0.25+0.48 y-0.29 y^2+0.049 y^3\right)\nn \\
F_{\Sigma ',\Delta }{}^{(p,n)}&=&e^{-2 y} \left(-0.088+0.17 y-0.081 y^2+0.011 y^3\right)\nn \\
F_{\Sigma ',\Delta }{}^{(n,p)}&=&e^{-2 y} \left(-0.020+0.031 y-0.0076 y^2-0.00027 y^3\right)\nn \\
F_{\Sigma ',\Delta }{}^{(n,n)}&=&e^{-2 y} \left(-0.0071+0.011 y-0.0030 y^2\right)\nn
\ea
\newpage
${}^{70}$Ge:
\ba
F_M{}^{(p,p)}&=&e^{-2 y} \left(1000.-2800. y+2900. y^2-1400. y^3+350. y^4-42. y^5+1.9 y^6-0.0027 y^7\right)\nn \\
F_M{}^{(p,n)}&=&e^{-2 y} \left(1200.-3500. y+3800. y^2-2000. y^3+530. y^4-70. y^5+3.8 y^6-0.034 y^7\right)\nn \\
F_M{}^{(n,n)}&=&e^{-2 y} \left(1400.-4300. y+4900. y^2-2700. y^3+780. y^4-110. y^5+7.2 y^6-0.11 y^7+0.00052 y^8\right)\nn \\
F_{\Phi''}{}^{(p,p)}&=&e^{-2 y} \left(36.-58. y+32. y^2-6.8 y^3+0.51 y^4-0.0028 y^5\right)\nn \\
F_{\Phi''}{}^{(p,n)}&=&e^{-2 y} \left(25.-45. y+29. y^2-7.8 y^3+0.91 y^4-0.034 y^5\right)\nn \\
F_{\Phi''}{}^{(n,n)}&=&e^{-2 y} \left(18.-35. y+25. y^2-8.4 y^3+1.3 y^4-0.088 y^5+0.0021 y^6\right)\nn \\
F_{M,\Phi''}{}^{(p,p)}&=&e^{-2 y} \left(-190.+420. y-330. y^2+110. y^3-17. y^4+1.0 y^5-0.0034 y^6\right)\nn \\
F_{M,\Phi''}{}^{(p,n)}&=&e^{-2 y} \left(-140.+320. y-280. y^2+110. y^3-22. y^4+2.0 y^5-0.063 y^6\right)\nn \\
F_{M,\Phi''}{}^{(n,p)}&=&e^{-2 y} \left(-230.+520. y-430. y^2+160. y^3-28. y^4+1.9 y^5-0.021 y^6\right)\nn \\
F_{M,\Phi''}{}^{(n,n)}&=&e^{-2 y} \left(-160.+400. y-370. y^2+160. y^3-35. y^4+3.5 y^5-0.14 y^6+0.0010 y^7\right)\nn \\
\ea
${}^{72}$Ge:
\ba
F_M{}^{(p,p)}&=&e^{-2 y} \left(1000.-2800. y+3000. y^2-1500. y^3+400. y^4-51. y^5+2.6 y^6-0.0069 y^7\right)\nn \\
F_M{}^{(p,n)}&=&e^{-2 y} \left(1300.-3700. y+4100. y^2-2200. y^3+600. y^4-82. y^5+4.5 y^6-0.017 y^7\right)\nn \\
F_M{}^{(n,n)}&=&e^{-2 y} \left(1600.-4800. y+5600. y^2-3100. y^3+910. y^4-130. y^5+7.8 y^6-0.039 y^7\right)\nn \\
F_{\Phi''}{}^{(p,p)}&=&e^{-2 y} \left(68.-110. y+64. y^2-16. y^3+1.4 y^4-0.010 y^5\right)\nn \\
F_{\Phi''}{}^{(p,n)}&=&e^{-2 y} \left(6.9-13. y+8.8 y^2-2.7 y^3+0.36 y^4-0.018 y^5\right)\nn \\
F_{\Phi''}{}^{(n,n)}&=&e^{-2 y} \left(0.71-1.5 y+1.2 y^2-0.42 y^3+0.075 y^4-0.0063 y^5+0.00020 y^6\right)\nn \\
F_{M,\Phi''}{}^{(p,p)}&=&e^{-2 y} \left(-260.+580. y-460. y^2+170. y^3-30. y^4+2.0 y^5-0.0094 y^6\right)\nn \\
F_{M,\Phi''}{}^{(p,n)}&=&e^{-2 y} \left(-27.+66. y-60. y^2+26. y^3-5.6 y^4+0.58 y^5-0.023 y^6\right)\nn \\
F_{M,\Phi''}{}^{(n,p)}&=&e^{-2 y} \left(-330.+760. y-650. y^2+250. y^3-47. y^4+3.4 y^5-0.020 y^6\right)\nn \\
F_{M,\Phi''}{}^{(n,n)}&=&e^{-2 y} \left(-34.+87. y-83. y^2+38. y^3-8.7 y^4+0.96 y^5-0.040 y^6+0.00010 y^7\right)\nn \\
\ea
${}^{74}$Ge:
\ba
F_M{}^{(p,p)}&=&e^{-2 y} \left(1000.-2800. y+2900. y^2-1400. y^3+350. y^4-41. y^5+1.9 y^6-0.0033 y^7\right)\nn \\
F_M{}^{(p,n)}&=&e^{-2 y} \left(1300.-4000. y+4400. y^2-2400. y^3+660. y^4-91. y^5+5.4 y^6-0.070 y^7\right)\nn \\
F_M{}^{(n,n)}&=&e^{-2 y} \left(1800.-5500. y+6700. y^2-3900. y^3+1200. y^4-190. y^5+14. y^6-0.33 y^7+0.0023 y^8\right)\nn \\
F_{\Phi''}{}^{(p,p)}&=&e^{-2 y} \left(36.-57. y+32. y^2-7.0 y^3+0.55 y^4-0.0035 y^5\right)\nn \\
F_{\Phi''}{}^{(p,n)}&=&e^{-2 y} \left(26.-52. y+36. y^2-11. y^3+1.5 y^4-0.073 y^5+0.00023 y^6\right)\nn \\
F_{\Phi''}{}^{(n,n)}&=&e^{-2 y} \left(20.-45. y+38. y^2-15. y^3+2.9 y^4-0.27 y^5+0.0093 y^6\right)\nn \\
F_{M,\Phi''}{}^{(p,p)}&=&e^{-2 y} \left(-190.+420. y-330. y^2+110. y^3-18. y^4+1.0 y^5-0.0042 y^6\right)\nn \\
F_{M,\Phi''}{}^{(p,n)}&=&e^{-2 y} \left(-140.+360. y-330. y^2+150. y^3-32. y^4+3.4 y^5-0.13 y^6+0.00012 y^7\right)\nn \\
F_{M,\Phi''}{}^{(n,p)}&=&e^{-2 y} \left(-250.+600. y-510. y^2+200. y^3-37. y^4+2.8 y^5-0.043 y^6+0.00012 y^7\right)\nn \\
F_{M,\Phi''}{}^{(n,n)}&=&e^{-2 y} \left(-190.+500. y-510. y^2+250. y^3-62. y^4+7.6 y^5-0.39 y^6+0.0047 y^7\right)\nn \\
\ea
${}^{76}$Ge:
\ba
F_M{}^{(p,p)}&=&e^{-2 y} \left(1000.-2800. y+2900. y^2-1400. y^3+340. y^4-40. y^5+1.8 y^6-0.0024 y^7\right)\nn \\
F_M{}^{(p,n)}&=&e^{-2 y} \left(1400.-4200. y+4800. y^2-2600. y^3+730. y^4-100. y^5+6.5 y^6-0.11 y^7\right)\nn \\
F_M{}^{(n,n)}&=&e^{-2 y} \left(1900.-6300. y+7800. y^2-4700. y^3+1500. y^4-260. y^5+21. y^6-0.66 y^7+0.0069 y^8\right)\nn \\
F_{\Phi''}{}^{(p,p)}&=&e^{-2 y} \left(31.-49. y+27. y^2-5.8 y^3+0.44 y^4-0.0024 y^5\right)\nn \\
F_{\Phi''}{}^{(p,n)}&=&e^{-2 y} \left(39.-78. y+54. y^2-17. y^3+2.3 y^4-0.11 y^5+0.00031 y^6\right)\nn \\
F_{\Phi''}{}^{(n,n)}&=&e^{-2 y} \left(50.-120. y+100. y^2-41. y^3+8.1 y^4-0.77 y^5+0.027 y^6\right)\nn \\
F_{M,\Phi''}{}^{(p,p)}&=&e^{-2 y} \left(-180.+390. y-300. y^2+100. y^3-16. y^4+0.89 y^5-0.0030 y^6\right)\nn \\
F_{M,\Phi''}{}^{(p,n)}&=&e^{-2 y} \left(-230.+580. y-550. y^2+240. y^3-53. y^4+5.5 y^5-0.22 y^6+0.00015 y^7\right)\nn \\
F_{M,\Phi''}{}^{(n,p)}&=&e^{-2 y} \left(-240.+590. y-520. y^2+210. y^3-38. y^4+3.0 y^5-0.061 y^6+0.00015 y^7\right)\nn \\
F_{M,\Phi''}{}^{(n,n)}&=&e^{-2 y} \left(-310.+870. y-910. y^2+460. y^3-120. y^4+15. y^5-0.85 y^6+0.014 y^7\right)\nn \\
\ea
${}^{73}$Ge:
\ba
F_M{}^{(p,p)}&=&e^{-2 y} \left(1000.-2800. y+2900. y^2-1500. y^3+380. y^4-47. y^5+2.3 y^6-0.0058 y^7\right)\nn \\
F_M{}^{(p,n)}&=&e^{-2 y} \left(1300.-3800. y+4200. y^2-2300. y^3+630. y^4-88. y^5+5.1 y^6-0.050 y^7\right)\nn \\
F_M{}^{(n,n)}&=&e^{-2 y} \left(1700.-5200. y+6100. y^2-3500. y^3+1100. y^4-160. y^5+11. y^6-0.17 y^7+0.0012 y^8\right)\nn \\
F_{\Sigma '}{}^{(p,p)}&=&e^{-2 y} \left(0.00020-0.00046 y+0.0015 y^2-0.0023 y^3+0.0021 y^4-0.00076 y^5+0.00010 y^6\right)\nn\\
F_{\Sigma '}{}^{(p,n)}&=&e^{-2 y} \left(0.012-0.053 y+0.072 y^2-0.067 y^3+0.038 y^4-0.012 y^5+0.0018 y^6-0.00014 y^7\right)\nn \\
F_{\Sigma '}{}^{(n,n)}&=&e^{-2 y} \left(0.74-4.7 y+12. y^2-13. y^3+6.9 y^4-2.1 y^5+0.35 y^6-0.031 y^7+0.0019 y^8\right)\nn \\
F_{\Sigma ''}{}^{(p,p)}&=&e^{-2 y} \left(0.00010-0.00096 y+0.0034 y^2-0.0036 y^3+0.0020 y^4-0.00046 y^5\right)\nn \\
F_{\Sigma ''}{}^{(p,n)}&=&e^{-2 y} \left(0.0061-0.039 y+0.058 y^2-0.054 y^3+0.026 y^4-0.0065 y^5+0.00082 y^6\right)\nn \\
F_{\Sigma ''}{}^{(n,n)}&=&e^{-2 y} \left(0.37-1.2 y+2.3 y^2-2.0 y^3+1.0 y^4-0.30 y^5+0.057 y^6-0.0060 y^7+0.00095 y^8\right)\nn \\
F_{\Delta}{}^{(p,p)}&=&e^{-2 y} \left(0.0069-0.012 y+0.0089 y^2-0.0031 y^3+0.00048 y^4\right)\nn \\
F_{\Delta}{}^{(p,n)}&=&e^{-2 y} \left(0.14-0.30 y+0.25 y^2-0.099 y^3+0.018 y^4-0.0013 y^5\right)\nn \\
F_{\Delta}{}^{(n,n)}&=&e^{-2 y} \left(3.0-7.2 y+6.9 y^2-3.1 y^3+0.72 y^4-0.080 y^5+0.0036 y^6\right)\nn \\
F_{\Phi''}{}^{(p,p)}&=&e^{-2 y} \left(51.-83. y+47. y^2-11. y^3+0.97 y^4-0.0073 y^5\right)\nn \\
F_{\Phi''}{}^{(p,n)}&=&e^{-2 y} \left(18.-34. y+24. y^2-7.2 y^3+0.99 y^4-0.049 y^5+0.00021 y^6\right)\nn \\
F_{\Phi''}{}^{(n,n)}&=&e^{-2 y} \left(6.6-15. y+13. y^2-5.5 y^3+1.1 y^4-0.12 y^5+0.0049 y^6\right)\nn \\
F_{M,\Phi''}{}^{(p,p)}&=&e^{-2 y} \left(-230.+500. y-400. y^2+140. y^3-23. y^4+1.5 y^5-0.0076 y^6\right)\nn \\
F_{M,\Phi''}{}^{(p,n)}&=&e^{-2 y} \left(-80.+200. y-190. y^2+85. y^3-19. y^4+2.1 y^5-0.089 y^6+0.00011 y^7\right)\nn \\
F_{M,\Phi''}{}^{(n,p)}&=&e^{-2 y} \left(-290.+690. y-590. y^2+230. y^3-42. y^4+3.1 y^5-0.035 y^6+0.00011 y^7\right)\nn \\
F_{M,\Phi''}{}^{(n,n)}&=&e^{-2 y} \left(-100.+270. y-280. y^2+130. y^3-33. y^4+4.0 y^5-0.20 y^6+0.0025 y^7\right)\nn \\
F_{\Sigma ',\Delta }{}^{(p,p)}&=&e^{-2 y} \left(-0.0012+0.0024 y-0.0033 y^2+0.0023 y^3-0.00088 y^4+0.00015 y^5\right)\nn \\
F_{\Sigma ',\Delta }{}^{(p,n)}&=&e^{-2 y} \left(-0.025+0.058 y-0.074 y^2+0.053 y^3-0.020 y^4+0.0036 y^5-0.00029 y^6\right)\nn \\
F_{\Sigma ',\Delta }{}^{(n,p)}&=&e^{-2 y} \left(-0.071+0.29 y-0.38 y^2+0.24 y^3-0.075 y^4+0.011 y^5-0.00066 y^6\right)\nn \\
F_{\Sigma ',\Delta }{}^{(n,n)}&=&e^{-2 y} \left(-1.5+6.5 y-10. y^2+7.1 y^3-2.6 y^4+0.49 y^5-0.048 y^6+0.0019 y^7\right)\nn
\ea
${}^{127}$I:
\ba
F_M{}^{(p,p)}&=&e^{-2 y} \left(2800.-10000. y+14000. y^2-9800. y^3+3800. y^4-840. y^5+100. y^6-6.3 y^7+0.15 y^8\right)\nn \\
F_M{}^{(p,n)}&=&e^{-2 y} \left(3900.-15000. y+23000. y^2-18000. y^3+7900. y^4-2000. y^5+290. y^6-23. y^7+0.75 y^8\right. \nn\\
&& \left. -0.0048 y^9\right)\nn \\
F_M{}^{(n,n)}&=&e^{-2 y} \left(5500.-23000. y+38000. y^2-32000. y^3+16000. y^4-4600. y^5+790. y^6-75. y^7+3.3 y^8 \right. \nn\\
 && \left. -0.041 y^9+0.00015 y^{10}\right)\nn \\
F_{\Sigma '}{}^{(p,p)}&=&e^{-2 y} \left(0.26-1.6 y+5.3 y^2-8.9 y^3+8.7 y^4-4.9 y^5+1.5 y^6-0.25 y^7+0.016 y^8\right)\nn\\
F_{\Sigma '}{}^{(p,n)}&=&e^{-2 y} \left(0.065-0.46 y+1.3 y^2-1.8 y^3+1.4 y^4-0.65 y^5+0.17 y^6-0.026 y^7+0.0020 y^8\right)\nn \\
F_{\Sigma '}{}^{(n,n)}&=&e^{-2 y} \left(0.016-0.13 y+0.37 y^2-0.48 y^3+0.34 y^4-0.14 y^5+0.033 y^6-0.0048 y^7+0.00041 y^8\right)\nn \\
F_{\Sigma ''}{}^{(p,p)}&=&e^{-2 y} \left(0.13-0.49 y+1.8 y^2-2.8 y^3+2.7 y^4-1.6 y^5+0.53 y^6-0.092 y^7+0.0067 y^8\right)\nn \\
F_{\Sigma ''}{}^{(p,n)}&=&e^{-2 y} \left(0.032-0.13 y+0.26 y^2-0.30 y^3+0.21 y^4-0.098 y^5+0.027 y^6-0.0042 y^7+0.00033 y^8\right)\nn \\
F_{\Sigma ''}{}^{(n,n)}&=&e^{-2 y} \left(0.0080-0.032 y+0.053 y^2-0.046 y^3+0.025 y^4-0.0086 y^5+0.0019 y^6-0.00026 y^7\right)\nn \\
F_{\Delta}{}^{(p,p)}&=&e^{-2 y} \left(0.54-1.3 y+1.6 y^2-1.2 y^3+0.51 y^4-0.11 y^5+0.0097 y^6\right)\nn \\
F_{\Delta}{}^{(p,n)}&=&e^{-2 y} \left(0.23-0.65 y+0.79 y^2-0.54 y^3+0.20 y^4-0.040 y^5+0.0039 y^6-0.00014 y^7\right)\nn \\
F_{\Delta}{}^{(n,n)}&=&e^{-2 y} \left(0.10-0.32 y+0.40 y^2-0.25 y^3+0.084 y^4-0.016 y^5+0.0018 y^6-0.00011 y^7\right)\nn \\
F_{\Phi''}{}^{(p,p)}&=&e^{-2 y} \left(200.-480. y+440. y^2-200. y^3+45. y^4-5.2 y^5+0.23 y^6\right)\nn \\
F_{\Phi''}{}^{(p,n)}&=&e^{-2 y} \left(95.-260. y+280. y^2-150. y^3+42. y^4-6.3 y^5+0.49 y^6-0.015 y^7\right)\nn \\
F_{\Phi''}{}^{(n,n)}&=&e^{-2 y} \left(44.-140. y+170. y^2-110. y^3+36. y^4-6.9 y^5+0.74 y^6-0.042 y^7+0.00095 y^8\right)\nn \\
F_{M,\Phi''}{}^{(p,p)}&=&e^{-2 y} \left(-750.+2200. y-2600. y^2+1500. y^3-440. y^4+73. y^5-6.0 y^6+0.18 y^7\right)\nn \\
F_{M,\Phi''}{}^{(p,n)}&=&e^{-2 y} \left(-350.+1200. y-1600. y^2+1000. y^3-370. y^4+76. y^5-8.7 y^6+0.51 y^7-0.012 y^8\right)\nn \\
F_{M,\Phi''}{}^{(n,p)}&=&e^{-2 y} \left(-1000.+3400. y-4400. y^2+2800. y^3-980. y^4+190. y^5-20. y^6+0.86 y^7-0.0059 y^8\right)\nn \\
F_{M,\Phi''}{}^{(n,n)}&=&e^{-2 y} \left(-490.+1800. y-2600. y^2+1900. y^3-780. y^4+190. y^5-25. y^6+1.9 y^7-0.060 y^8\right. \nn\\
 && \left. +0.00038 y^9\right)\nn \\
F_{\Sigma ',\Delta }{}^{(p,p)}&=&e^{-2 y} \left(-0.37+1.6 y-3.1 y^2+3.4 y^3-2.1 y^4+0.73 y^5-0.13 y^6+0.0086 y^7\right)\nn \\
F_{\Sigma ',\Delta }{}^{(p,n)}&=&e^{-2 y} \left(-0.16+0.75 y-1.5 y^2+1.5 y^3-0.82 y^4+0.26 y^5-0.047 y^6+0.0043 y^7-0.00015 y^8\right)\nn \\
F_{\Sigma ',\Delta }{}^{(n,p)}&=&e^{-2 y} \left(-0.093+0.48 y-0.85 y^2+0.79 y^3-0.43 y^4+0.13 y^5-0.021 y^6+0.0017 y^7\right)\nn \\
F_{\Sigma ',\Delta }{}^{(n,n)}&=&e^{-2 y} \left(-0.040+0.22 y-0.43 y^2+0.38 y^3-0.19 y^4+0.053 y^5-0.0090 y^6+0.00088 y^7\right)\nn\ea

${}^{128}$Xe:
\ba
F_M{}^{(p,p)}&=&e^{-2 y} \left(2900.-11000. y+15000. y^2-11000. y^3+4200. y^4-950. y^5+120. y^6-7.8 y^7+0.20 y^8\right)\nn \\
F_M{}^{(p,n)}&=&e^{-2 y} \left(4000.-15000. y+24000. y^2-19000. y^3+8300. y^4-2100. y^5+320. y^6-25. y^7+0.85 y^8 \right. \nn\\
&& \left. -0.0055 y^9\right)\nn \\
F_M{}^{(n,n)}&=&e^{-2 y} \left(5500.-23000. y+38000. y^2-32000. y^3+16000. y^4-4600. y^5+790. y^6-75. y^7+3.3 y^8 \right. \nn\\
&& \left. -0.041 y^9+0.00015 y^{10}\right)\nn \\
F_{\Phi''}{}^{(p,p)}&=&e^{-2 y} \left(180.-440. y+410. y^2-180. y^3+42. y^4-4.9 y^5+0.22 y^6\right)\nn \\
F_{\Phi''}{}^{(p,n)}&=&e^{-2 y} \left(91.-250. y+270. y^2-140. y^3+40. y^4-6.2 y^5+0.48 y^6-0.014 y^7\right)\nn \\
F_{\Phi''}{}^{(n,n)}&=&e^{-2 y} \left(44.-140. y+170. y^2-110. y^3+36. y^4-6.8 y^5+0.74 y^6-0.042 y^7+0.00095 y^8\right)\nn \\
F_{M,\Phi''}{}^{(p,p)}&=&e^{-2 y} \left(-730.+2200. y-2500. y^2+1500. y^3-450. y^4+76. y^5-6.4 y^6+0.21 y^7\right)\nn \\
F_{M,\Phi''}{}^{(p,n)}&=&e^{-2 y} \left(-360.+1200. y-1600. y^2+1100. y^3-390. y^4+81. y^5-9.5 y^6+0.57 y^7-0.014 y^8\right)\nn \\
F_{M,\Phi''}{}^{(n,p)}&=&e^{-2 y} \left(-1000.+3300. y-4200. y^2+2700. y^3-940. y^4+180. y^5-19. y^6+0.85 y^7-0.0058 y^8\right)\nn \\
F_{M,\Phi''}{}^{(n,n)}&=&e^{-2 y} \left(-490.+1800. y-2600. y^2+1900. y^3-780. y^4+190. y^5-25. y^6+1.9 y^7-0.060 y^8\right. \nn\\
&& \left. +0.00038 y^9\right)\nn \\
\ea
${}^{130}$Xe:
\ba
F_M{}^{(p,p)}&=&e^{-2 y} \left(2900.-11000. y+15000. y^2-10000. y^3+4100. y^4-910. y^5+110. y^6-7.0 y^7+0.17 y^8\right)\nn \\
F_M{}^{(p,n)}&=&e^{-2 y} \left(4100.-16000. y+25000. y^2-19000. y^3+8600. y^4-2200. y^5+330. y^6-26. y^7+0.91 y^8\right. \nn\\
&& \left. -0.0076 y^9\right)\nn \\
F_M{}^{(n,n)}&=&e^{-2 y} \left(5800.-24000. y+41000. y^2-36000. y^3+18000. y^4-5200. y^5+910. y^6-89. y^7+4.2 y^8\right. \nn\\
&& \left. -0.068 y^9+0.00034 y^{10}\right)\nn \\
F_{\Phi''}{}^{(p,p)}&=&e^{-2 y} \left(150.-370. y+330. y^2-150. y^3+34. y^4-3.8 y^5+0.17 y^6\right)\nn \\
F_{\Phi''}{}^{(p,n)}&=&e^{-2 y} \left(120.-350. y+370. y^2-190. y^3+54. y^4-8.2 y^5+0.63 y^6-0.019 y^7\right)\nn \\
F_{\Phi''}{}^{(n,n)}&=&e^{-2 y} \left(100.-320. y+390. y^2-240. y^3+80. y^4-15. y^5+1.7 y^6-0.094 y^7+0.0021 y^8\right)\nn \\
F_{M,\Phi''}{}^{(p,p)}&=&e^{-2 y} \left(-670.+2000. y-2300. y^2+1300. y^3-400. y^4+65. y^5-5.4 y^6+0.17 y^7\right)\nn \\
F_{M,\Phi''}{}^{(p,n)}&=&e^{-2 y} \left(-540.+1800. y-2400. y^2+1600. y^3-580. y^4+120. y^5-14. y^6+0.81 y^7-0.019 y^8\right)\nn \\
F_{M,\Phi''}{}^{(n,p)}&=&e^{-2 y} \left(-940.+3100. y-4000. y^2+2500. y^3-900. y^4+180. y^5-18. y^6+0.84 y^7-0.0075 y^8\right)\nn \\
F_{M,\Phi''}{}^{(n,n)}&=&e^{-2 y} \left(-760.+2800. y-4100. y^2+3000. y^3-1200. y^4+300. y^5-41. y^6+3.1 y^7-0.10 y^8\right. \nn\\
 && \left. +0.00085 y^9\right)\nn \\
\ea
${}^{132}$Xe:
\ba
F_M{}^{(p,p)}&=&e^{-2 y} \left(2900.-11000. y+15000. y^2-10000. y^3+4000. y^4-880. y^5+110. y^6-6.4 y^7+0.15 y^8\right) \nn \\
F_M{}^{(p,n)}&=&e^{-2 y} \left(4200.-17000. y+26000. y^2-20000. y^3+9000. y^4-2300. y^5+340. y^6-27. y^7+0.98 y^8\right. \nn\\
&& \left. -0.0095 y^9\right)\nn \\
F_M{}^{(n,n)}&=&e^{-2 y} \left(6100.-26000. y+44000. y^2-39000. y^3+20000. y^4-5900. y^5+1000. y^6-100. y^7+5.3 y^8\right. \nn\\
 && \left. -0.099 y^9+0.00061 y^{10}\right)\nn \\
F_{\Phi''}{}^{(p,p)}&=&e^{-2 y} \left(130.-320. y+290. y^2-130. y^3+29. y^4-3.2 y^5+0.14 y^6\right)\nn \\
F_{\Phi''}{}^{(p,n)}&=&e^{-2 y} \left(150.-430. y+460. y^2-240. y^3+67. y^4-10. y^5+0.77 y^6-0.023 y^7\right)\nn \\
F_{\Phi''}{}^{(n,n)}&=&e^{-2 y} \left(180.-570. y+700. y^2-430. y^3+140. y^4-27. y^5+3.0 y^6-0.17 y^7+0.0038 y^8\right)\nn \\
F_{M,\Phi''}{}^{(p,p)}&=&e^{-2 y} \left(-620.+1900. y-2100. y^2+1200. y^3-360. y^4+59. y^5-4.7 y^6+0.14 y^7\right)\nn \\
F_{M,\Phi''}{}^{(p,n)}&=&e^{-2 y} \left(-720.+2400. y-3200. y^2+2100. y^3-760. y^4+160. y^5-18. y^6+1.0 y^7-0.024 y^8\right)\nn \\
F_{M,\Phi''}{}^{(n,p)}&=&e^{-2 y} \left(-900.+3000. y-3800. y^2+2500. y^3-880. y^4+170. y^5-18. y^6+0.85 y^7-0.0092 y^8\right)\nn \\
F_{M,\Phi''}{}^{(n,n)}&=&e^{-2 y} \left(-1000.+3900. y-5700. y^2+4200. y^3-1700. y^4+420. y^5-59. y^6+4.5 y^7-0.16 y^8\right. \nn\\
 && \left. +0.0015 y^9\right)\nn \\
\ea
${}^{134}$Xe:
\ba
F_M{}^{(p,p)}&=&e^{-2 y} \left(2900.-11000. y+15000. y^2-10000. y^3+3700. y^4-770. y^5+85. y^6-4.5 y^7+0.098 y^8\right. \nn\\
 && \left. -0.00028 y^9\right)\nn \\
F_M{}^{(p,n)}&=&e^{-2 y} \left(4300.-17000. y+27000. y^2-21000. y^3+9200. y^4-2300. y^5+340. y^6-26. y^7+0.93 y^8\right. \nn\\
 && \left. -0.011 y^9\right)\nn \\
F_M{}^{(n,n)}&=&e^{-2 y} \left(6400.-28000. y+48000. y^2-43000. y^3+22000. y^4-6600. y^5+1200. y^6-120. y^7+6.6 y^8 \right. \nn\\
 && \left. -0.15 y^9+0.0012 y^{10}\right)\nn \\
F_{\Phi''}{}^{(p,p)}&=&e^{-2 y} \left(80.-190. y+180. y^2-77. y^3+17. y^4-1.9 y^5+0.085 y^6-0.00064 y^7\right)\nn \\
F_{\Phi''}{}^{(p,n)}&=&e^{-2 y} \left(180.-490. y+520. y^2-270. y^3+75. y^4-11. y^5+0.85 y^6-0.026 y^7\right)\nn \\
F_{\Phi''}{}^{(n,n)}&=&e^{-2 y} \left(380.-1200. y+1500. y^2-910. y^3+310. y^4-59. y^5+6.3 y^6-0.35 y^7+0.0075 y^8\right)\nn \\
F_{M,\Phi''}{}^{(p,p)}&=&e^{-2 y} \left(-480.+1500. y-1700. y^2+920. y^3-270. y^4+41. y^5-3.1 y^6+0.094 y^7-0.00047 y^8\right)\nn \\
F_{M,\Phi''}{}^{(p,n)}&=&e^{-2 y} \left(-1100.+3600. y-4700. y^2+3000. y^3-1100. y^4+210. y^5-23. y^6+1.3 y^7-0.027 y^8\right)\nn \\
F_{M,\Phi''}{}^{(n,p)}&=&e^{-2 y} \left(-720.+2400. y-3100. y^2+2000. y^3-720. y^4+140. y^5-15. y^6+0.75 y^7-0.012 y^8\right)\nn \\
F_{M,\Phi''}{}^{(n,n)}&=&e^{-2 y} \left(-1600.+5900. y-8700. y^2+6500. y^3-2700. y^4+660. y^5-92. y^6+7.1 y^7-0.26 y^8\right. \nn\\
 && \left. +0.0030 y^9\right)\nn \\
\ea
${}^{136}$Xe:
\ba
F_M{}^{(p,p)}&=&e^{-2 y} \left(2900.-11000. y+15000. y^2-10000. y^3+3700. y^4-770. y^5+85. y^6-4.5 y^7+0.097 y^8 \right. \nn\\
&& \left. -0.00028 y^9\right) \nn\\
F_M{}^{(p,n)}&=&e^{-2 y} \left(4400.-18000. y+28000. y^2-22000. y^3+9700. y^4-2500. y^5+360. y^6-28. y^7+1.0 y^8\right. \nn\\
&& \left. -0.012 y^9\right)\nn \\
F_M{}^{(n,n)}&=&e^{-2 y} \left(6700.-30000. y+51000. y^2-46000. y^3+24000. y^4-7300. y^5+1300. y^6-140. y^7+7.6 y^8\right. \nn\\
 && \left. -0.17 y^9+0.0014 y^{10}\right)\nn \\
F_{\Phi''}{}^{(p,p)}&=&e^{-2 y} \left(81.-200. y+180. y^2-78. y^3+17. y^4-1.9 y^5+0.088 y^6-0.00065 y^7\right)\nn \\
F_{\Phi''}{}^{(p,n)}&=&e^{-2 y} \left(180.-510. y+540. y^2-280. y^3+78. y^4-12. y^5+0.90 y^6-0.028 y^7+0.00011 y^8\right)\nn \\
F_{\Phi''}{}^{(n,n)}&=&e^{-2 y} \left(400.-1300. y+1600. y^2-960. y^3+320. y^4-62. y^5+6.7 y^6-0.38 y^7+0.0085 y^8\right)\nn \\
F_{M,\Phi''}{}^{(p,p)}&=&e^{-2 y} \left(-490.+1500. y-1700. y^2+930. y^3-270. y^4+42. y^5-3.2 y^6+0.095 y^7-0.00048 y^8\right)\nn \\
F_{M,\Phi''}{}^{(p,n)}&=&e^{-2 y} \left(-1100.+3700. y-4800. y^2+3100. y^3-1100. y^4+220. y^5-24. y^6+1.3 y^7-0.029 y^8\right)\nn \\
F_{M,\Phi''}{}^{(n,p)}&=&e^{-2 y} \left(-740.+2500. y-3300. y^2+2100. y^3-760. y^4+150. y^5-16. y^6+0.82 y^7-0.013 y^8\right)\nn \\
F_{M,\Phi''}{}^{(n,n)}&=&e^{-2 y} \left(-1600.+6200. y-9200. y^2+6900. y^3-2900. y^4+710. y^5-100. y^6+7.8 y^7-0.29 y^8\right. \nn\\
 && \left. +0.0034 y^9\right)\nn \\
\ea
${}^{129}$Xe:
\ba
F_M{}^{(p,p)}&=&e^{-2 y} \left(2900.-11000. y+15000. y^2-10000. y^3+3900. y^4-840. y^5+98. y^6-5.6 y^7+0.12 y^8\right) \nn \\
F_M{}^{(p,n)}&=&e^{-2 y} \left(4000.-16000. y+24000. y^2-19000. y^3+8200. y^4-2100. y^5+300. y^6-23. y^7+0.77 y^8\right. \nn\\
&& \left. -0.0065 y^9\right)\nn \\
F_M{}^{(n,n)}&=&e^{-2 y} \left(5600.-24000. y+39000. y^2-34000. y^3+17000. y^4-4900. y^5+850. y^6-82. y^7+3.9 y^8\right. \nn\\
 && \left. -0.065 y^9+0.00034 y^{10}\right)\nn \\
F_{\Sigma '}{}^{(p,p)}&=&e^{-2 y} \left(0.00042-0.0019 y+0.0065 y^2-0.012 y^3+0.017 y^4-0.012 y^5+0.0041 y^6-0.00063 y^7\right)\nn\\
F_{\Sigma '}{}^{(p,n)}&=&e^{-2 y} \left(0.014-0.080 y+0.23 y^2-0.41 y^3+0.43 y^4-0.24 y^5+0.070 y^6-0.010 y^7+0.00058 y^8\right)\nn \\
F_{\Sigma '}{}^{(n,n)}&=&e^{-2 y} \left(0.49-3.3 y+8.8 y^2-12. y^3+9.8 y^4-4.5 y^5+1.2 y^6-0.16 y^7+0.010 y^8\right)\nn \\
F_{\Sigma ''}{}^{(p,p)}&=&e^{-2 y} \left(0.00021-0.0015 y+0.0043 y^2-0.0063 y^3+0.0049 y^4-0.0020 y^5+0.00042 y^6\right)\nn \\
F_{\Sigma ''}{}^{(p,n)}&=&e^{-2 y} \left(0.0072-0.038 y+0.082 y^2-0.097 y^3+0.068 y^4-0.027 y^5+0.0061 y^6-0.00073 y^7\right)\nn \\
F_{\Sigma ''}{}^{(n,n)}&=&e^{-2 y} \left(0.25-0.84 y+1.4 y^2-1.5 y^3+0.94 y^4-0.37 y^5+0.089 y^6-0.012 y^7+0.00076 y^8\right)\nn \\
F_{\Delta}{}^{(p,p)}&=&e^{-2 y} \left(0.038-0.090 y+0.091 y^2-0.048 y^3+0.014 y^4-0.0021 y^5+0.00012 y^6\right)\nn \\
F_{\Delta}{}^{(p,n)}&=&e^{-2 y} \left(-0.0041-0.012 y+0.012 y^2+0.00056 y^3-0.0032 y^4+0.00062 y^5\right)\nn \\
F_{\Delta}{}^{(n,n)}&=&e^{-2 y} \left(0.00046+0.0038 y+0.0091 y^2+0.0054 y^3-0.00062 y^5\right)\nn \\
F_{\Phi''}{}^{(p,p)}&=&e^{-2 y} \left(130.-300. y+280. y^2-130. y^3+30. y^4-3.5 y^5+0.16 y^6\right)\nn \\
F_{\Phi''}{}^{(p,n)}&=&e^{-2 y} \left(120.-330. y+360. y^2-190. y^3+55. y^4-8.4 y^5+0.64 y^6-0.019 y^7\right)\nn \\
F_{\Phi''}{}^{(n,n)}&=&e^{-2 y} \left(120.-360. y+440. y^2-270. y^3+93. y^4-18. y^5+1.9 y^6-0.10 y^7+0.0021 y^8\right)\nn \\
F_{M,\Phi''}{}^{(p,p)}&=&e^{-2 y} \left(-610.+1800. y-2100. y^2+1200. y^3-360. y^4+59. y^5-4.8 y^6+0.14 y^7\right)\nn \\
F_{M,\Phi''}{}^{(p,n)}&=&e^{-2 y} \left(-580.+2000. y-2600. y^2+1700. y^3-600. y^4+120. y^5-14. y^6+0.76 y^7-0.016 y^8\right)\nn \\
F_{M,\Phi''}{}^{(n,p)}&=&e^{-2 y} \left(-840.+2800. y-3500. y^2+2300. y^3-810. y^4+160. y^5-17. y^6+0.78 y^7-0.0074 y^8\right)\nn \\
F_{M,\Phi''}{}^{(n,n)}&=&e^{-2 y} \left(-800.+3000. y-4300. y^2+3100. y^3-1300. y^4+310. y^5-42. y^6+3.1 y^7-0.10 y^8\right. \nn\\
 && \left. +0.00085 y^9\right)\nn \\
F_{\Sigma ',\Delta }{}^{(p,p)}&=&e^{-2 y} \left(-0.0040+0.014 y-0.033 y^2+0.040 y^3-0.024 y^4+0.0076 y^5-0.0011 y^6\right)\nn \\
F_{\Sigma ',\Delta }{}^{(p,n)}&=&e^{-2 y} \left(0.00044+0.00083 y-0.0010 y^2+0.0067 y^3-0.00090 y^4-0.0018 y^5+0.00037 y^6\right)\nn \\
F_{\Sigma ',\Delta }{}^{(n,p)}&=&e^{-2 y} \left(-0.14+0.61 y-1.1 y^2+0.95 y^3-0.47 y^4+0.13 y^5-0.019 y^6+0.0011 y^7\right)\nn \\
F_{\Sigma ',\Delta }{}^{(n,n)}&=&e^{-2 y} \left(0.015+0.012 y-0.13 y^2+0.11 y^3+0.018 y^4-0.030 y^5+0.0050 y^6+0.00051 y^7\right. \nn\\
&& \left. -0.00012 y^8\right)\nn
\ea
${}^{131}$Xe:
\ba
F_M{}^{(p,p)}&=&e^{-2 y} \left(2900.-11000. y+15000. y^2-10000. y^3+3800. y^4-810. y^5+92. y^6-5.1 y^7+0.11 y^8\right)\nn \\
F_M{}^{(p,n)}&=&e^{-2 y} \left(4200.-16000. y+25000. y^2-20000. y^3+8600. y^4-2200. y^5+310. y^6-24. y^7+0.83 y^8\right. \nn\\
&& \left. -0.0082 y^9\right)\nn \\
F_M{}^{(n,n)}&=&e^{-2 y} \left(5900.-25000. y+43000. y^2-37000. y^3+19000. y^4-5500. y^5+980. y^6-97. y^7+4.9 y^8\right. \nn\\
 && \left. -0.096 y^9+0.00061 y^{10}\right)\nn \\
F_{\Sigma '}{}^{(p,p)}&=&e^{-2 y} \left(0.00012-0.00089 y+0.0015 y^2+0.0015 y^3-0.00069 y^4-0.0012 y^5+0.00080 y^6\right. \nn\\
 && \left. -0.00016 y^7\right)\nn\\
F_{\Sigma '}{}^{(p,n)}&=&e^{-2 y} \left(0.0045-0.039 y+0.095 y^2-0.038 y^3-0.077 y^4+0.087 y^5-0.035 y^6+0.0059 y^7\right. \nn\\
 && \left. -0.00035 y^8\right)\nn \\
F_{\Sigma '}{}^{(n,n)}&=&e^{-2 y} \left(0.18-1.6 y+5.8 y^2-9.7 y^3+9.1 y^4-4.9 y^5+1.4 y^6-0.21 y^7+0.012 y^8\right)\nn \\
F_{\Sigma ''}{}^{(p,p)}&=&e^{-2 y} \left(0.00013 y^2-0.00062 y^3+0.00088 y^4-0.00053 y^5+0.00015 y^6\right)\nn \\
F_{\Sigma ''}{}^{(p,n)}&=&e^{-2 y} \left(0.0023+0.0032 y-0.011 y^2-0.00077 y^3+0.019 y^4-0.018 y^5+0.0066 y^6-0.0011 y^7\right)\nn \\
F_{\Sigma ''}{}^{(n,n)}&=&e^{-2 y} \left(0.088+0.30 y-0.23 y^2-0.47 y^3+1.2 y^4-1.1 y^5+0.44 y^6-0.086 y^7+0.0067 y^8\right)\nn \\
F_{\Delta}{}^{(p,p)}&=&e^{-2 y} \left(0.022-0.054 y+0.053 y^2-0.026 y^3+0.0071 y^4-0.00098 y^5\right)\nn \\
F_{\Delta}{}^{(p,n)}&=&e^{-2 y} \left(0.11-0.28 y+0.31 y^2-0.19 y^3+0.062 y^4-0.010 y^5+0.00073 y^6\right)\nn \\
F_{\Delta}{}^{(n,n)}&=&e^{-2 y} \left(0.56-1.4 y+1.8 y^2-1.3 y^3+0.51 y^4-0.11 y^5+0.0097 y^6-0.00015 y^7\right)\nn \\
F_{\Phi''}{}^{(p,p)}&=&e^{-2 y} \left(100.-250. y+230. y^2-100. y^3+24. y^4-2.7 y^5+0.12 y^6\right)\nn \\
F_{\Phi''}{}^{(p,n)}&=&e^{-2 y} \left(150.-400. y+430. y^2-230. y^3+64. y^4-9.9 y^5+0.75 y^6-0.022 y^7\right)\nn \\
F_{\Phi''}{}^{(n,n)}&=&e^{-2 y} \left(200.-640. y+780. y^2-480. y^3+160. y^4-31. y^5+3.3 y^6-0.18 y^7+0.0038 y^8\right)\nn \\
F_{M,\Phi''}{}^{(p,p)}&=&e^{-2 y} \left(-550.+1700. y-1900. y^2+1100. y^3-320. y^4+51. y^5-4.0 y^6+0.11 y^7\right)\nn \\
F_{M,\Phi''}{}^{(p,n)}&=&e^{-2 y} \left(-770.+2600. y-3400. y^2+2200. y^3-790. y^4+160. y^5-17. y^6+0.96 y^7-0.021 y^8\right)\nn \\
F_{M,\Phi''}{}^{(n,p)}&=&e^{-2 y} \left(-790.+2600. y-3400. y^2+2200. y^3-770. y^4+150. y^5-16. y^6+0.77 y^7-0.0086 y^8\right)\nn \\
F_{M,\Phi''}{}^{(n,n)}&=&e^{-2 y} \left(-1100.+4100. y-5900. y^2+4400. y^3-1800. y^4+440. y^5-61. y^6+4.5 y^7-0.16 y^8\right. \nn\\
 && \left. +0.0015 y^9\right)\nn \\
F_{\Sigma ',\Delta }{}^{(p,p)}&=&e^{-2 y} \left(0.0016-0.0081 y+0.0047 y^2+0.0049 y^3-0.0061 y^4+0.0023 y^5-0.00039 y^6\right)\nn \\
F_{\Sigma ',\Delta }{}^{(p,n)}&=&e^{-2 y} \left(0.0080-0.041 y+0.027 y^2+0.020 y^3-0.038 y^4+0.020 y^5-0.0042 y^6+0.00032 y^7\right)\nn \\
F_{\Sigma ',\Delta }{}^{(n,p)}&=&e^{-2 y} \left(0.063-0.37 y+0.72 y^2-0.68 y^3+0.35 y^4-0.098 y^5+0.014 y^6-0.00077 y^7\right)\nn \\
F_{\Sigma ',\Delta }{}^{(n,n)}&=&e^{-2 y} \left(0.31-1.9 y+3.8 y^2-4.1 y^3+2.5 y^4-0.87 y^5+0.15 y^6-0.011 y^7\right)\nn
\ea

\section{Bound State Scattering Amplitudes: Momentum Space}
\label{app:nuclmat}

The calculation of scattering matrix elements for dark matter
off of atomic nuclei starting with the nucleon-level Lagrangian is
in general a complicated and difficult calculation, involving
contributions from many different spin configurations
of a multi-body system.  However, in practice, at the end of the day
most amplitudes depend on the complicated nucleus state
through only a handful of quantities, such as the total atomic number,
charge distribution, average nucleon spin, etc.  
The calculation of matrix elements for scattering of atomic nuclei
in terms of these macroscopic quantities is a fairly straightforward
application of quantum mechanics of bound states, which
we now discuss.  In the body of the paper, we have worked mostly with a partial
wave basis of operators and a nuclear shell mode basis of states, which has
several advantages.  However, there are some advantages of a momentum-space
basis for the nuclei as bound states of nucleons, not least of which is the fact
that such a basis is more familiar in particle physics.  Thus, it is somewhat
simpler to make connection with standard formula for field operators and
matrix elements.  Furthermore, it is typically easier to separate out the
center-of-mass motion of the nucleus in such a basis, which can make couplings
to the coherent motion of nucleus easier to identify.  

Our starting point is the expression for the target non-relativistic bound
state at rest:
\be
|T(0) \> &=& \sqrt{2 m_T} \sum_{s,s'} \int \frac{d^3 k}{(2\pi)^3} 
  \psi_{s,s'}(k) \frac{1}{\sqrt{2m_1 2m_2}} | k,s; -k,s' \> .
\ee
For simplicity, we are considering the bound state to be made of two particles, of mass $m_1$ and $m_2$, and the total target mass $m_T$ is approximately $m_1 + m_2$, up to a relatively small binding energy. The generalization to more particles is straightforward and in fact not very different from focusing on one constituent at a time and grouping the remaining constituents into a second much bigger constituent.  The dependence on and sum over spins $s,s'$ of the constituents will be left implicit from now on. 
The expression for the bound state $|T(k_{\rm out})\>$ with non-vanishing momentum $k_{\rm out}$ is obtained simply by shifting the velocity by $v= k_{\rm out}/m_T$:
\be
|T(k_{\rm out}) \> 
&=& \sqrt{2 m_T} \int \frac{d^3 k d^3 r}{(2\pi)^3} e^{i k \cdot r} \psi(r) \frac{1}{\sqrt{2 m_1 2 m_2}} | k + m_1 v, -k + m_2 v \> ,
\ee
where we are using the same symbol $\psi$ for the wavefunction and its Fourier transform.

Scattering matrix elements can be computed by evaluating matrix elements of interactions on the states $|T(k')\>$ constructed above.
 Our primary concern will be to understand contributions to $v$ from the internal motions of the nucleus, so
 we will consider the example where we are calculating matrix elements
of the operator
$\vec{v}^\perp N^- N^+(y) $.
\footnote{ In order for the interaction to satisfy Galilean invariance, it is important that 
$\vec{v}^\perp$
in the Lagrangian is an operator that acts non-trivially on the fields $N^\pm$, so that it gets contributions not only from the nucleus center-of-mass velocities
$v_{\rm in}, v_{\rm out}$, but also from the internal momenta $k,k'$ of the nucleon states $|k+m_1 v\>, |-k+m_2 v\>$ within the nucleus.  Explicitly, $v^\perp N^- N^+ = i \frac{ 1}{2m_N}(N^- \partial N^+ - \partial N^- N^+)$. 
}
Since we are dealing with non-relativistic physics, we
have separated out the field operator $N(y)$ into its positive- and negative-frequency parts.  They satisfy $[ N^-(y), a^\dagger_N(k)]=0, [N^-(y) , a_N(k)]  = -\frac{1}{\sqrt{2m_N}}e^{-i k \cdot y}$, and similarly for $N^+=N^{-\dagger}$.  Let us take particle 2 to be $N$, and particle 1 to be some other particle.  Then,
\be
 \< T(k_{\rm in}) | \vec{v}^\perp N^- N^+(y) | T(k_{\rm out}) \>& =&
2 m_T \int \frac{d^3 k d^3 k' d^3 r d^3 r'}{2 m_1 2 m_N} e^{i k \cdot r + i (-k + m_N v_{\rm out}) \cdot y } \psi(r) e^{-i k' \cdot r' + i (k' - m_N v_{\rm in}) \cdot y} \psi^*(r') \nn\\
  && \times  (\frac{k' + k}{m_N}  -(v_{\rm in} + v_{\rm out}) ) 2 m_1 \delta^{(3)}((k+m_1 v_{\rm out})- (k'+m_1 v_{\rm in})).
\ee
The $\delta$-function inside the integrand imposes the relation $k-k' = m_1 (v_{\rm in} - v_{\rm out})$, so
\begin{equation}
k-k' + m_N (v_{\rm in} - v_{\rm out}) = (m_1 + m_N) (v_{\rm in} - v_{\rm out})
\approx  m_T (v_{\rm in } - v_{\rm out}) \approx k_{\rm in} - k_{\rm out} =q,
\end{equation}
where the approximation $m_1 + m_N \approx m_T$ is valid as long as the binding energy of the bound state is small. 
Then, making the following change of variables,
\be
& k = \half ( \tilde{k} + \frac{m_1}{m_T} q), \ \ \ k' = \half (\tilde{k} - \frac{m_1}{m_T} q ), \ \ \ r-r' = x, \ \ \ \frac{r+r'}{2} = X,
\ee
the scattering matrix element reduces to
\be
 \< T(k_{\rm in}) | \vec{v}^\perp N^- N^+(y)| T(k_{\rm out}) \>& =&
\frac{-1}{m_N} e^{i q\cdot y} \int \frac{1}{8} d^3 \tilde{k} d^3 x d^3 X  e^{i \tilde{k} \cdot \frac{x}{2} + i q \frac{m_1}{m_T}\cdot X } \nn\\
 & \times & \left(m_T (v_{\rm in} + v_{\rm out}) \psi^*(X-\frac{x}{2}) \psi(X+ \frac{x}{2})  -i \frac{m_T}{m_N}\psi^*(X - \frac{x}{2})2 \lrpar_X \psi(X + \frac{x}{2}) \right).\nn\\
\ee
By integration by parts, we have traded a $\tilde{k}$ for a derivative with respect to $x$, which is then an asymmetric derivative $\lrpar_X = \frac{1}{2} (\lpar_X - \rpar_X)$ with respect to $X$ acting on the wavefunctions.  Doing the $d^3 \tilde{k}$ and $d^3 x$ integration is now trivial. 
We can further simplify the expression by making the approximation $m_N \ll m_1 \sim m_T$, which is accurate to about $\CO(m_N/m_T) \sim \CO(1/Z)$ for our physical cases of interest. 
Defining the density $\rho(X) = \psi^*(X) \psi(X)$, we have
\ba
&&\< T(k_{\rm in}) |  \vec{v}^\perp N^- N^+(y)| T(k_{\rm out}) \>=\\
 && \ \ \ \ \
  \frac{-e^{i q \cdot y}}{m_N} \left( m_T (v_{\rm in} + v_{\rm out}) \int d^3 X e^{i q \cdot X} \rho(X)  
 -i \frac{m_T}{m_N} \int d^3 X e^{i q \cdot X} \psi^*(X) 2\lrpar_X \psi(X) \right),\nn
\ea
  The first
term is recognized as just the bound state momentum times a form factor
\be
m_T (\vec{v}_{\rm in} + \vec{v}_{\rm out}) \int d^3 X e^{i q \cdot X} \rho(X) &=&
  \vec{K}_T F(q^2),
\ee
which reproduces the well-known fact that the form factor in the spin-independent case is the Fourier transform of the density.
The second term is more complicated, and can be evaluated in an expansion in small momentum transfer $q$.  Let us define the probability
current $\vec{J} \equiv \psi^*(X)2\lrpar_X\psi(X)$.  
This is conserved $ \partial_i J_i =0 $, and as a consequence $J^i = \partial_j(X^i J^j)$ is a total derivative and therefore its spatial average vanishes.
So, at $q=0$, we get no contribution from $J^i$.  The leading non-vanishing term in the small $q$ expansion is 
$\sim \int d^3 X i q^i  X^i J^j$. First, note that $X^{(i} J^{j)} = \frac{1}{2}\partial_k (X^i X^j J^k)$ is a total derivative, which must
have vanishing spatial average, 
so $\int d^3 X X^i J^j$ is anti-symmetric in its indices. Consequently, 
\be
\frac{m_T}{m_N} q^i \int d^3 X X^i J^j &=&
 i \frac{m_T}{m_N} q^i \epsilon^{ijk} \int d^3 X \psi^* (X) (X \times p)^k \psi(X)
  =i \frac{m_T}{m_N} q^i L_N^k \epsilon^{ijk}. 
\ee
We thus have for the full matrix element that
\begin{equation}
\< T(k_{\rm in}) | v^{\perp i} N^- N^+(y) | T(k_{\rm out}) \> =
  \frac{-e^{i q \cdot y} }{m_N} \left( m_T (v_{\rm in}^i + v_{\rm out}^i)  F(q^2) + i\frac{m_T}{m_N}\epsilon^{ijk} L_N^j q^k S(q^2)\right),
\end{equation}
where $S(q^2)$ is some form factor encoding the remaining  subleading $q^2$ dependence.
The coefficient $\frac{m_T}{m_N}$ of the final term is the effective $A$ enhancement factor discussed
in the body of the paper.  In the relevant case of atomic nuclei, where many $N$ particles are present
in the bound state, this amplitude should be summed over all such particles. This results
in an additional coherence factor for the first (spin-independent) term, and sends $L_N \rightarrow
\< L_N \>$ for the second (spin-dependent) term.

\section{Non-relativistic Matching}
\label{app:matching}

Let us reduce the full set of relativistic operators into non-relativistic form.  For this purpose, we will label
the incoming(outgoing) momentum of the dark matter $\chi$ to be $p^\mu$($p'^\mu$), and the incoming (outgoing)
momentum of the nucleon $N$ to be $k^\mu$($k'^\mu$).  The momentum transfer $q$ is defined to be
$q^\mu= p'^\mu - p^\mu = k'^\mu - k^\mu$.  We will also define $P^\mu = p^\mu + p'^\mu$ and $K^\mu = k^\mu + k'^\mu$.
The velocity $\vec{v} = \vec{v}_{\chi, in} - \vec{v}_{N,in} =  \frac{\vec{p}}{m_\chi} - \frac{\vec{k}}{m_N} = -\frac{\vec{q}}{2\mu_N}
  + \frac{\vec{P}}{2 m_\chi} - \frac{\vec{K}}{2m_N}$ is defined so that it is a kinematic invariant ($\mu_N$ is the reduced dark-matter-nucleon
  mass).  
Since there are
2 scalar operators ($\bar{\chi} \chi$ and $ \bar{\chi} \gamma^5 \chi$) and 4 vector operators
($P^\mu \bar{\chi} \chi$, $P^\mu \bar{\chi} \gamma^5 \chi$, $\bar{\chi} i \sigma^{\mu\nu} q_\nu \chi$, and
$\bar{\chi} \gamma^\mu \gamma^5 \chi$), there are $2^2 + 4^2 = 20$ possible combinations of operators: 

\begin{center}
\begin{tabular}{c | c | c}
\hline
$\bar{\chi} \chi \bar{N} N $ & $4 m_\chi m_N 1_\chi 1_N$ & $ 4 m_\chi m_N \CO_1$ \\
$i \bar{\chi} \chi \bar{N} \gamma^5 N$ & $-4 m_\chi i q \cdot S_N$ &  $-4 m_\chi \CO_{10} $ \\
$i \bar{\chi} \gamma^5 \chi \bar{N} N $ & $4 m_N i q \cdot S_\chi $ & $  4 m_N \CO_{11} $ \\
$ \bar{\chi} \gamma^5 \chi \bar{N} \gamma^5 N$ & $ -4  q \cdot S_\chi q \cdot S_N$ & $ -4 \CO_6$ \\
$P^\mu \bar{\chi} \chi K_\mu \bar{N} N $ & $(4 m_\chi m_N)^2 1_\chi 1_N $ & $ (4 m_\chi m_N)^2 \CO_1$\\
$P^\mu \bar{\chi} \chi \bar{N} i \sigma_{\mu \alpha} q^\alpha N $  &  $ -(4 m_\chi^2) q^2 + 16 m_N m_\chi^2 i v^\perp \cdot (q \times S_N) $ &  $-4 m_\chi^2 q^2 \CO_1 - 16 m_N m_\chi^2 \CO_3$ \\
$P^\mu \bar{\chi} \chi \bar{N} \gamma_\mu \gamma^5 N $ & $-16 m_N m_\chi^2 v^\perp \cdot S_N $ & $ -16 m_N m_\chi^2 \CO_7 $ \\
$i P^\mu \bar{\chi} \chi K_\mu \bar{N} \gamma^5 N $ & $-16 m_\chi^2 m_N i  q \cdot S_N $ & $ -16 m_\chi^2 m_N \CO_{10} $ \\
$\bar{\chi} i \sigma^{\mu\nu} q_\nu \chi K_\mu \bar{N} N $ & $(2m_N)^2 q^2 -16 m_N^2 m_\chi i v^\perp \cdot (q \times S_\chi) $ & $4 m_N^2 q^2 \CO_1 +16 m_N^2 m_\chi \CO_5$ \\
$\bar{\chi} i \sigma^{\mu\nu} q_\nu \chi \bar{N} i \sigma_{\mu\alpha} q^\alpha N $ & $16 m_\chi m_N(q \times S_\chi)\cdot (q \times S_N)$ & $16 m_N m_\chi (q^2 \CO_4 - \CO_6)$ \\
$\bar{\chi} i \sigma^{\mu\nu} q_\nu \chi \bar{N} \gamma^\mu \gamma^5 N $ & $ 16 m_N m_\chi i S_N \cdot (q \times S_\chi) $ & $-16 m_N m_\chi \CO_9$ \\
$i \bar{\chi} i \sigma^{\mu\nu} q_\nu \chi K_\mu \bar{N} \gamma^5 N $ & $4m_N(-q^2 +4 m_\chi  i v^\perp \cdot (q \times S_\chi)) i  q \cdot S_N $ & $4m_N \CO_{10}  (-q^2 - 4 m_\chi \CO_5) $ \\
$\bar{\chi} \gamma^\mu \gamma^5 \chi K_\mu \bar{N} N $ & $16 m_N^2 m_\chi v^\perp \cdot S_\chi $ & $ 16 m_N^2 m_\chi \CO_8 $ \\
$\bar{\chi} \gamma^\mu \gamma^5 \chi \bar{N} i \sigma_{\mu\alpha} q^\alpha N $ & $16m_\chi m_N i S_\chi \cdot (q \times S_N) $ & $ 16 m_\chi m_N \CO_9$ \\
$\bar{\chi} \gamma^\mu \gamma^5 \chi \bar{N} \gamma^\mu \gamma^5 N $ & $ -16 m_N m_\chi S_\chi  \cdot S_N $ & $-16 m_N m_\chi \CO_4$\\
$i \bar{\chi} \gamma^\mu \gamma^5 \chi K^\mu \bar{N} \gamma^5 N $ & $-16 m_\chi m_N v^\perp \cdot S_\chi i q \cdot S_N $ & $ -16 m_\chi m_N \CO_{10} \CO_8$ \\
$i P^\mu \bar{\chi} \gamma^5 \chi K_\mu \bar{N} N $ & $ 16 m_N^2 m_\chi i q \cdot S_\chi $ & $ 16 m_N^2 m_\chi \CO_{11} $\\
$iP^\mu \bar{\chi} \gamma^5 \chi \bar{N} i \sigma_{\mu \alpha} q^\alpha N $ & $(4 m_\chi) (iq \cdot S_\chi) (-q^2 +4 m_N i v^\perp \cdot (q \times S_N)) $ &  $4 m_\chi \CO_{11} (-q^2 - 4 m_N  \CO_3 ) $ \\
$i P^\mu \bar{\chi} \gamma^5 \chi \bar{N} \gamma_\mu \gamma^5 N $ & $ -16 m_\chi m_N (i q \cdot S_\chi) v^\perp \cdot S_N $ &  $- 16 m_\chi m_N \CO_{11} \CO_7$ \\
$P^\mu \bar{\chi} \gamma^5 \chi K_\mu \bar{N} \gamma^5 N $ & $ - 16 m_N m_\chi q\cdot S_\chi q \cdot S_N$ & $-16 m_N m_\chi \CO_6$ \\
\hline
\end{tabular}
\end{center}

Many combinations have not been included since they are equivalent by the equations of motion, and the above terms
tend to give simpler non-relativistic pieces.  The most commonly used such combination is the vector interaction $\gamma^\mu$,
which can be written in terms of the above by using the Gordon identity:
\ba
\bar{N} \gamma^\mu N &=& \frac{1}{2 m_N} \bar{N} \left( K^\mu + i \sigma^{\mu\nu} q_\nu \right) N .
\ea
Note that every non-relativistic operator occurs in the above table, except for $\CO_2$,
which appear if there are cancellations in the leading 
pieces, for instance through the linear combination $(4 m_N m_\chi \bar{\chi} \chi \bar{N} N - P^\mu \bar{\chi} \chi K_\mu \bar{N} N)$.

\parskip 0pt

\section{Single-particle Operators} 
Four of the six operators introduced in section 3 are familiar from standard treatments of semi-leptonic electroweak interactions \cite{Walecka75,Donnelly79}, 
$M_{JM}(q \vec{x}), ~\Delta_{JM}(q \vec{x}),~
\Sigma^\prime_{JM}(q \vec{x}),~ \mathrm{and~}\Sigma^{\prime \prime}_{JM}(q \vec{x})$.
The matrix elements of these operators between single-particle harmonic oscillator states, the most common basis
for nuclear physics calculations, can be evaluated analytically, yielding explicit forms for the nuclear form factors governing DM scattering.  A Mathematica script \cite{Cecilia08} and tables \cite{Donnelly79} are available.  The remaining two operators are a symmetrized form of $\Phi^\prime_{JM}$ and $\Phi^{\prime \prime}_{JM}$,  operators originally
introduced by Serot \cite{Serot78} in his treatment of $1/M^2$
corrections to currents and charges.
(Operators that differ from their standard forms because of symmetrization are denoted here by a tilde.  The underlying currents from which these operators
are derived have a simple behavior under exchange of bra and ket that, unfortunately, is sometimes lost in the conventional
definitions of the weak nuclear operators.   The re-definitions used here simply restore that symmetry, which is important in dividing elastic operators into those
that either break or preserve time-reversal invariance.)
The single-nucleon matrix elements of these operators are real, with the conventions of \cite{Donnelly79}, and transform as follows
\begin{equation}
\langle n (l ~1/2) j || \hat{O}_J || n^\prime (l^\prime ~1/2) j^\prime \rangle = (-1)^{j^\prime - j} \langle n^\prime (l^\prime~ 1/2) j^\prime || \hat{O}_J || n (l ~1/2) j \rangle,
~~~\hat{O} = M,~ \Delta,~\Sigma^\prime,~\Sigma^{\prime \prime},~\tilde{\Phi}^\prime,~\Phi^{\prime \prime}
\end{equation}
where $n$, $l$, $s=1/2$, and $j$ are the nodal quantum number, orbital angular momentum, spin, and total angular momentum of the nucleon state.
This interchange property is related to the time-reversal character of the operator, as we discuss below.  A Mathematica script, 
generalized from \cite{Cecilia08}, to evaluate harmonic oscillator matrix elements of these and all other operators discussed in this appendix is available from the authors.\\
 
\subsection{ Matrix Elements and the One-body Density Matrix}
 Here we address the evaluation of the nuclear matrix elements, including their isospin dependence and the restrictions imposed by time reversal and parity.
 We represent operators as  $ a_I \hat{O}^I_J$, where $a_I$ is the overall coupling strength.  All
 nuclear matrix elements appearing in the final results of section 3 are singly reduced -- reduced
 in angular momentum only.  Thus making the suppressed isospin now explicit, and assuming that the nuclear
 ground state has good isospin (thus ignoring charge symmetry breaking)
 \begin{eqnarray}
\langle J_i; T_i M_{T_i}||  \sum_{i=1}^A a_I  \hat{O}_J(i)~ || ~J_i; T_i M_{T_i} \rangle \equiv  \langle J_i; T_i M_{T_i}~ || ~\sum_{i=1}^A (a_I^{T=0}+ a_I^{T=1} \tau_3(i))~ \hat{O}_J^I(i) ~ || ~J_i; T_i M_{T_i} \rangle && \nonumber \\
  ={a_I^{T=0} \over [T_i] } \langle J_i; T_i ~\vdots \vdots~ \sum_{i=1}^A  \hat{O}_J^I(i) ~ \vdots \vdots ~J_i; T_i  \rangle+ {a_I^{T=1} \over [T_i]} {M_{T_i} \over \sqrt{T_i(T_i+1)}}\langle J_i; T_i ~ \vdots \vdots~ \sum_{i=1}^A \hat{O}_J^I(i) \tau(i)~ \vdots \vdots ~J_i; T_i M_{T_i} \rangle~~~~&&
 \end{eqnarray}
 where $\vdots \vdots$ denotes a many-body matrix element reduced in both angular momentum and isospin.  If the operator $O_J^I$ is a one-body operator, 
 then
 \begin{eqnarray}
 \label{eq:densitymatrix}
 \langle J_i; T_i ~\vdots \vdots~ \sum_{i=1}^A  \hat{O}_J^I(i) ~ \vdots \vdots ~J_i; T_i  \rangle &=& \sum_{|\alpha|,|\beta|} \Psi^{J;T=0}_{i;i} (|\alpha| ,|\beta|)~ \langle |\alpha| ~\vdots \vdots \hat{O}_J \vdots \vdots~ |\beta| \rangle \nonumber \\
 &=& \sqrt{2} \sum_{|\alpha|,|\beta|} \Psi^{J;T=0}_{i;i} (|\alpha|,|\beta|)~\langle |\alpha|~ || \hat{O}_J ||~ |\beta| \rangle \nonumber \\
  \langle J_i; T_i ~\vdots \vdots~ \sum_{i=1}^A  \hat{O}_J^I(i) \tau(i) ~ \vdots \vdots ~J_i; T_i  \rangle &=& \sum_{|\alpha| ,|\beta|} \Psi^{J;T=1}_{i;i}(|\alpha| ,|\beta|)~ \langle |\alpha| ~\vdots \vdots \hat{O}_J \tau \vdots \vdots~ |\beta| \rangle \nonumber \\
  &=& \sqrt{6} \sum_{|\alpha|,|\beta|} \Psi^{J;T=1}_{i,i}(|\alpha|,|\beta|)~ \langle |\alpha| ~|| \hat{O}_J ||~ |\beta| \rangle
  \nonumber \\
  ~~
  \end{eqnarray}
  where $\langle |\alpha| ~|| \hat{O}_J ||~ |\beta| \rangle$ and $\langle |\alpha|~ \vdots \vdots \hat{O}_J \tau \vdots \vdots~ |\beta| \rangle$ represent single-particle space/spin matrix
  elements reduced in angular momentum and space/spin/isospin matrix elements reduced in both angular momentum and isospin.  The one-body density
  matrix coefficients $ \Psi^{J;T}_{f;i} (|\alpha|,|\beta|)$ are labeled by the transition
  $i \rightarrow f$, angular momentum $J$ and isospin $T$ carried by the transition operator, and by the nonmagnetic single-particle quantum numbers
  $| \alpha |$ and $| \beta |$.  (That is, $\beta = \{n,l,j,m_j\}$ for a space/spin matrix element or $\{n,l,j,m_j,m_t\}$ for a space/spin/isospin matrix element,
  while $| \beta | = \{n,l,j\}$.) The sums extend over a 
  complete set of single-particle matrix elements.   Equation (\ref{eq:densitymatrix}) is exact for a one-body
  operator, regardless of the complexity of the many-body wave functions.  It factors the many-body matrix element of such an operator into a sum 
  single-particle matrix elements multiplying numerical coefficients, the density matrix.   All of the operator physics is contained in the former, while all of
  the nuclear structure physics is isolated in the latter.\\
  
  The singly-reduced matrix elements $\langle \alpha || \hat{O}_J || \beta \rangle$ needed for the present dark matter elastic scattering can be conveniently
  evaluated for a harmonic oscillator single-particle
  basis, where $| \beta \rangle = | n(l~1/2)j m_j \rangle$.  The results can be evaluated analytically, producing form factors of the form $e^{-y} p(y)$ where $y=(qb/2)^2$,
  $b$ is the oscillator parameter, and $p(y)$ is a polynomial.\\
  
  \noindent
  \subsubsection{Symmetry constraints}
  The density matrix is defined in second-quantization as
  \begin{equation}
  \Psi^{J;T}_{f;i}(|\alpha|,|\beta|) = {1 \over [J] [T]} \langle J_f; T_f~\vdots \vdots \left[ a^\dagger_{|\alpha|} \otimes \tilde{a}_{|\beta|} \right]_{J;T} \vdots \vdots~J_i;T_i \rangle
  \end{equation}
  where $\tilde{a}_{|\beta|, m_j ,m_t }\equiv (-1)^{j_\beta -m_j + 1/2-m_t} a_{|\beta|,-m_j,-m_t}$.  With the conventions used here \cite{Walecka75,Donnelly79} the density
  matrix is real for parity- and time-reversal-symmetric interactions.  Consequently, for elastic transitions it follows that
  \begin{equation}
    \Psi^{J;T}_{i;i}(|\alpha|,|\beta|) =  \Psi^{J;T}_{i;i}(|\beta|,|\alpha|) (-1)^{j_\alpha - j_\beta}
  \end{equation}
  so that Eq. (\ref{eq:densitymatrix}) can be rewritten as
  \begin{eqnarray}
 && \langle J_i; T_i ~\vdots \vdots~ \sum_{i=1}^A  \hat{O}_J^I(i) ~ \vdots \vdots ~J_i; T_i  \rangle = \nonumber \\
 && \sqrt{2} \sum_{|\alpha|,|\beta|} \Psi^{J;T=0}_{i;i} (|\alpha|,|\beta|)~{1\over 2} \left( \langle |\alpha|~ || \hat{O}_J ||~ |\beta| \rangle +(-1)^{j_\alpha-j_\beta} \langle |\beta|~ || \hat{O}_J ||~ |\alpha| \rangle \right) \nonumber \\
 && \langle J_i; T_i ~\vdots \vdots~ \sum_{i=1}^A  \hat{O}_J^I(i) \tau(i) ~ \vdots \vdots ~J_i; T_i  \rangle = \nonumber \\
 &&  \sqrt{6} \sum_{|\alpha|,|\beta|} \Psi^{J;T=1}_{i,i}(|\alpha|,|\beta|) ~{1 \over 2} \left(  \langle |\alpha| ~|| \hat{O}_J ||~ |\beta| \rangle+(-1)^{j_\alpha-j_\beta} \langle |\beta| ~|| \hat{O}_J ||~ |\alpha| \rangle \right)
  \end{eqnarray}
  Now all of the operators as conventionally defined for semi-leptonic weak interactions \cite{Walecka75,Donnelly79,Serot78} have real matrix elements with \cite{Walecka75,Donnelly79}
  phase conventions.   The contributing operators used here have been defined so that they operate symmetrically on bra and ket.  Consequently it can be 
  shown that the single-particle matrix elements transform as \cite{Donnelly79}
  \begin{equation}
   \langle |\alpha| ~|| \hat{O}_J ||~ |\beta| \rangle=(-1)^\lambda \langle |\beta| ~|| \hat{O}_J ||~ |\alpha| \rangle 
   \end{equation}
   where $\lambda$ is either $j_\alpha + j_\beta$ or $j_\alpha - j_\beta$.  This gives us the immediate result that the only electroweak operators that can
   contribute to elastic transitions are those transforming with $\lambda= j_\alpha - j_\beta$, which determines the subset of operators introduced
   at the beginning of this appendix:
   $M_J,~\Delta_J,~\Sigma^\prime_J,~\Sigma^{\prime \prime}_J,~\tilde{\Phi}^\prime_J,~\mathrm{and~}\Phi^{\prime \prime}$.   If we consider the constraints
   of parity separately, this set is further restricted to the even multipoles of $M_J,~\tilde{\Phi}^\prime_J,~\mathrm{and~}\Phi^{\prime \prime}$ and the
   odd multipoles of $\Sigma^\prime_J,~\Sigma^{\prime \prime}_J,~\mathrm{and~}\tilde{\Phi}^\prime_J$.  This end result does not depend on conventions.\\
   
   \noindent
 \subsubsection{Nuclear structure}
 While in principle Eq. (\ref{eq:densitymatrix}) is an exact expression for the many-body matrix elements of the contributing
   operators, in practice, for the complex nuclei of interest to dark matter experiments, the density matrix must be taken from a nuclear model.
   This entails a truncation of the sums over $|\alpha|$ and $|\beta|$ to some finite Hilbert space, e.g., the active valence shells of a shell model,
   as well as the use of an effective interaction within that valence space, to correct the bare nuclear Hamiltonian for
   the effects of the truncation.  The field lacks exact techniques for calculating the effective interaction or the corresponding corrections
   to the operators: in practice, most work involves phenomenological adjustments.   Some qualitative checks on the resulting density 
   matrix can be made by calculating observables other than
   dark matter responses, such as electron scattering elastic charge and (if $J_i \ge 1/2$) magnetic form factors (or, in the long wavelength limit,
   the nuclear magnetic moment).
   
In the present study, the density matrices needed for the targets we considered came from shell-model calculations
employing moderately large bases and effective interactions that were either determined from global fits to data
or based on theoretical input, such as realistic g-matrices, supplemented by empirically terms, adjusted to experiment.
Potential tests of these density matrices include observables such as charge and magnetic elastic form factors and
magnetic and quadrupole moments.  \\

\subsubsection{ Non-recoil Contributions}
The effective theory immediately identifies the Galilean-invariant operators, which in the point nucleus limit define
all responses connected with target recoil.  As explained in section 3, this leaves contributions associated with
$v_N^\perp$, that is, connected with the intrinsic degrees of freedom within the nucleus that would
normally be described by introducing the (A-1) Jacobi coordinates.  Under the assumption that the dark matter 
couples to to nucleus through local one-body interactions
with the nucleons within the nucleus, the desired non-recoil intrinsic operators can generally be derived.  A 
specific example was given in section 3, where the recoil and intrinsic components of the axial charge operator
were explicitly identify.  Similar identifications are possible for all of the interactions where this separation 
should be done.

In fact, common practice is to avoid such a separation because the explicit intrinsic operators tend to be far more
complicated than their original one-body forms.  Instead, the separation is addressed through the nuclear
wave functions.   While the wave functions are also evaluated in an over-complete basis consisting of 
single-particle Slater determinants, numerical means are available to remove the center-of-mass motion in the
harmonic oscillator shell model.  This is done by projecting out the center-of-mass motion by adding a large
multiple of the center-of-mass Hamiltonian to the interaction.  The resulting wave functions then have the
center-of-mass in a 1s state: additional ``copies" of intrinsic states where the center-of-mass is in an excited
state are ``blown out."   This technique requires that calculation be done in separable bases: some of the
calculations reported here were done in such bases, but others required basis truncations that destroy
the separability.  Furthermore, even if the separation is done exactly, the use of the one-body operator forms
between such states does not provide the exact non-recoil result.  Although center-of-mass components in the
one-body operator cannot generate excitations, they do generate an unwanted center-of-mass form factor.

In fact, all of these considerations are rather technical given other uncertainties in nuclear wave functions.
The identification of the Galilean-invariant recoil corrections -- a trivial task in the effective theory -- is in most
applications more important than whether the nuclear physics calculations are done with operators that 
are precisely intrinsic.  The discussion of section 3 provided an illustrative example:  the entire axial-charge
contribution to the elastic scattering of dark matter was due to $v_T^\perp$:  there are no  contributions from
$v_N^\perp$ due to the combined constraints of time reversal invariance and parity.
 
 \subsection{Operators Needed for Inelastic DM Scattering}
 Certain target nuclei may have thresholds for exciting low-lying states that are sufficiently low that
 dark matter interactions can be inelastic.  The full scattering probability, analogous to Eq. (\ref{eq:Ham}), will
 be presented elsewhere.   The initial and final nuclear states are then distinct, symmetry under time reversal no
longer constrains operators or their multi-polarities (though parity selection rules remain, depending on the definite parity of the final state, and are used in some simplifications below).  Here we point out, for those familiar with standard formulations
of semi-leptonic weak interactions, that a larger set of
multipole operators must be defined, including the remaining three of the seven basic electroweak operators of \cite{Walecka75,Donnelly79}
\begin{eqnarray}
\Delta^\prime_{JM}(q \vec{x}) &\equiv& -i \left\{ {1 \over q} \vec{\nabla} \times \vec{M}_{JJ}^M(q \vec{x}) \right\} \cdot {1 \over q} \vec{\nabla} \nonumber \\
\Sigma_{JM}(q \vec{x}) &\equiv& \vec{M}_{JJ}^M(q \vec{x}) \cdot \vec{\sigma} \nonumber \\
\tilde{\Omega}_{JM}(q \vec{x}) &\equiv& \Omega_{JM}(q \vec{x}) + {1 \over 2} \Sigma_{JM}^{\prime \prime}(q \vec{x}) = M_{JM}(q \vec{x}) \vec{\sigma} \cdot {1 \over q} \vec{\nabla} + {1 \over 2} \left\{ {1 \over q} \vec{\nabla} M_{JM}(q \vec{x}) \right\} \cdot \vec{\sigma}.
\end{eqnarray}
(Note that the symmetrized operator $\tilde{\Omega}$ is identical to $\Omega^\prime$ defined and tabulated in \cite{Donnelly79}.  We introduce the
alternative notation to avoid confusion, as $\Omega^\prime$ is defined in \cite{Serot78} as another operator.   Here we consistently designate with a tilde 
all operators that require symmetrization to restore a simple behavior of matrix elements
under interchange of bra and ket.)  In addition, two symmetrized operators related to Serot's $\Phi_{JM}$  and $\Delta^{\prime \prime}_{JM}$ are also needed
\begin{eqnarray}
\tilde{\Phi}_{JM}(q \vec{x}) &\equiv& \Phi_{JM}(q \vec{x}) - {1 \over 2} \Sigma^\prime_{JM}(q \vec{x}) = i \vec{M}_{JJ}^M(q \vec{x}) \cdot \left( \vec{\sigma} \times {1 \over q} \vec{\nabla} \right) +{i \over 2} \left\{ {1 \over q} \vec{\nabla} \times \vec{M}_{JJ}^M(q \vec{x}) \right\} \cdot \vec{\sigma} \nonumber \\
\tilde{\Delta}^{\prime \prime}_{JM}(q \vec{x}) &\equiv& \Delta^{\prime \prime}_{JM}(q \vec{x}) -{1 \over 2} M_{JM}(q \vec{x}) = \left\{ {1 \over q} \vec{\nabla} M_{JM}(q \vec{x}) \right\} \cdot {1 \over q} \vec{\nabla} -{1 \over 2} M_{JM}(q \vec{x})
\end{eqnarray}
The single-nucleon matrix elements of these operators are again real, under the conventions of \cite{Donnelly79}, but transform with a different sign from those
needed for elastic DM scattering,
\begin{equation}
\langle n (l ~1/2) j || \hat{O}_J || n^\prime (l^\prime ~1/2) j^\prime \rangle = (-1)^{j^\prime + j} \langle n^\prime (l^\prime~ 1/2) j^\prime || \hat{O}_J || n (l ~1/2) j \rangle,~~\hat{O} = \Delta^\prime,~\Sigma,~\tilde{\Omega},~\tilde{\Phi},~\tilde{\Delta}^{\prime \prime}.
\end{equation}
This sign is related to operator behavior under time reversal, and thus is the reason that these operators do not appear
in our treatment of elastic DM interactions.

\bibliography{DMref}{}
\bibliographystyle{utphys}

\end{document}